\documentclass[3p,10pt]{elsarticle}
\usepackage{hyperref}
\usepackage{xcolor}     
\usepackage{lineno}
\usepackage{enumitem}
\biboptions{sort&compress}

\usepackage{booktabs}
\usepackage{braket}
\usepackage{tabularx}
\usepackage{multirow}

\usepackage{siunitx}
\usepackage{amsmath, amssymb} 

\usepackage{bm} 
\usepackage[version=4]{mhchem}
\usepackage{placeins}

\usepackage[font=small,labelfont=bf,labelsep=period,format=plain]{caption}
\begin{document}
\begin{frontmatter}
\title{Exceptional Points in Photonics: \\ From Non-Hermitian Physics to Applications}

\author[1,2]{Fan Zhang}
\author[1,2]{Nikolay Solodovchenko}
\author[3,4]{Dmitrii N. Maksimov}
\author[1]{Xuchen Wang}
\author[1]{Mingzhao Song}
\author[5]{Filippo Capolino}
\author[6]{C. T. Chan}
\author[1,2]{Andrey Bogdanov\corref{cor1}}

\cortext[cor1]{Corresponding author: bogdan.taurus@gmail.com}
\affiliation[1]{organization={Qingdao Innovation and Development Center, Harbin Engineering University},
            addressline={Sansha road 1777}, 
            city={Qingdao},
            postcode={266000}, 
            state={Shandong},
            country={China}}
\affiliation[2]{organization={School of Physics and Engineering, ITMO University},
            city={St. Petersburg},
            postcode={197101}, 
            country={Russia}}
\affiliation[3]{organization={IRC SQC, Siberian Federal University},
            city={Krasnoyarsk},
            postcode={660041}, 
            country={Russia}}
\affiliation[4]{organization={LV Kirensky Institute of Physics, Federal Research Center KSC SB RAS},
            city={Krasnoyarsk},
            postcode={660036}, 
            country={Russia}}
\affiliation[5]{organization={Department of Electrical Engineering and Computer Science, University of California},
            city={Irvine},
            postcode={92697}, 
            state={California},
            country={USA}}
\affiliation[6]{organization={Department of Physics, The Hong Kong University of Science and Technology},
            city={Hong Kong},
            country={China}}
\begin{abstract}
Open photonic systems provide a versatile platform for non-Hermitian physics, enabling control over complex spectra, transport, and light–matter interactions. Exceptional points (EPs), at which eigenvalues and eigenvectors coalesce and the governing operator becomes defective, play a central role because they combine branch-point spectral topology, nonanalytic perturbative response, and controllable eigenstate conversion. This Review provides a unified framework for EP photonics by systematically distinguishing exceptional degeneracies according to the underlying operator, spectral variable, boundary conditions, and experimentally accessible observables. We discuss Hamiltonian EPs, absorbing EPs associated with scattering zeros, real-frequency scattering-matrix and Jones-matrix EPs, Bloch and Floquet EPs, and Liouvillian EPs in open quantum systems. We review their spectral topology, static and dynamical encircling, higher-order exceptional structures, and coexistence with bound states in the continuum, together with applications in sensing, lasing, coherent absorption, directional scattering, polarization and wavefront control, nonlinear optics, optical storage, nonreciprocal photonics, and quantum photonics. We also critically assess the current limitations, practical challenges, and future perspectives of EP-based photonic technologies, with particular attention to robustness, noise, scalability, and experimentally measurable performance.
\end{abstract}


\end{frontmatter}

\tableofcontents
\newpage
\section{Introduction}
\label{sec: Introduction}
The spectral theory of linear operators serves as the mathematical foundation for our understanding of quantum mechanics~\cite{vonNeumann1955mathematical, dirac1958principles} and classical wave physics~\cite{courant1953methods,fedorov1968theory}. Unitary time evolution in an isolated quantum system imposes Hermiticity on the Hamiltonian, ensuring real eigenvalues and an orthogonal basis of eigenstates~\cite{landau2013quantum}. However, realistic physical systems are invariably open, exchanging energy, matter, or information with their environment~\cite{rotter2009non, moiseyev2011non}, a context effectively described by non-Hermitian Hamiltonians. The most striking features of this non-Hermitian regime are exceptional points (EPs)~\cite{miri2019exceptional}, singular degeneracies at which both eigenvalues and eigenvectors coalesce, leading to a defect in basis completeness and the formation of a Jordan block structure, rendering the underlying linear operator non-diagonalizable. While originally identified as mathematical curiosities in perturbation theory~\cite{Kato1966Perturbation}, their physical implications have been articulated in a series of foundational analyses~\cite{heiss2004exceptional,berry2004physics,heiss2012physics}. Over the past decade, EPs have transitioned from theoretical abstractions to the central theme of non-Hermitian photonics, fueled by the unique ability of optical platforms to control gain, loss, and synthetic gauge fields~\cite{feng2017non, el2018non, ozdemir2019parity}.

Photonics has proven to be an ideal platform for exploring EP physics because Maxwell’s equations in conducting or amplifying media map onto Schrödinger-like wave equations with complex potentials~\cite{el2007theory,makris2008beam}. This correspondence has enabled the realization of parity-time (PT) symmetry in optical systems, where balanced gain and loss distributions allow for controlled phase transitions across the EPs~\cite{guo2009observation,ruter2010observation}. The rapidly expanding range of photonic implementations has also produced a proliferation of quantities described as exceptional degeneracies. Depending on the physical formulation, defectiveness may occur in a propagation generator, an effective resonant Hamiltonian defined by outgoing boundary conditions~\cite{peng2014parity}, an incoming solution associated with a scattering zero~\cite{sweeney2019perfectly, wang2021coherent}, a real-frequency scattering~\cite{chong2011pt, lin2011unidirectional} or Jones response matrix~\cite{lawrence2014manifestation}, a Bloch~\cite{zhen2015spawning} or Floquet~\cite{longhi2017floquet,park2022revealing,liu2024floquet} representation, or a Liouvillian superoperator governing open quantum dynamics~\cite{minganti2019quantum}. These descriptions can coexist in the same structure, although their exceptional degeneracies need not occur at the same frequency or at the same values of the control parameters. Identifying the defective operator, its spectral variables, and its experimentally accessible observables is therefore essential for comparing results across photonic platforms.

Beyond their local spectral singularity properties, EPs behave as topological defects in parameter space. Around an EP, eigenvalues and eigenvectors form a multi-sheeted Riemann structure, so adiabatic encircling can result in a state permutation rather than merely accumulate a geometric phase~\cite{heiss1999phases, dembowski2001experimental}. This nontrivial monodromy underlies topological state transfer and robust mode switching. EP topology also introduces dynamical encircling, where finite-time evolution and the breakdown of adiabaticity allow for chiral state selection determined by the direction of encirclement~\cite{doppler2016dynamically,xu2016topological}. In periodic systems, exceptional singularities can further extend into momentum space, forming exceptional lines, rings and higher-dimensional structures~\cite{bergholtz2021exceptional}. These features connect EP physics to non-Hermitian band topology and to departures from conventional Hermitian bulk-boundary correspondence, including point-gap winding, non-Bloch spectra, and the non-Hermitian skin effect~\cite{yao2018edge,kunst2018biorthogonal}.

The same spectral and topological properties have motivated a broad range of EP-enabled functionalities, but they also require careful performance criteria. Sensing and metrology provide a clear example~\cite{chen2017exceptional,hodaei2017enhanced,lai2019observation,hokmabadi2019non}. Perturbation theory near an EP predicts root-law scaling of complex eigenfrequency splittings, which theoretically offers infinite responsivity to infinitesimal perturbations~\cite{wiersig2014enhancing,wiersig2016sensors,wiersig2020review}. Yet enhanced eigenvalue responsivity does not by itself translate to superior measurement precision~\cite{langbein2018no,lau2018fundamental}. Because the eigenbasis collapses at an EP, mode non-orthogonality can amplify noise through the Petermann factor~\cite{petermann1979calculated,wang2020petermann} and reduce the signal-to-noise ratio~\cite{loughlin2024exceptional,mortensen2018fluctuations}. Rigorous assessment of EP-based sensors therefore requires moving beyond simple eigenvalue splitting to a full analysis of the estimator performance and noise processes.

A related recurring issue concerns reciprocity and directionality. EP-enabled structures can exhibit strongly asymmetric responses, including unidirectional reflectionless behavior~\cite{lin2011unidirectional,regensburger2012parity,feng2013experimental} and extreme polarization selectivity~\cite{lawrence2014manifestation}. These effects are sometimes discussed together with nonreciprocal transport. In linear time-invariant systems without explicit time-reversal-symmetry breaking, however, Lorentz reciprocity still constrains the scattering response~\cite{potton2004reciprocity, jalas2013and}. It is therefore important to distinguish reciprocal asymmetric transmission from strict nonreciprocity. Throughout this Review, we use nonreciprocity in its electrodynamic sense, and otherwise refer to directional or asymmetric response under reciprocity constraints.

\begin{figure}[htbp]
\centering
\includegraphics[width=\textwidth]{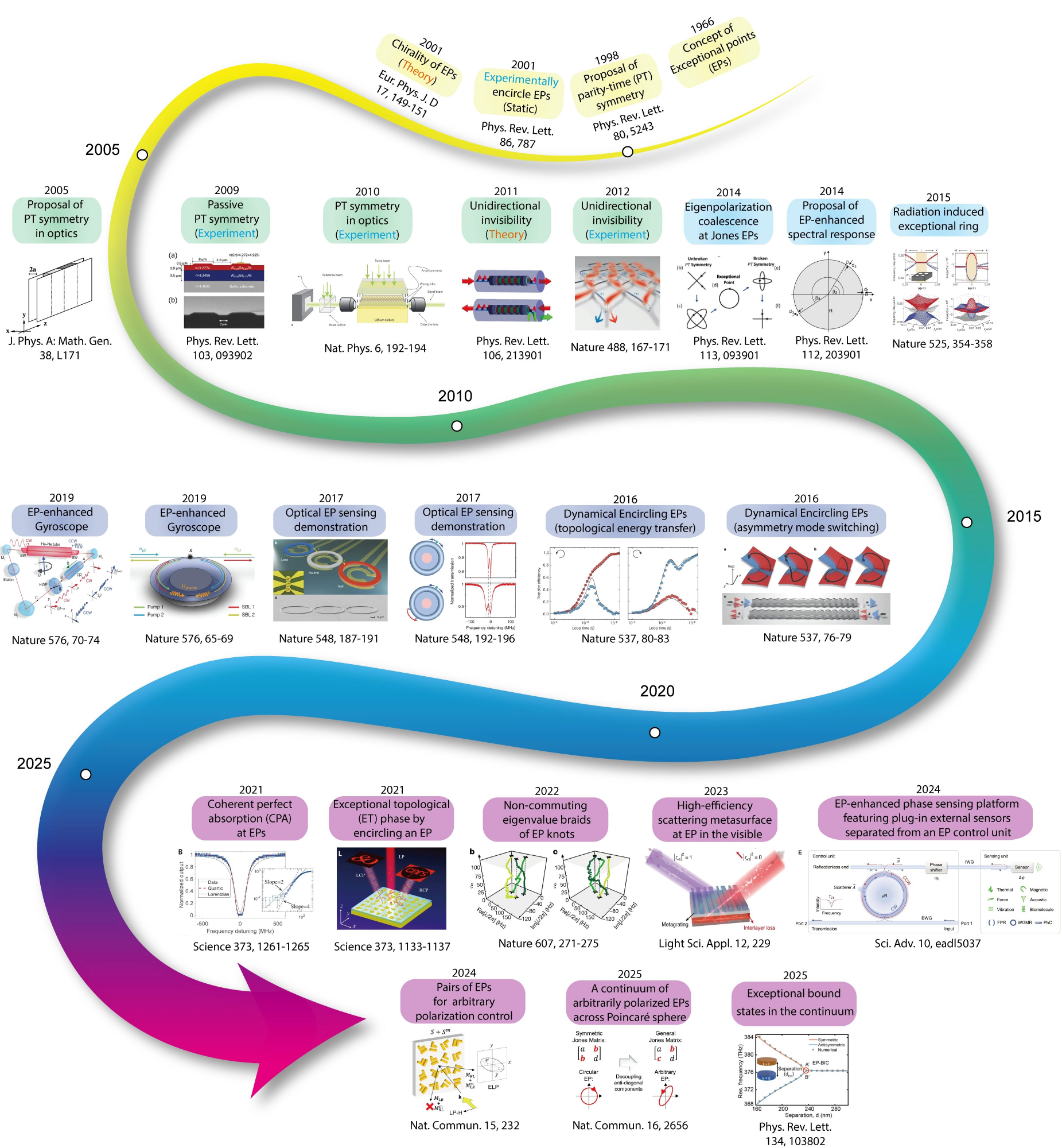}
\caption{\textbf{Historical development of EPs in optics and photonics.}
Selected milestones showing how EPs evolved from singular degeneracies in non-Hermitian operator theory to experimentally accessible photonic platforms and device concepts. The timeline starts with Kato's perturbation-theory formulation of EPs~\cite{Kato1966Perturbation}, the proposal of PT-symmetric Hamiltonians~\cite{bender1998real}, early work on EP chirality~\cite{heiss2001chirality}, and static encircling experiments in microwave cavities~\cite{dembowski2001experimental}. It then follows optical PT proposals and demonstrations~\cite{ruschhaupt2005physical,guo2009observation,ruter2010observation}, unidirectional invisibility~\cite{lin2011unidirectional,regensburger2012parity}, polarization-space EPs~\cite{lawrence2014manifestation}, EP-enhanced spectral response~\cite{wiersig2014enhancing}, exceptional rings in photonic-crystal slabs ~\cite{zhen2015spawning}, dynamical encircling~\cite{doppler2016dynamically,xu2016topological}, optical sensing~\cite{chen2017exceptional,hodaei2017enhanced} and gyroscope~\cite{lai2019observation,hokmabadi2019non} demonstrations, coherent perfect absorption~\cite{wang2021coherent} and exceptional-topological metasurfaces~\cite{song2021plasmonic}, non-commuting eigenvalue braids~\cite{patil2022measuring}, visible-frequency scattering metasurfaces~\cite{he2023scattering}, externalized phase-sensing architectures~\cite{mao2024exceptional}, arbitrary-polarization EP metasurfaces~\cite{yang2024creating,qin2025sphere}, and exceptional bound states in the continuum~\cite{canos2025exceptional}.} 
\label{fig:Intoduction}
\end{figure}

Authoritative reviews have established the foundations of EP optics~\cite{miri2019exceptional,ozdemir2019parity} and surveyed the broader development of non-Hermitian photonics across classical, semiclassical, and quantum regimes~\cite{feng2017non, el2018non,wang2023non,xiao2025non}. Existing surveys have emphasized non-Hermitian band topology and point-gap winding~\cite{ashida2020non, bergholtz2021exceptional, ding2022non}, as well as nanoscale~\cite{li2023exceptional} metasurface~\cite{xiao2025non} and engineered material~\cite{qin2026photonic} implementations where resonant and scattering EPs are distinguished. Building on this literature, the chronology in Fig.~\ref{fig:Intoduction} illustrates EP photonics from operator singularities in perturbation theory~\cite{Kato1966Perturbation} and PT-symmetric Hamiltonians~\cite{bender1998real} to optical realizations~\cite{guo2009observation,ruter2010observation}, scattering~\cite{lin2011unidirectional} and polarization~\cite{lawrence2014manifestation} EPs, static~\cite{dembowski2001experimental} and dynamical~\cite{doppler2016dynamically,xu2016topological} encircling, and recent directions including sensing~\cite{chen2017exceptional,hodaei2017enhanced,mao2024exceptional}, coherent absorption~\cite{sweeney2019perfectly,wang2021coherent}, noncommuting eigenvalues braids~\cite{patil2022measuring}, arbitrary-polarization EPs~\cite{yang2024creating,qin2025sphere}, and exceptional bound states in the continuum~\cite{canos2025exceptional}. The present review identifies, for each setting, the defective operator, the boundary condition or evolution law that fixes the relevant spectrum, and the observable from which the degeneracy is inferred. This perspective allows propagation, resonance, absorption, scattering, polarization, periodic-system, and open-quantum EPs to be compared within a common narrative while retaining their distinct physical meanings. 

Several developments make this comparison especially timely. In polarization optics, Jones-matrix engineering has extended exceptional polarization states from isolated circular eigenpolarizations~\cite{lawrence2014manifestation} to pairs~\cite{yang2024creating} and continuous families spanning arbitrary points on the Poincar\'e sphere~\cite{qin2025sphere}. Hybrid exceptional bound states in the continuum~\cite{canos2025exceptional} connect non-Hermitian defectiveness with radiative confinement. Open quantum systems now define and reconstruct exceptional degeneracies in Liouvillian~\cite{minganti2018spectral,abo2024experimental}, non-Markovian~\cite{lin2025non,longhi2026non}, and quantum-channel generators~\cite{wong2026non}. At the device level, programmable modulation~\cite{ding2024electrically}, externalized sensing architectures~\cite{mao2024exceptional}, and integrated dielectric metasurfaces~\cite{yi2025metasurfaceEP} are moving EP research toward reconfigurable photonic functions.

The review is organized as follows. 
Section~\ref{sec: Theoretical Foundations of Exceptional Points} introduces the mathematical foundations of EPs, including defective degeneracies, Jordan structure, Puiseux expansions, biorthogonality, phase rigidity, and symmetry constraints. 
Section~\ref{sec: Optical and Photonic Realizations of Exceptional Points} surveys classical photonic realizations according to the operator or response representation used to define the EP, including propagation and resonant Hamiltonians, absorbing and scattering formulations, Jones matrices, and Bloch and Floquet descriptions. 
Section~\ref{sec: Exceptional Points in Topological Photonics} discusses the spectral topology of EPs, static and dynamical encircling, higher-order and higher-dimensional exceptional geometries, and their coexistence with bound states in the continuum.
Section~\ref{sec: Exceptional Points in Quantum Optics and Photonics} extends the discussion to open quantum systems and distinguishes effective-Hamiltonian EPs from Liouvillian EPs and their corresponding observables.
Section~\ref{sec: Applications of Exceptional Points} reviews representative applications in sensing, lasing, coherent absorption, directional scattering, polarization and wavefront control, nonlinear optics, storage, nonreciprocity, and quantum photonics.
Section~\ref{sec: Conclusion and Outlook} summarizes open challenges in robustness, fabrication tolerance, scalable integration, and quantum-limited measurement.

\newpage 
\section{Theoretical Foundations of Exceptional Points}
\label{sec: Theoretical Foundations of Exceptional Points}

The ubiquity of EPs across diverse physical systems stems from their mathematical origin in the spectral theory of non-self-adjoint operators. This section establishes a mathematical language used throughout the review. We first formalize EPs as non-diagonalizable spectral degeneracies and summarize their local analytic structure. We then examine how analytic perturbations unfold a defective eigenvalue into Puiseux-series branches. Next, we formulate left and right eigenvectors, biorthogonal normalization, eigenvalue conditioning, the Petermann factor, and phase rigidity, and clarify their relation to spectral sensitivity and observable response. We finally discuss how symmetry constrains and reduces the codimension of exceptional degeneracies, and stabilizes higher-order exceptional structures. Although most derivations are presented for finite-dimensional linear operators, we indicate where generalized or nonlinear eigenvalue formulations are required for the photonic problems considered in later sections.

\subsection{EPs as defective degeneracies in linear operators}

EPs were formalized in Toshio Kato's perturbation theory~\cite{Kato1966Perturbation} as parameter values at which the eigenvalues and spectral projectors of a linear operator cannot be continued as single-valued holomorphic functions of the control parameters. Consider a finite-dimensional operator $T\in\mathbb{C}^{n\times n}$, under a small generic perturbation, $T$ becomes:
\begin{equation}
    T(\varepsilon) = T + \varepsilon T'
\end{equation}
where $\varepsilon$ is a small scalar perturbation parameter, $\varepsilon T'$ represents the perturbation and $T(0)=T$ is the unperturbed operator. To fix ideas, following Ref.~\cite{wiersig2020review}, we first consider the universal case of an isolated two-fold degeneracy. The diagonalizable degeneracies, which are often called diabolic points (DPs) in physics~\cite{berry1984diabolical} when arising from Hermitian limits, do not exhibit branching in eigenvalues or eigenvectors under generic perturbations. The $2\times2$-matrix $T_{2\times2}$ can be diagonalized to
\begin{equation} \label{eq:matrix_DP}
    T_{2\times2}|_{\rm{DP}}= \begin{pmatrix}
        \lambda_{\rm{DP}} & 0        \\
        0         & \lambda_{\rm{DP}}
    \end{pmatrix}
\end{equation}
with coalesced complex eigenvalue $\lambda_{\rm{DP}}$. Obviously, there are two linearly independent eigenvectors which can be chosen to be $(1,0)^{\rm{T}}$ and $(0,1)^{\rm{T}}$. At an EP, the operator is non-diagonalizable and possesses a nontrivial nilpotent part $A_0$:
\begin{equation} \label{eq:matrix_EP}
    T_{2\times2}|_{\rm{EP}}= \begin{pmatrix}
        \lambda_{\rm{EP}} & A_0        \\
        0         & \lambda_{\rm{EP}}
    \end{pmatrix}
\end{equation}
with coalesced complex eigenvalue $\lambda_{\rm{EP}}$. It can be brought to a Jordan normal form $J$ via a similarity transformation by an invertible matrix $V$:
\begin{equation} \label{eq:2x2_Jordan}
    T_{2\times2}|_{\rm{EP}} = V J_{2\times2}|_{\rm{EP}} V^{-1}, \qquad
    J_{2\times2}|_{\rm{EP}} = \begin{pmatrix}
        \lambda_{EP} & 1        \\
        0         & \lambda_{EP}
    \end{pmatrix}
\end{equation}
In contrast to the matrix in Eq.~\ref{eq:matrix_DP}, the eigenvectors of the matrix in Eq.~\ref{eq:matrix_EP} also coalesce into $(1,0)^{\rm{T}}$. In this case, the algebraic multiplicity $m^{(a)}=2$ exceeds its geometric multiplicity $m^{(g)}=1$.

With the general $2\times2$ perturbation
\begin{equation}
    T' = \begin{pmatrix}
        \lambda_1 & A_1        \\
        B_1         & \lambda_2
    \end{pmatrix},
\end{equation}
it follows directly that the eigenvalue splitting of $T_{2\times2}|_{\rm{DP}}$ is proportional to $\varepsilon$,
\begin{equation} \label{eq:DP_splitting}
    \Delta\lambda_{\rm{DP}}=\varepsilon\sqrt{(\lambda_2-\lambda_1)^2+4A_1B_1}.
\end{equation}
While the splitting of $T_{2\times2}|_{\rm{EP}}$ is 
\begin{align} \label{eq:EP_splitting}
    \Delta\lambda_{\rm{EP}}
    &=\sqrt{\varepsilon}\sqrt{\varepsilon(\lambda_2-\lambda_1)^2+4A_0B_1+4\varepsilon A_1B_1} \\
    &=\sqrt{\varepsilon}\sqrt{4A_0B_1}+\mathcal{O}(\varepsilon),
\end{align}
which is proportional to $\sqrt{\varepsilon}$ for small $|\varepsilon|$ provided that $B_1 \neq 0$. This is called the square-root spectral response of a second-order EP ($\rm{EP}_2$) due to the defectiveness of the Jordan block.

\subsection{Jordan normal form of EPs} \label{sec:Jordan normal form of EPs}
The above $2\times2$ model straightforwardly illustrates the singular feature of a generic isolated $\rm{EP}_2$. However, as cutting-edge photonic systems become increasingly complex, more intricate non-Hermitian degeneracies often emerge. Here, Eq.~\ref{eq:2x2_Jordan} can be generalized into multiple Jordan block forms, as illustrated in Fig.~\ref{fig:math}(A), to describe the general algebraic characteristics of an arbitrary finite-dimensional operator $T\in\mathbb{C}^{n\times n}$ at EPs~\cite{ashida2020non}.
\begin{equation} \label{eq:Jordan normal form}
    T = V 
    \left[
    \bigoplus_{j=1}^{J}\bigoplus_{\alpha=1}^{m_j^{(g)}} J_{n_{j\alpha}}\!\left(\lambda_j\right)
    \right]
    V^{-1}
\end{equation}
where $\{ \lambda_j\}_{j=1}^{J}$ are distinct eigenvalues. For a fixed $j$, the index $\alpha=1,2,\dots,m_j^{g}$ labels the Jordan blocks associated with the eigenvalue $\lambda_j$, where $m_j^{(g)}$ is the geometric multiplicity of $\lambda_j$. The matrix $J_{n_{j\alpha}}\!\left(\lambda_j\right)$ denotes a Jordan block of size $n_{j\alpha}$ associated with eigenvalue $\lambda_j$, explicitly
\begin{equation} \label{eq:Jordan block}
    J_{n_{j\alpha}}(\lambda_j) =
    \begin{pmatrix}
        \lambda_j & 1        & 0        & \cdots & 0 \\
        0         & \lambda_j & 1        & \ddots & \vdots \\
        \vdots    & \ddots   & \ddots   & \ddots & 0 \\
        \vdots    &          & \ddots   & \lambda_j & 1 \\
        0         & \cdots   & \cdots   & 0        & \lambda_j
    \end{pmatrix}
    \in \mathbb{C}^{\,n_{j\alpha}\times n_{j\alpha}}
\end{equation}
here $n_{j\alpha}\in\mathbb{N}$ is the size of the $\alpha$-th Jordan block for $\lambda_j$. The algebraic multiplicity of $\lambda_j$ is $m_j^{(a)} = \sum_{\alpha=1}^{m_j^{(g)}} n_{j\alpha}$, and the total dimension satisfies $\sum_{j=1}^{J} \sum_{\alpha=1}^{m_j^{(g)}} n_{j\alpha} = n$.
Each Jordan block decomposes into a diagonal part proportional to the identity and a strictly upper-triangular nilpotent part.
\begin{equation}
J_{n_{j\alpha}}\!\left(\lambda_j\right)
= \lambda_j I_{n_{j\alpha}} + N_{n_{j\alpha}}
\end{equation}
where $I_{n_{j\alpha}}$ is the identity matrix and $N_{n_{j\alpha}}$ is a nilpotent matrix with ones on the superdiagonal and zeros elsewhere. 

\subsection{Puiseux expansion of eigenvalues at EPs}
For a Hamiltonian subject to a small perturbation, conventional Hermitian degeneracies (i.e. DPs) admit a regular Taylor expansion, so the eigenvalue splitting remains linear in the perturbation to leading order $\Delta\lambda_{\rm{DP}}\sim\varepsilon$, as shown in Eq.~\ref{eq:DP_splitting}. By contrast, at an EP the Hamiltonian becomes defective and is locally described by a Jordan form [Eq.~(\ref{eq:Jordan normal form})-(\ref{eq:Jordan block})]. For a Jordan block of size $N$, the perturbed eigenvalues generally follow a Puiseux expansion~\cite{Kato1966Perturbation} (a fractional series expansion)
\begin{equation} \label{eq:Puiseux expansion}
    \lambda_h(\varepsilon) = 
    \lambda_{\mathrm{EP}} + 
    \sum_{k=1}^{\infty} \lambda_k\,
    \omega^{hk}\,
    \varepsilon^{k/N},
    \qquad
    h=0,1,\ldots,N-1 
\end{equation}
where $\omega=e^{2\pi i/N}$ represents the $N$-th root of unity, which labels the $N$ Puiseux branches of the eigenvalues, each $h\in\{0,1,\cdots,N-1\}$ is the corresponding $h$-th eigenvalue [yellow in Fig.~\ref{fig:math}(B)]. For a given perturbation strength near $\rm{EP}_N$, if one defines the complex eigenvalue splitting as the maximum pairwise separation among the N branches~\cite{kullig2023higher}:
\begin{equation}
\Delta \lambda_{\mathrm{EP}}(\varepsilon)\equiv 
\max_{h\neq h'} \left|\lambda_h(\varepsilon)-\lambda_{h'}(\varepsilon)\right|,
\label{eq:max_delta_omega}
\end{equation}
then a generic perturbation gives $\Delta\lambda_{\mathrm{EP}}\sim\varepsilon^{1/N}$ to leading order [blue in Fig.~\ref{fig:math}(B)]. Fig.~\ref{fig:math}(B) visualizes the $N$ Puiseux branches and the $\varepsilon^{1/N}$ splitting. This nonanalytic spectral susceptibility motivates EP-based sensing, although its metrological practicality depends on the constraints discussed in Sec.~\ref{Section:sensing}. Eq.~\ref{eq:EP_splitting} gives an $\rm{EP_{2}}$ example for square-root response. Eq.~\ref{eq:Puiseux expansion} also implies an $N$-sheeted Riemann surface and monodromy under parameter encircling with the emergence of branch-point analytic structure. In this process, each loop around the EP transfers the eigenvalue to a different branch, and only after completing the full $N$ encircling cycles does the eigenvalue return to its original value. Fig.~\ref{fig:math}(D) shows the associated sheet permutation. This intrinsic topology will be discussed in the subsequent Sec.~\ref{sec: Exceptional Points in Topological Photonics} of this review.

Remark (second-type singularities). Kato's taxonomy~\cite{Kato1966Perturbation}  also admits singularities of the similarity transformation and of the nilpotent part even without eigenvalue coalescence (sometimes termed “second-type EPs” in the mathematical literature~\cite{ashida2020non}). These objects can be relevant in permanently degenerate families, but they will be explicitly distinguished from the defective coalescences that are conventionally meant by “EP” in physics. We therefore treat them as mathematical spectral singularities and reserve the term “EP” in the main text for defective degeneracies unless stated otherwise.

\subsection{Biorthogonality of eigenvectors}

\begin{figure}[hbtp]
\centering
\includegraphics[width=0.8\textwidth]{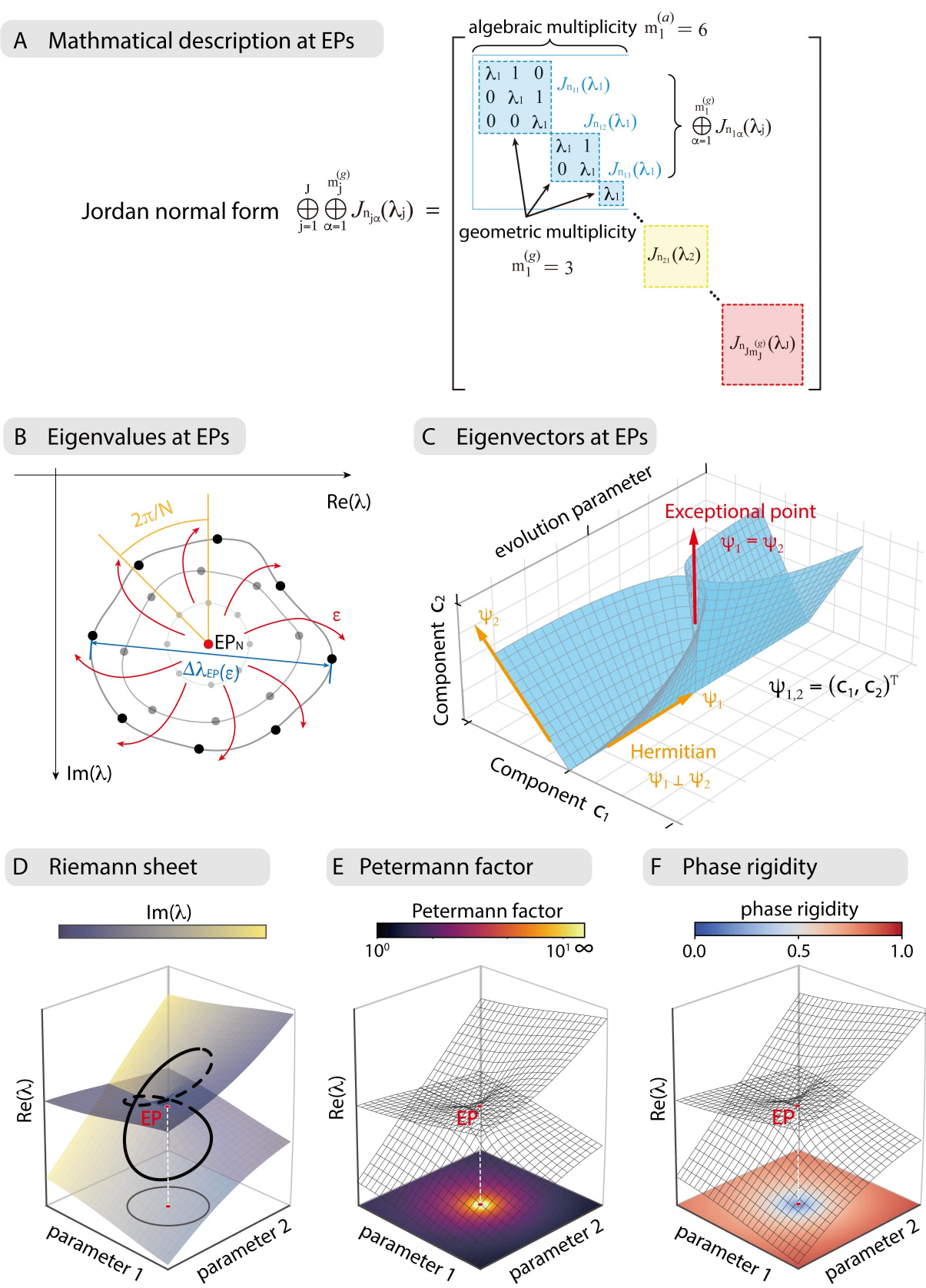}
\caption{\textbf{Mathematical and physical signatures of EPs in non-Hermitian systems}
(\textbf{A}) Illustration of Jordan normal form (Eq.~\ref{eq:Jordan normal form}). Blue blocks show a local Jordan normal form with eigenvalue $\lambda_1$ where algebraic multiplicity ($m_1^{(a)}=6$) exceeds the geometric multiplicity ($m_1^{(g)}=3$). Adapted from~\cite{ashida2020non}.
(\textbf{B}) Puiseux expansion of eigenvalue at an EP (Eq.~\ref{eq:Puiseux expansion}), where trajectories exhibit root-law complex eigenvalue splitting response. Adapted from~\cite{kullig2023higher}.
(\textbf{C}) Evolution of the directions of two eigenvectors $\psi_{\rm{1,2}}=(c_1, c_2)^{\rm{T}}$ in two dimensions. In a Hermitian system, the two eigenvectors are orthogonal, and they coalesce at an EP. Adapted from~\cite{chen2020revealing}.
(\textbf{D}) Riemann-sheet structure around an EP exhibits branch-point behavior and mode permutation.
(\textbf{E}) Petermann factor diverges near an EP (Eq.~\ref{eq:petermann factor}), reflecting enhanced mode non-orthogonality.
(\textbf{F}) Phase rigidity decreases and vanishes at the EP (Eq.~\ref{eq:phase rigidity}), indicating eigenvector self-coalescence.}
\label{fig:math}
\end{figure}

In standard quantum mechanics, closed systems are described by Hermitian operators~\cite{landau2013quantum}, $\hat{O}=\hat{O}^\dagger$. Hermiticity guarantees a real energy spectrum and a complete set of eigenstates. However, realistic physical systems are invariably open: gain, loss, or open boundaries generically break self-adjointness. Such systems are therefore described by complex (non-Hermitian) operators, $\hat{H} \ne \hat{H}^\dagger$, whose spectral properties must be formulated in terms of biorthogonal eigenvectors, defined independently through 
\begin{equation}
    \hat{H}\,\ket{\psi_n^{R}} = \lambda_n\,\ket{\psi_n^{R}},
    \qquad
    \hat{H^\dagger}\,\ket{\psi_n^{L}} = \lambda_n^*\,\ket{\psi_n^{L}}
\end{equation}
rather than by a single orthonormal basis~\cite{brody2014biorthogonal}. One convenient gauge choice fixing the inherent normalization freedom of non-Hermitian eigenvectors is the symmetric normalization:
\begin{equation}
    \ket{\bar{\psi}_n^{R}} =
    \frac{\ket{\psi_n^{R}}}
    {\sqrt{\langle \psi_n^{L} \mid \psi_n^{R} \rangle}},
    \qquad
    \bra{\bar{\psi}_n^{L}} =
    \frac{\bra{\psi_n^{L}}}
    {\sqrt{\langle \psi_n^{L} \mid \psi_n^{R} \rangle}}
\end{equation}
Although these normalized eigenvectors are not orthogonal under the standard inner product, they satisfy the biorthonormal condition $\langle \bar{\psi}_m^{L} |\bar{\psi}_n^{R}\rangle = \delta_{mn}$ away from degeneracies. The conventional resolution of the identity operator is then replaced by the biorthogonal completeness basis 
\begin{equation}
    \mathbb{I}=\sum_n \ket{\bar{\psi}_n^{R}}\bra{\bar{\psi}_n^{L}},
\end{equation}
and the operator admits the spectral decomposition 
\begin{equation}
    \hat{H}=\sum_n \lambda_n \ket{\bar{\psi}_n^{R}}\bra{\bar{\psi}_n^{L}}
\end{equation}
At EPs, this diagonalizable biorthogonal relation breaks down, eigenvectors coalesce into a single self-orthogonal state where $\langle \psi_{\rm EP}^{L} |\psi_{\rm EP}^{R}\rangle=0$, as illustrated in Fig.~\ref{fig:math}(C). In this case, the spectral decomposition must be replaced by a Jordan normal form, as discussed in Sec.~\ref{sec:Jordan normal form of EPs}.

\subsection{Petermann factor and phase rigidity}
The nonorthogonality of the right and left eigenstates can be quantified by a number of equivalent quantities, such as the Petermann factor~\cite{petermann1979calculated,siegman1989excess1,siegman1989excess2,wiersig2023petermann}
\begin{equation}
\label{eq:petermann factor}
    K_n :=
    \frac{\langle \psi_n^{R}\mid \psi_n^{R}\rangle\,\langle \psi_n^{L}\mid \psi_n^{L}\rangle}
    {\left|\langle \psi_n^{L}\mid \psi_n^{R}\rangle\right|^2},
\end{equation}
which has been introduced to quantify the linewidth broadening resulting from quantum excess noise in lasers~\cite{zhang2018phonon} and laser gyroscopes~\cite{wang2020petermann}, or the phase rigidity~\cite{bulgakov2006phase,schomerus2024eigenvalue}
\begin{equation}
\label{eq:phase rigidity}
    r_n :=
    \frac{\left|\langle \psi_n^{L}\mid \psi_n^{R}\rangle\right|}
    {\sqrt{\langle \psi_n^{L}\mid \psi_n^{L}\rangle\,\langle \psi_n^{R}\mid \psi_n^{R}\rangle}}
\end{equation}
with $0 \leq r_n = 1/\sqrt{K_n} \leq 1$. In the normal limit one may choose $\ket{\psi_n^{L}}=\ket{\psi_n^{R}}$ and obtain $r_n=1$. For generic non-Hermitian systems, $r_n<1$ reflects the non-orthogonality and mixing of eigenstates. In particular, approaching an EP the coalescing eigenstate becomes self-orthogonal, $\langle \psi_m^{L} |\psi_n^{R}\rangle = 0$, so that $K_n$ [Eq.~\ref{eq:petermann factor}] diverges [Fig.~\ref{fig:math}(E)] and $r_n$ [Eq.~\ref{eq:phase rigidity}] vanishes [Fig.~\ref{fig:math}(F)], signaling the failure of the biorthogonal normalization and the onset of strong spectral responsivity. For an EP of order $N$, perturbation theory generically predicts a power-law vanishing $r\propto |\delta|^{(N-1)/N}$ as the control parameter $\delta$ approaches the EP~\cite{heiss2008chirality}, which provides a characteristic of both the presence and the order of EPs.

\subsection{Symmetry-constrained approaches to EPs}

Even though symmetry is not a prerequisite for EPs, it constrains the characteristic polynomial and can reduce the number of independent parameters required to realize or stabilize EPs~\cite{mandal2021symmetry}. In a generic complex non-Hermitian $n$-level system, an EP of order $n$ has real codimension $2(n-1)$~\cite{delplace2021symmetry}. Therefore, higher-order EPs are generally difficult to stabilize in low-dimensional parameter spaces. Anti-linear symmetries such as PT symmetry~\cite{bender2019pt,bender2024pt} (within the 38-fold classification of non-Hermiticity based on generalized internal symmetries and the distinct topology of complex-energy gaps~\cite{kawabata2019symmetry}) can reduce this codimension. PT-symmetric photonics has therefore become one of the most extensively explored experimental platforms for EP physics, because it can be implemented through gain-loss engineering and complex refractive-index control~\cite{ruter2010observation}.

PT symmetry was originally introduced to extend quantum mechanics into the complex domain~\cite{bender1998real}. More generally, for a finite-dimensional diagonalizable Hamiltonian, a completely real spectrum is equivalent to the existence of a positive-definite metric \(\eta>0\) satisfying $H^\dagger\eta=\eta H$~\cite{scholtz1992quasi}. Ordinary Hermiticity corresponds to the special case \(\eta=\mathbb I\). By contrast, pseudo-Hermiticity with an indefinite metric constrains the spectrum to be real or complex-conjugate paired~\cite{mostafazadeh2010pseudo}, but does not by itself guarantee spectral reality~\cite{mostafazadeh2001pseudo,mostafazadeh2002pseudo}. However, symmetry is not a prerequisite for EPs. The appeal of PT symmetry is its simplicity, as it can be physically realized by engineering balanced gain-loss structures, which has led to fruitful interactions between theoretical proposals and experimental demonstrations~\cite{feng2017non,el2018non}.

\begin{table}[h]
    \small 
    \centering
    \caption{Representative $2\times2$ non-Hermitian Hamiltonians under PT symmetry and extensions}
    \begin{tabular}{llll}
        \toprule
         Name & Symmetry condition & Matrix form & Coupling type\\
        \midrule
         PT symmetry & $[\mathcal{PT},\mathcal{H}]=0$ & $\begin{pmatrix} x+iy & \kappa \\ \kappa & x-iy\end{pmatrix}$ 
         \quad & Dispersive \\\\

         Anti-PT symmetry & $\{\mathcal{PT},\mathcal{H}\}=0$ & $\begin{pmatrix} x-iy & i\kappa \\ i\kappa & -x-iy\end{pmatrix}$ & Dissipative \\\\

         Anyonic-PT symmetry & $\mathcal{PT}\mathcal{H}=e^{2i\varphi}\mathcal{H}\mathcal{PT}$ & $\begin{pmatrix} x-iy & \kappa e^{-i\varphi} \\ \kappa e^{-i\varphi} & e^{-2i\varphi}(x+iy)\end{pmatrix}$ & Hybrid \\\\  

         Generic non-Hermitian & None & $\begin{pmatrix} x_1-iy_1 & \kappa_{12} \\ \kappa_{21} & x_2-iy_2\end{pmatrix}$ & Asymmetric \\\\
        \bottomrule
    \end{tabular}
    \par\vspace{1ex}
    \footnotesize Here $x_j-i y_j$ denotes a generic complex diagonal element of the chosen operator. Its physical meaning depends on the physical problem being addressed. For a resonant-frequency problem, $x_j=\omega_j$ represents the resonant frequencies. For a propagation problem, $x_j=\beta_j$ represents the propagation constant. The imaginary part $y_j$ represents the corresponding attenuation or amplification. The coefficients $\kappa_{ij}$ represent the coupling strength.
    \label{tab: PT symmetry}
\end{table}

A PT symmetric Hamiltonian commutes with the combined parity-time operator $[PT, H]=0$, i.e. $\mathcal{HPT}=\mathcal{PTH}$, deriving from the invariance of the Hamiltonian under the combined parity (\(\mathcal{P}\)) and time-reversal (\(\mathcal{T}\)) operations. The parity operator \(\mathcal{P}\) means spatial inversion, reversing the signs of canonical operators (position operator $\hat{x} \rightarrow -\hat{x}$ and momentum operator $\hat{p} \rightarrow -\hat{p}$). It is a linear operator subject to $\mathcal{P}^2=\mathbb{I}$. The time reversal operator \(\mathcal{T}\) means reversion of time and is anti-linear, because it reverses the signs of the momentum operator and simultaneously takes the complex conjugate $(\hat{p} \rightarrow -\hat{p}$,  $i \rightarrow -i)$. PT symmetry requires spatial reflection to exchange gain and loss regions while maintaining balanced dissipation profiles. However, the strict condition of PT symmetry restricts the role of EPs to the transition point between purely real eigenvalues and complex spectra in canonical two-mode systems. To broaden this framework, generalizations have been proposed in terms of both diagonal and off-diagonal elements of the Hamiltonian. For diagonal elements, the sign can be altered to include only loss terms, leading to passive PT symmetry, characterized by a global decay factor $\chi$, expressed as $\mathcal{H}_{\rm eff}^{\rm passive}=\mathcal{H}_{\rm eff}^{\rm PT}-i\chi\mathbb{I}$~\cite{guo2009observation,ozdemir2019parity}. As for the off-diagonal elements, which usually represent coupling, they can be extended to mixed coupling that includes both dispersive and dissipative interactions, known as anyonic PT symmetry $\mathcal{H}_{\rm eff}^{\rm anyonic}\mathcal{PT}=e^{2i\varphi}\mathcal{PT}\mathcal{H}_{\rm eff}^{\rm anyonic}$\cite{longhi2019anyonic,arwas2022anyonic}. When the phase $\varphi$ takes the values of 0 or $\pi$, the specific cases correspond to the standard PT symmetry. The cases $\varphi=\pm \pi/2$ correspond to anti-PT symmetry~\cite{ge2013antisymmetric,yang2017anti}. Besides a Hamiltonian formulation, PT symmetry can also be described by the complex scattering matrix $\mathcal{S}$, with the condition generalized at arbitrary frequency $\omega$ through the relation $(\mathcal{PT})\mathcal{S}(\omega^*)(\mathcal{PT})=\mathcal{S}^{-1}(\omega)$~\cite{chong2011pt, ge2012conservation}. The first three rows of Table~\ref{tab: PT symmetry} summarize representative matrices under PT symmetry and its generalizations.

\subsection{EPs in asymmetrically coupled systems}
\label{sec:asymmetric coupling}

In the generic case, the two complex diagonal elements and the two directional coupling amplitudes can be controlled independently, as shown in the last row of
Table~\ref{tab: PT symmetry}. The eigenvalues are
\begin{equation}
\lambda_{\pm}
=
\frac{\xi_1+\xi_2}{2}
\pm
\frac{1}{2}
\sqrt{
(\xi_1-\xi_2)^2+
4\kappa_{12}\kappa_{21}
}.
\label{eq:generic_asymmetric_eigenvalues}
\end{equation}
where $\xi_j=x_j-iy_j$, a second-order EP occurs when the discriminant $(\xi_1-\xi_2)^2+
4\kappa_{12}\kappa_{21} =0$.

The bidirectionally asymmetric coupling is defined in a specified and physically
normalized basis by $\kappa_{12}\neq\kappa_{21}$ with
$\kappa_{12}\kappa_{21}\neq0$. The limiting case of unidirectional coupling is obtained for $\kappa_{12}\neq0$ and $\kappa_{21}=0$, or vice versa. If the diagonal elements additionally coalesce, $\xi_1=\xi_2=\xi_0$, then
\begin{equation}
\mathcal{H}
=
\xi_0 I+
\begin{pmatrix}
0 & \kappa_{12}\\
0 & 0
\end{pmatrix}.
\label{eq:generic_unidirectional_EP}
\end{equation}
The repeated eigenvalue has algebraic multiplicity two and geometric
multiplicity one, and the matrix is similar to a second-order Jordan
block. This inequality is basis dependent, whereas the discriminant and Jordan structure are invariant under similarity transformations. The relevant basis may consist of clockwise and counterclockwise whispering-gallery resonator~\cite{wiersig2008asymmetric,peng2016chiral,lee2023chiral}, the local guided modes in coupled waveguides~\cite{guo2009observation,ruter2010observation} or other physically distinguishable states.

\newpage 
\section{Classical Photonic Realizations of Exceptional Points}
\label{sec: Optical and Photonic Realizations of Exceptional Points}
The quantities introduced in the previous section identify the local mathematical signatures of an EP but do not specify which physical operator becomes defective. This distinction is important in photonics because the relevant operator depends on the boundary conditions imposed on the electromagnetic problem and on the observable used to probe the system. The same structure can support inequivalent EPs under different experimental protocols, including effective Hamiltonian EPs, absorbing operator EPs, scattering matrix EPs, Jones matrix EPs, and Floquet or Bloch EPs. Accordingly, this section organizes classical photonic realizations according to the operator in which a defective degeneracy occurs.

\subsection{Propagation-Hamiltonian EPs in transversely structured optical media}
\label{subsec: 3.1}
The earliest optical realizations of EP physics were closely connected to the PT symmetry framework in coupled-waveguide systems~\cite{ruter2010observation}, where Maxwell's equations reduce to a Schrödinger-like evolution equation whose evolution variable is the propagation coordinate $z$ under a paraxial approximation~\cite{makris2008beam}. In this setting, the relevant operator is a propagation Hamiltonian, and its eigenvalues are the effective complex propagation constants of the supported supermodes.

\begin{figure}[htbp]
\includegraphics[width=0.8\textwidth]{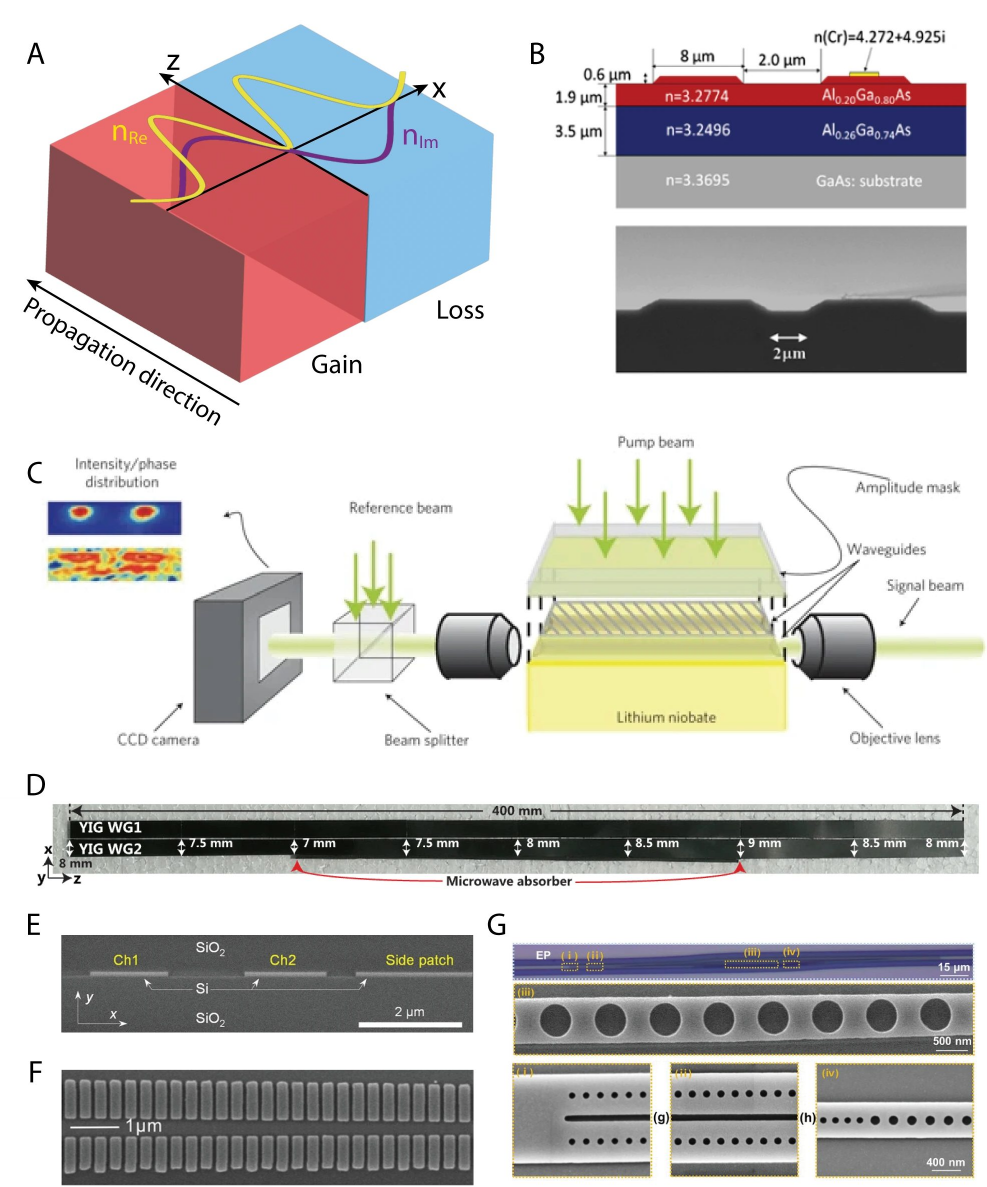}
\centering
\caption{\textbf{Concept and representative implementations of propagation-operator EPs in transversely structured optical media}
(\textbf{A}) Schematic of a transverse PT-symmetric refractive-index profile, with an even real part and an odd imaginary part, light propagates along $z$ direction.
(\textbf{B}) Passive multilayer $\ce{Al_xGa_{1-x}As}$ waveguide coupler with a $\ce{Cr}$ stripe on the right waveguide, reproduced from Ref.~\cite{guo2009observation}.
(\textbf{C}) Actively controlled $\ce{Fe}$-doped $\ce{LiNbO3}$ waveguide array and interferometric readout, reproduced from Ref.~\cite{ruter2010observation}.
(\textbf{D}) Yttrium-iron-garnet waveguide coupler with a localized microwave absorber, reproduced from Ref.~\cite{zhang2018dynamically}.
(\textbf{E}) Silicon channel-waveguide coupler in which an auxiliary slab patch provides continuum-mediated leakage, reproduced from Ref.~\cite{yoon2018time}. 
(\textbf{F}) Silicon subwavelength-grating waveguide doublet with corrugation-induced radiation loss, reproduced from Ref.~\cite{liu2020efficient}.
(\textbf{G}) Silicon-on-insulator coupled waveguides with longitudinally patterned nanoholes, reproduced from Ref.~\cite{li2025high-Performance}.
}
\label{fig:3.1_prop_Hamiltonian_EPs}
\end{figure}

The frequency-domain Helmholtz equation 
\begin{equation} \label{eq:Helmholtz equation}
    \left[ \nabla^2 + k_0^2 n^2(\bm{r},\omega)\right]\bm{E}(\bm{r},\omega) = 0
\end{equation}
with $k_0=\omega/c$, describes the spatial field distribution of a monochromatic scalar optical field in an inhomogeneous optical medium. When the refractive index $n(\bm{r},\omega)$ is complex, optical gain and loss are incorporated directly into the wave equation. The PT symmetry condition becomes $n(x) = n^*(-x)$, namely, an even symmetric real part $n_{\rm Re}(x) = n_{\rm Re}(-x)$ and an odd symmetric imaginary part $n_{\rm Im}(x) = -n_{\rm Im}(-x)$. Fig.~\ref{fig:3.1_prop_Hamiltonian_EPs}(A) defines the transverse PT-symmetry index profile. Physically this condition corresponds to a spatially balanced distribution of gain and loss.
For a refractive-index modulation that is transverse to the propagation direction ($z$-axis), the electric field can be expressed as a rapidly oscillating carrier $e^{ikz}$ multiplied by a slowly varying envelope $\psi(x,z)$,
\begin{equation}
    \bm{E}({\bm{r}}) = \psi(x,z)e^{ikz}
\end{equation}
where $k=k_0n_0$ is the reference propagation constant in a uniform background medium of refractive index $n_0$. Substituting this ansatz into Eq.~\ref{eq:Helmholtz equation} and applying the paraxial approximation, $|\dfrac{\partial^2 \psi}{\partial z^2}|\ll |2ik\dfrac{\partial \psi}{\partial z}|$, one obtains~\cite{makris2008beam}
\begin{equation}
     i\,\frac{\partial}{\partial z}\psi(x,z) = \left[-\frac{1}{2k}\,\frac{\partial^2}{\partial x^2} + V(x)\right]\psi(x,z)    
\end{equation}
where
\begin{equation}
     V(x)=\frac{k^2-k_0^2n^2(x)}{2k}\simeq -k_0\Delta n(x),
\end{equation}
under the weakly guiding limit ($n(x)=n_0+\Delta n(x)$ with $|\Delta n(x)|\ll n_0$). Thus, the effective propagation Hamiltonian of the paraxial wave equation is 
\begin{equation} \label{eq:propagation Hamiltonian}
   \hat{\mathcal{H}}_{\mathrm{prop}}=-\dfrac{1}{2k}\,\dfrac{\partial^2}{\partial x^2} - k_0\Delta n(x),
\end{equation}
with the operator in Eq.~\ref{eq:propagation Hamiltonian} governing longitudinal modal transport. Its eigenvalues correspond to effective complex propagation constants: the real part determines the phase accumulation along $z$, whereas the imaginary part describes net amplification or attenuation.


However, rewriting the optical wave equation in Schrödinger-like form does not by itself imply the presence of an EP. A propagation-Hamiltonian EP appears only when the operator in Eq.~\ref{eq:propagation Hamiltonian} becomes defective, so that two eigenvalues and their corresponding transverse supermodes coalesce simultaneously. Physically, this means that the system loses a complete eigenmode basis and that the field evolution can no longer be written as a superposition of two independent exponential transport channels.

A more concise and sufficient description is obtained by reducing the continuous wave equation to a finite-dimensional coupled-mode model. For the simplest two-waveguide PT-symmetric dimer, the effective propagation Hamiltonian can be written as
\begin{align} \label{eq:propagation Hamiltonian2}
\hat{\mathcal{H}}_{\mathrm{prop}}
= \begin{pmatrix}
\beta_0+i\gamma& \kappa \\
\kappa& \beta_0-i\gamma
\end{pmatrix} 
\end{align}
where $\beta_0$ is the reference propagation constant, $\gamma$ characterizes the gain-loss contrast, and $\kappa$ is the evanescent coupling coefficient. The corresponding supermode propagation constants are $\beta_{\pm}=\beta_0\pm\sqrt{\kappa^2-\gamma^2}$, so that an EP occurs at $|\kappa|=|\gamma|$. More generally, the real and imaginary detunings can be controlled through the structural and material parameters of the individual waveguides, whereas $\kappa$ is determined primarily by the transverse spacing and modal overlap. This relative separability of coupling and non-Hermiticity is one reason why transversely structured coupled waveguides provide a suitable platform for engineering propagation-Hamiltonian EPs.

An early experimental realization of propagation-Hamiltonian EPs was reported in passive multilayer $\ce{Al_xGa_{1-x}As}$ coupled waveguides, where a chromium stripe deposited on one waveguide introduced a controllable loss imbalance~\cite{guo2009observation}, as shown in Fig.~\ref{fig:3.1_prop_Hamiltonian_EPs}(B). Increasing the stripe width drove the coupled supermodes through an EP and produced loss-induced transparency through modal bifurcation and field localization in the lower-loss channel. A related effect was theoretically analyzed for active coupled-waveguide systems~\cite{ramezani2010unidirectional} and experimentally observed in the $\ce{Fe}$-doped $\ce{LiNbO3}$ active platform shown in Fig.~\ref{fig:3.1_prop_Hamiltonian_EPs}(C)~\cite{ruter2010observation}. In that experiment, optical pumping and $\ce{Ti}$ in-diffusion produced the required complex transverse index profile, while interferometric imaging resolved the corresponding intensity and phase distributions. This counterintuitive phenomenon also anticipates the development of topologically asymmetric mode switching schemes associated with EPs~\cite{doppler2016dynamically}. 

In practical implementations, inter-waveguide coupling is governed by controlled geometric perturbations along the propagation direction, while the degree of loss imbalance can be precisely engineered through diverse mechanisms, compared in Figs.~\ref{fig:3.1_prop_Hamiltonian_EPs}(D-G). In the yttrium-iron-garnet coupler shown in Fig.~\ref{fig:3.1_prop_Hamiltonian_EPs}(D), a microwave absorber attached to one waveguide produces a spatially localized loss imbalance~\cite{zhang2018dynamically}. The silicon structure in Fig.~\ref{fig:3.1_prop_Hamiltonian_EPs}(E) instead couples one channel waveguide to an auxiliary slab-waveguide patch, so that leakage into the slab acts as a controllable continuum-mediated decay channel~\cite{yoon2018time}; related continuum-coupling strategies were developed in Refs.~\cite{li2020hamiltonian,li2022riemann}. The subwavelength-grating coupler in Fig.~\ref{fig:3.1_prop_Hamiltonian_EPs}(F) introduces radiation loss through periodic corrugations~\cite{liu2020efficient}, whereas the silicon-on-insulator device in Fig.~\ref{fig:3.1_prop_Hamiltonian_EPs}(G) uses arrays of etched nanoholes to control the local attenuation and coupling profile~\cite{li2025high-Performance}. Deposited metallic absorbers provide an additional route to longitudinally programmable loss~\cite{shu2022fast}.

When the complex coupling and detuning are varied continuously along the propagation coordinate $z$, the same coupled-waveguide architecture implements a dynamical loop in parameter space. Such longitudinal modulation enables direction-dependent state conversion through non-adiabatic evolution near an EP~\cite{doppler2016dynamically,zhang2019dynamically2}. The development of this approach includes early theoretical waveguide proposals~\cite{hassan2017chiral}, compact silicon implementations based on radiative-loss modulation~\cite{liu2020efficient}, accelerated encircling protocols using deposited absorbers~\cite{shu2022fast}, integrated anti-PT configurations~\cite{feng2022harnessing}, and hybrid III-V/silicon switching networks designed for scalable photonic integration~\cite{feng2025non}.

Because the defective operator directly governs evolution along the propagation direction, propagation-Hamiltonian EPs are observed through output modal evolution and field redistribution. This makes transversely structured waveguides a particularly suitable platform for studying EP topology, asymmetric mode transfer and selection. From a device perspective, these systems provide a practical route to selective guiding or suppression of specific modes, compact chiral mode converters~\cite{liu2022chip}, and scalable non-Hermitian switching networks~\cite{feng2025non}.


\subsection{Resonant-Hamiltonian EPs in open photonic structures}
\label{subsec: 3.2}
Open resonant structures are commonly described by temporal coupled-mode dynamics for a small number of resonant modal amplitudes~\cite{fan2003temporal,suh2004temporal}. Their source-free eigenstates are quasi-normal modes (QNMs), namely solutions of the source-free Maxwell equations at complex eigenfrequencies that satisfy purely outgoing boundary conditions~\cite{lalanne2018light}. The corresponding eigenvalues are complex resonant frequencies $\widetilde{\omega}_{n}=\omega_n-i\gamma_n$, where $\omega_n$ gives the oscillation frequency and $\gamma_n$ determines the total decay rate, including both radiative and nonradiative contributions. A resonant-Hamiltonian EP occurs when two complex eigenfrequencies and their associated QNM profiles coalesce, rendering the source-free resonant operator defective. 


For a generic two-resonance system, the resonant effective Hamiltonian can be written as
\begin{align} \label{eq:resonant Hamiltonian}
\hat{\mathcal{H}}_{\mathrm{res}}
=
\begin{pmatrix}
\omega_1-i\gamma_1& \kappa_{\rm{12}} \\
\kappa_{\rm{21}}& \omega_2-i\gamma_2
\end{pmatrix} 
\end{align}
where $\omega_{1,2}$ are the uncoupled resonant frequencies and $\gamma_{1,2}$ are the total decay rates. If an optical gain is present, it may be incorporated as a negative contribution to $\gamma_j$. The off-diagonal terms $\kappa_{\rm{12}}$ and $\kappa_{\rm{21}}$ are generally complex. Their real parts describe coherent or dispersive coupling, which may arise from near-field overlap or scatter-induced backscattering, whereas their imaginary parts represent dissipative or radiative coupling, often mediated by a common radiation continuum or external ports. The complex eigenfrequencies of the coupled system are 
\begin{equation}
    \widetilde{\omega}_{\pm}=\frac{\widetilde{\omega}_1+\widetilde{\omega}_2}{2}\pm\sqrt{\left(\frac{\widetilde{\omega}_1-\widetilde{\omega}_2}{2}\right)^2+\kappa_{\rm{12}}\kappa_{\rm{21}}},
\end{equation}
The EP condition is
\begin{equation}
    \left(\dfrac{\widetilde{\omega}_1-\widetilde{\omega}_2}{2}\right)^2+\kappa_{\rm{12}}\kappa_{\rm{21}}=0.
\end{equation}
\begin{figure}[htbp]
\includegraphics[width=0.85\textwidth]{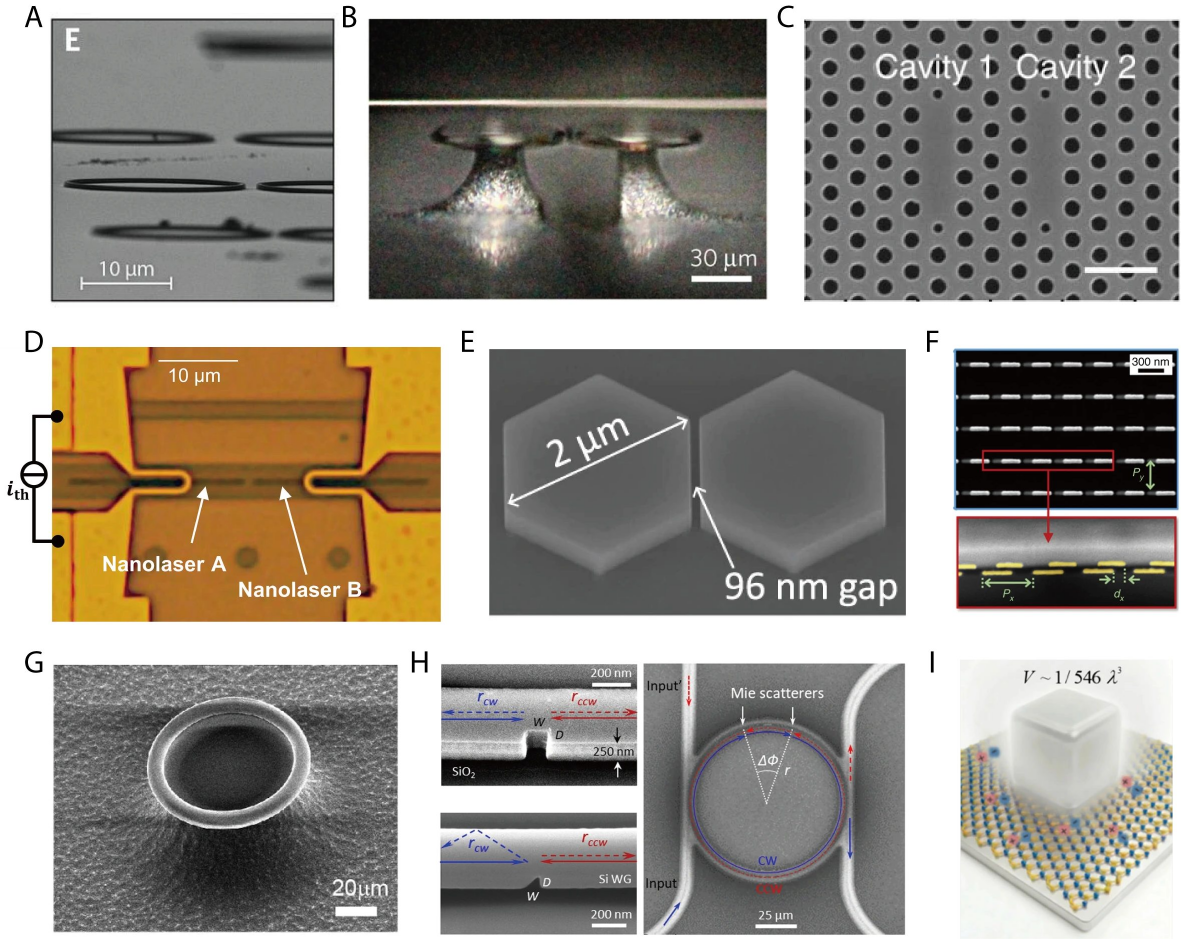}
\centering
\caption{\textbf{Representative platforms of resonant-Hamiltonian EPs} 
(\textbf{A-F}) Coupled-resonator implementations:
(\textbf{A}) Coupled $\ce{InGaAsP}$ microrings, reproduced from Ref.~\cite{hodaei2014parity}.
(\textbf{B}) Active-passive silica microtoroids, reproduced from Ref.~\cite{peng2014parity}.
(\textbf{C}) Graphene-loaded coupled photonic crystal cavities, reproduced from Ref.~\cite{kim2016direct}.
(\textbf{D}) Coupled nanolasers with integrated thermoresistive gold nanowires, reproduced from Ref.~\cite{madiot2024harnessing}.
(\textbf{E}) Optically pumped $\ce{InP}$ microdisks, reproduced from Ref.~\cite{fischer2024controlling}.
(\textbf{F}) Bilayer plasmonic metallic bar metasurface, reproduced from Ref.~\cite{park2020symmetry}.
(\textbf{G,H}) Single-resonator implementations based on controllable coupling between clockwise and counterclockwise modes:
(\textbf{G}) A microtoroid with external scatterers, reproduced from Ref.~\cite{chen2017exceptional}.
(\textbf{H}) A silicon microring with lithographically defined Mie scatterers, reproduced from Ref.~\cite{lee2023chiral}.
(\textbf{I}) Plexciton hybrid structure composed of a single plasmonic silver nanocube-on-mirror cavity coupled to a monolayer of $\ce{WS2}$, reproduced from Ref.~\cite{ji2026enhanced}.
}
\label{fig:3.2_res_Hamiltonian_EPs}
\end{figure}


Representative resonant-Hamiltonian EP platforms are organized in Fig.~\ref{fig:3.2_res_Hamiltonian_EPs} according to the modal basis on which the defective Hamiltonian is constructed. The first class consists of coupled-resonator systems, in which each resonator contributes one or more resonant modes, and the inter-resonator coupling is controlled by the cavity separation or evanescent field overlap. The coupled $\ce{InGaAsP}$ quantum well microrings in Fig~\ref{fig:3.2_res_Hamiltonian_EPs}(A) use evanescent overlap across the inter-ring gap to control coherent coupling~\cite{hodaei2014parity}. This concept can be extended to ternary microring laser systems by introducing a third neutral resonator, enabling the realization of a third-order EP~\cite{hodaei2017enhanced}. The active-passive silica microtoroid pair in Fig~\ref{fig:3.2_res_Hamiltonian_EPs}(B) provides independent control of resonance detuning and inter-cavity coupling~\cite{peng2014parity}. A similar strategy can also be implemented in purely passive coupled silica WGM microtoroids, where additional loss is introduced by a chromium–coated silica-nanofiber tip~\cite{peng2014loss}. Additional control can be obtained by introducing local scatterers into coupled ring resonators, for which the perturbation strengths and relative scatterer positions provide extra degrees of freedom to tune the EP condition and the chirality of the coalesced states \cite{Ramezanpour2021Perturbed}.

Coupled photonic-crystal nanocavities provide more integrated semiconductor implementations. A monolayer graphene sheet partially covering only one $\ce{InGaAsP}$ photonic-crystal slab in Fig.~\ref{fig:3.2_res_Hamiltonian_EPs}(C) introduces asymmetric absorption and thus creates a controlled gain-loss contrast~\cite{kim2016direct}. Electrically pumped photonic-crystal coupled-nanocavity lasers based on air-suspended $\ce{InP}$ slabs provide another route, where diagonally patterned $p$ and $n$-doped layers and contact pads allow independent current injection into adjacent nanocavities, thereby tuning the imaginary potential contrast~\cite{takata2021observing}. Real-frequency detuning between the two cavities can be independently controlled by varying the dimensions of additional detuning holes~\cite{ji2023tracking}. More generally, when the coupling is complex, both real and imaginary detunings must be controlled. For example, gold resistive nanowires fabricated above individual nanocavities can act as micro-heaters, producing a thermo-optic redshift of the cavity resonances, while an external continuous-wave pump can independently tune the carrier density and gain level in each nanocavity shown in Fig.~\ref{fig:3.2_res_Hamiltonian_EPs}(D)~\cite{madiot2024harnessing}. In bottom-up epitaxially grown coupled $\ce{InP}$ microdisks, a digital micromirror device can shape the pump into two independently controlled semicircular spots, allowing programmable control of the pump power delivered to each microdisk shown in Fig.~\ref{fig:3.2_res_Hamiltonian_EPs}(E)~\cite{fischer2024controlling}. At a smaller length scale, the bilayer plasmonic metasurface in Fig.~\ref{fig:3.2_res_Hamiltonian_EPs}(F) realizes a resonant-Hamiltonian EP either through two optically inequivalent metallic nanorod arrays or by placing nominally identical resonators in distinct environments~\cite{park2020symmetry}.

The second class consists of single resonant structures supporting multiple internal resonances. In Fig.~\ref{fig:3.2_res_Hamiltonian_EPs}(G), controllable Rayleigh scatterers, such as nanoprobes~\cite{zhu2010controlled} or nanotips~\cite{peng2016chiral, chen2017exceptional}, can couple the clockwise (CW) and counterclockwise (CCW) whispering-gallery modes of one microtoroid, allowing the two backscattering amplitudes to be tuned toward an EP. Beyond externally introduced scatterers, lithographically patterned Mie scatterers can provide deterministic control. Fig.~\ref{fig:3.2_res_Hamiltonian_EPs}(H) shows asymmetric triangular and symmetric rectangular Mie scatterers that control the amplitude and phase of CW-CCW coupling in a silicon microring~\cite{lee2023chiral}. Dynamical tuning can be further achieved by applying a voltage to a local heater defined along the microring perimeter. The thermo-optic effect of silicon then alters the local refractive index and tunes the optical path difference between the two scatterers~\cite{lee2025chiral}. Fabry-Pérot (FP) cavities offer another example of single-cavity multimode resonant systems. In a single magneto-optical FP cavity incorporating a tunable liquid-crystal absorber, electrically controlled linear dichroism can be used to balance coherent polarization coupling against dissipation and drive the system to a resonant-Hamiltonian EP, resulting in a loss-enhanced magneto-optical response~\cite{ruan2025observation}. Isolated subwavelength Mie resonators with multipolar electromagnetic resonances provide another route to control radiative leakage. Mirror-symmetry-breaking induced by bianisotropy~\cite{CanosValero2024Jan} and geometric tuning~\cite{Solodovchenko2024Feb} can control both modal hybridization and radiation loss. Isolated dielectric ring resonators have provided an experimental platform for observing resonant-Hamiltonian EPs~\cite{zhang2025non} and Fermi-arc-like connections between them~\cite{solodovchenko2025experimental}.


A third modal basis is based on strong light-matter coupling, which creates hybridized polaritonic systems. These systems exploit the intrinsic loss imbalance between photonic or plasmonic resonances and matter excitations, whereas the coupling is set by the light-matter interaction strength. The optically anisotropic lead-halide perovskite crystals embedded in optical microcavities form hybrid exciton-polariton EPs~\cite{su2021direct}. Under nonlinear optical pumping, related polaritonic platforms can further support polariton lasing~\cite{masharin2023room} and giant ultrafast all-optical modulation~\cite{masharin2024giant}. Another example is provided by vibrational polaritons formed by coupling terahertz photons to collective intermolecular vibrations of organic $\alpha$-lactose crystals, where electrical control enables dynamic tuning in the terahertz parameter space~\cite{ergoktas2022topological}. The plexcitonic platform in Fig.~\ref{fig:3.2_res_Hamiltonian_EPs}(I) couples a deeply subwavelength plasmonic nanocavity to an excitonic transition in monolayer $\ce{WS2}$. The unequal photonic and excitonic decay rates supply the non-Hermitian contrast, while the light–matter interaction provides the off-diagonal coupling required to reach the hybrid resonant-Hamiltonian EP~\cite{ji2026enhanced}.

Because QNMs have complex eigenfrequencies and cannot be directly excited as steady-state fields, experiments usually access resonant-Hamiltonian EPs through real-frequency input-output spectra. Temporal coupled-mode theory (TCMT)~\cite{haus1984waves,haus2002coupled,fan2003temporal,suh2004temporal} describes the internal resonant Hamiltonian coupled to external scattering channels. For a vector of resonant amplitudes $\boldsymbol{\alpha}$ coupled to incoming and outgoing waves $\mathbf{s}_{+}$ and $\mathbf{s}_{-}$, one may write
\begin{equation} \label{eq:TCMT}
\left\{
\begin{aligned}
\frac{d\boldsymbol{\alpha}}{dt} &= -i\left(\boldsymbol{\Omega}-i\boldsymbol{\Gamma}_{\mathrm{rad}}-i\boldsymbol{\Gamma}_{\mathrm{nr}}\right)\boldsymbol{\alpha} + \mathcal{K}^{\mathrm T}\mathbf{s}_{+} \\
\mathbf{s}_{-} &= \mathcal{C}\mathbf{s}_{+} + \mathcal{D}\boldsymbol{\alpha}
\end{aligned}
\right.
\end{equation}

where $\boldsymbol{\Omega}$ is the Hermitian frequency matrix, $\boldsymbol{\Gamma}_{\mathrm{rad}}$ and $\boldsymbol{\Gamma}_{\mathrm{nr}}$ describe radiative and nonradiative decay, respectively, and $\mathcal{C}$, $\mathcal{D}$, and $\mathcal{K}$ characterize direct scattering, resonant out-coupling, and resonant in-coupling. Under harmonic excitation, $\boldsymbol{\alpha}(t)=\boldsymbol{\alpha}(\omega)e^{-i\omega t}$, the scattering matrix becomes
\begin{equation}
    \mathcal{S}(\omega) = \mathcal{C}+i\mathcal{D}\left(\omega\mathbb{I}-\hat{\mathcal{H}}_{\mathrm{res}}\right)^{-1}\mathcal{K}^T
\end{equation}
with
\begin{equation}
    \hat{\mathcal{H}}_{\mathrm{res}}=\boldsymbol{\Omega}-i\left(\boldsymbol{\Gamma}_{\mathrm{rad}}+\boldsymbol{\Gamma}_{\mathrm{nr}}\right).
\end{equation}
For purely outgoing boundary conditions, $\mathbf{s}_{+}=0$, and a nontrivial resonant amplitude $\boldsymbol{\alpha}$ exists when 
\begin{equation}
\det\left(
\omega\mathbb{I}
-\hat{\mathcal{H}}_{\mathrm{res}}
\right)=0.
\end{equation}
The roots are the eigenvalues of $\mathcal{H}_{\mathrm{res}}$ and also appear as poles of the analytically continued scattering matrix. A resonant-Hamiltonian EP manifests itself equivalently as a coalescence of scattering poles and, generically, a higher-order pole.
\subsection{Absorbing EPs associated with coalescing scattering zeros}
\label{subsec: 3.3}
\begin{figure}[htbp]
\includegraphics[width=0.9\textwidth]{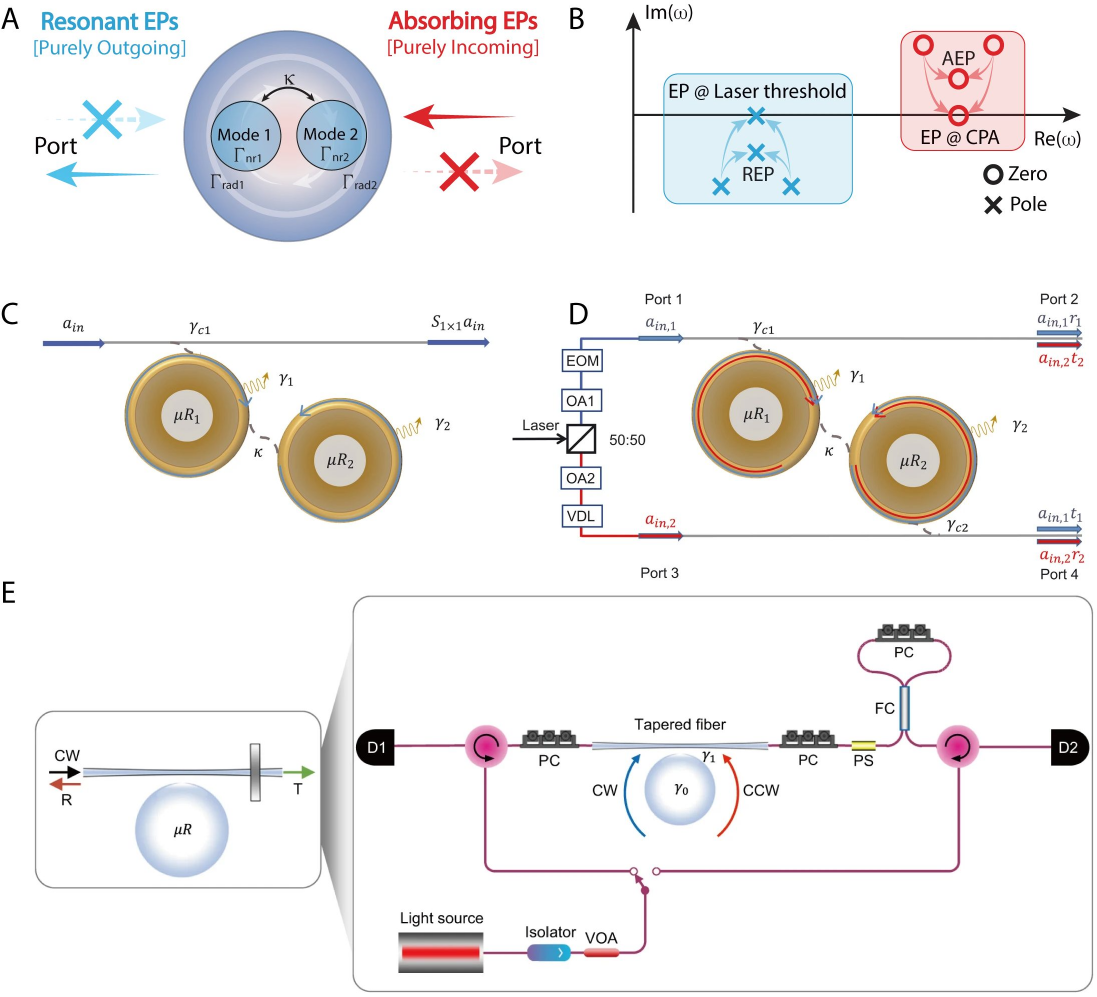}
\centering
\caption{\textbf{Concept and representative implementations of absorbing EPs}
(\textbf{A}) Comparison of a resonant EP defined by purely outgoing solutions and an absorbing EP defined by purely incoming solutions.
(\textbf{B}) Schematic motion of scattering poles (crosses) and zeros (open circles) in the complex-frequency plane. REP and AEP denote resonant and absorbing EPs, respectively.
(\textbf{C}) One-channel CPA-EP formed by two coupled microtoroids and one fiber taper, reproduced from Ref.~\cite{wang2021coherent}.
(\textbf{D}) Two-channel CPA-EP setup with independent amplitude, phase, and delay control of the coherent inputs, reproduced from Ref.~\cite{wang2021coherent}.
(\textbf{E}) Waveguide-coupled WGM resonator with a one-sided reflector, realizing directional feedback and an absorbing exceptional surface, reproduced from Ref.~\cite{soleymani2022chiral}.
}
\label{fig:3.3_abs_EPs}
\end{figure}

The pole description captures source-free resonant dynamics under purely outgoing boundary conditions. The same input-output formalism also contains scattering zeros, which correspond to coherent incident wavefronts that generate no outgoing field. A defective degeneracy of these zeros defines an absorbing EP governed by purely incoming boundary conditions~\cite{sweeney2019perfectly}.
The distinction between resonant and absorbing EPs is summarized in Fig.~\ref{fig:3.3_abs_EPs}(A) and Fig.~\ref{fig:3.3_abs_EPs}(B). Fig.~\ref{fig:3.3_abs_EPs}(A) shows that resonant poles satisfy purely outgoing boundary conditions, whereas absorbing zeros satisfy purely incoming boundary conditions. 

The purely incoming problem is obtained by imposing $\mathbf{s}_{-}=0$ in the TCMT equations. Eq.~\ref{eq:TCMT} reduces to $\dot{\boldsymbol{\alpha}} = -i(\hat{\mathcal{H}}_{\mathrm{res}}-i\mathcal{K}^T\mathcal{C}^{-1}\mathcal{D})\boldsymbol{\alpha}$. Under the standard TCMT constraints associated with reciprocity and power conservation in external channels~\cite{suh2004temporal}, the effective operator that governs the purely incoming solutions is 
\begin{equation}
\label{eq:absorbing operator}
    \hat{\mathcal{H}}_{\mathrm{abs}} \equiv \hat{\mathcal{H}}_{\mathrm{res}}-i\mathcal{K}^T\mathcal{C}^{-1}\mathcal{D} 
    = \boldsymbol{\Omega}+i\boldsymbol{\Gamma}_{\mathrm{rad}}-i\boldsymbol{\Gamma}_{\mathrm{nr}}
\end{equation}
Thus, incoming boundary conditions reverse the sign of the radiative decay, while nonradiative absorption remains dissipative. Fig.~\ref{fig:3.3_abs_EPs}(B) shows their analytic continuation in the complex-frequency plane. Crosses denote poles, and open circles denote zeros. A resonant EP corresponds to defective pole coalescence, while an absorbing EP corresponds to defective zero coalescence of $\hat{\mathcal{H}}_{\mathrm{abs}}$. When the coalesced zero reaches the real-frequency axis, the system realizes a coherent perfect absorption EP (CPA-EP)~\cite{sweeney2019perfectly,wang2021coherent}.

The generic single-channel CPA-EP implementation depicted in Fig.~\ref{fig:3.3_abs_EPs}(C) comprises two coupled silica microtoroids interfaced with a single fiber-taper waveguide~\cite{wang2021coherent}. The inter-cavity gap controls the resonator coupling, whereas the taper-resonator gap regulates the external decay rate. Simultaneous tuning of these parameters induces two scattering zeros to converge at defective degeneracy on the real-frequency axis.
The two-channel CPA-EP configuration illustrated in Fig.~\ref{fig:3.3_abs_EPs}(D) incorporates a second fiber taper and consequently necessitates coherent control of both incident amplitudes and their relative phase. Variable optical attenuators equilibrate the input amplitudes, and an electro-optic phase modulator regulates the relative phase between the two incident waves. When the input state is matched to the zero eigenchannel of the scattering matrix, the outgoing field vanishes and the system displays strongly phase-sensitive coherent perfect absorption~\cite{wang2021coherent}. 

Figure \ref{fig:3.3_abs_EPs}(E) shows a different single WGM resonator implementation in which one end of the coupling waveguide is terminated by a tunable fiber-loop reflector~\cite{soleymani2022chiral}. The resulting directional feedback modifies the coupling between the clockwise and counterclockwise resonant modes and produces a continuous exceptional surface of absorbing states. Because the feedback is direction dependent, the input from one side can match the absorbing eigenchannel while the reverse illumination experiences a different absorption response.

Absorbing EPs can also be extended beyond linear optical microcavities. In a nonlinear microwave resonator coupled to two interrogating antennas, the scattering zeros depend on the incident intensity as well as the relative phase and amplitude of the input waves~\cite{suwunnarat2022non}. By tuning the total incident power and input phase relation near a critical-coupling condition, the system can reach a self-induced near-perfect absorption state associated with nonlinear scattering-zero degeneracies. 

\subsection{Scattering matrix EPs}
\label{subsec: 3.4}
The descriptions of resonant-Hamiltonian EPs and absorbing EPs both rely on analytic continuation into the complex-frequency plane. In many optical experiments, however, the directly measured object is not the poles or zeros, but a finite-frequency response matrix that connects incident channels with outgoing channels at a prescribed real frequency. This induces another class of photonic EPs, where the defective operator is the response matrix itself.

For a general finite scattering or polarization system, the input-output relation can be written as 
\begin{equation}
    \ket{s_{-}}=\mathcal{R}\ket{s_{+}},
\end{equation}
where $\mathcal{R}$ may denote a multi-port scattering matrix, a diffraction-channel matrix on a metasurface, or a Jones matrix of polarization states. The right eigenvectors of $\mathcal{R}$ represent invariant incident wavefronts (or eigenchannels) that preserve their spatial profile after scattering. When a response-matrix EP occurs, the distinct incident wavefront profiles merge into a single specific excited state, rendering the external scattering basis defective.

\begin{figure}[htbp]
\includegraphics[width=0.85\textwidth]{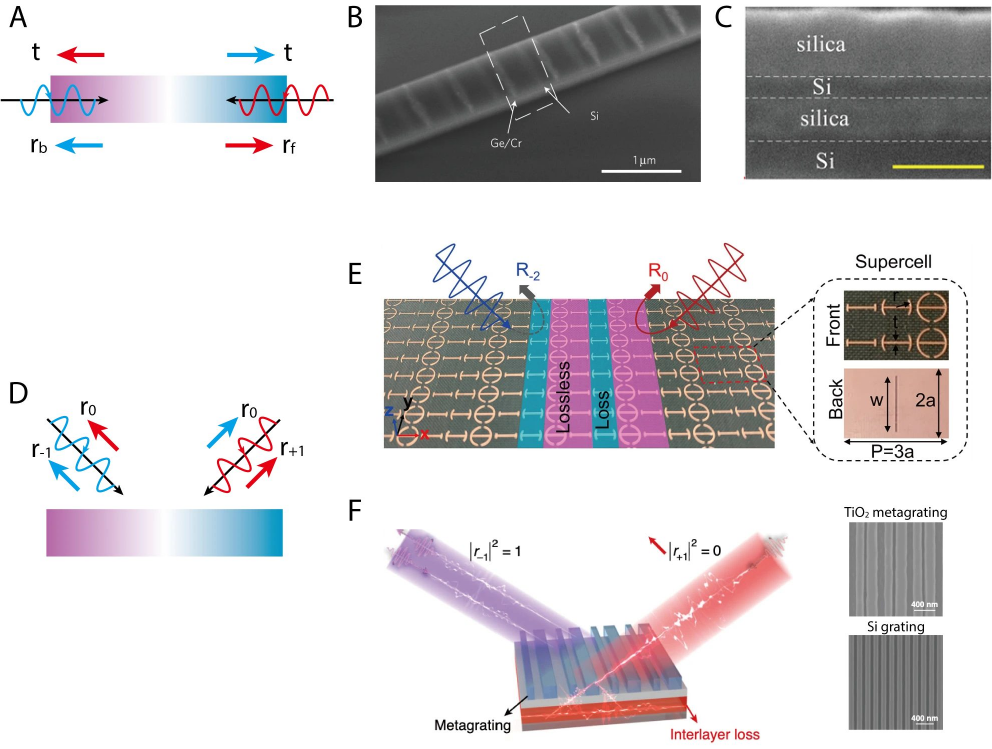}
\centering
\caption{\textbf{Concept and representative implementations of scattering EPs}
(\textbf{A}) Schematic of reciprocal one-dimensional scattering structure in a directional-channel basis.
(\textbf{B}) Integrated passive complex grating formed on a silicon waveguide with periodically placed $\ce{Ge}/\ce{Cr}$~\cite{feng2013experimental}.
(\textbf{C}) Planar lossy dielectric multilayer composed of alternating $\ce{Si}$ and silica films~\cite{feng2013demonstration}.
(\textbf{D}) Schematic of reflective gradient metasurfaces in diffraction-channel basis.
(\textbf{E}) Microwave loss-assisted metasurface based on patterned metallic supercells, where lossless and lossy regions are arranged within each period~\cite{dong2020loss}.
(\textbf{F}) Visible-frequency bilayer metagrating architecture with an upper $\ce{TiO2}$ diffractive layer and a lower lossy $\ce{Si}$ grating~\cite{he2023scattering}.
}
\label{fig:3.4_scattering_EPs}
\end{figure}

A simple case is a reciprocal two-port scattering system. After a conventional permutation of the output channels, the scattering matrix can be written in a transmission-reflection basis as~\cite{chong2011pt,lin2011unidirectional} 
\begin{equation} \label{eq: permuted scattering matrix}
    \mathcal{S}=
\begin{pmatrix} t & r_{b} \\ r_{f} & t \end{pmatrix},
\end{equation}
where $t$ is the reciprocal transmission coefficient, $r_{f}$ and $r_{b}$ are the reflection coefficients for the incidence of forward (left-side) and backward (right-side), respectively. The scattering eigenvalues are
\begin{equation}
    \lambda_{\pm}^{S}=t \pm \sqrt{r_{f}r_{b}}.
\end{equation}
Thus, a scattering-matrix EP occurs when one reflection coefficient vanishes while the other remains finite ($r_{f}=0$, $r_{b}\neq0$, or vice versa). The essential condition of scattering EPs is single-sided reflectionlessness. This differs from ordinary bidirectional impedance matching, for which both reflections vanish and $\mathcal{S}=t\mathbb{I}$. Fig.~\ref{fig:3.4_scattering_EPs} compares two channel representations of a real-frequency scattering EP.

Figure \ref{fig:3.4_scattering_EPs}(A) uses the two sides of a one-dimensional structure as the external channels, so the defect is identified by vanishing reflection from one side. Longitudinally patterned PT-symmetric optical media provided one of the first physical settings for this type of scattering EP~\cite{ruschhaupt2005physical}. In these structures, the complex refractive index is modulated along the propagation direction, and a balanced complex index grating can suppress reflection from one side while retaining or enhancing it from the opposite side~\cite{lin2011unidirectional}. The integrated silicon grating shown in Fig.~\ref{fig:3.4_scattering_EPs}(B) implements this mechanism using a silicon waveguide embedded in $\ce{SiO2}$, where the real-index modulation is provided by the patterned silicon waveguide and the loss modulation is introduced by periodically arranged $\ce{Ge}$-$\ce{Cr}$ absorbers~\cite{feng2013experimental}. Related integrated complex-grating designs further developed the same guided-mode Bragg scattering and distributed absorption strategy~\cite{zhao2016metawaveguide}.

Another implementation uses a multilayer planar stack composed of alternating thin films of absorbing amorphous $\ce{Si}$ and low-loss $\ce{SiO2}$, as shown in Fig.~\ref{fig:3.4_scattering_EPs}(C)~\cite{feng2013demonstration}. The film thicknesses are chosen so that the accumulated propagation phase and absorption cancel reflection for one illumination direction while maintaining finite reflection from the opposite direction. A related free-space realization uses a three-layer $\ce{Au}$-polymer-$\ce{Au}$ heterostructure, in which the thermally deformable polymer layer is perturbed by temperature, breaking the EP condition, and thus causing a drastically enhanced reflection coefficient~\cite{zhao2018exceptional}.

The same scattering response matrix idea can be extended from one-dimensional two-port structures to metasurfaces, where the external channels are diffraction orders. Angularly asymmetric radiation in free space can be achieved by tailoring local loss profiles in phase-gradient metasurfaces~\cite{wang2018extreme,li2022controlling}, and the same physics can be formulated in terms of scattering-matrix EPs when the relevant diffraction channels become defective~\cite{wang2019extremely,dong2020loss}. For a reflective gradient metasurface operated near a Littrow or retroreflection condition, the truncated two-channel reflection matrix is defined in the diffraction-channel basis shown in Fig.~\ref{fig:3.4_scattering_EPs}(D), and can be written as 
\begin{equation} \label{eq: diffraction scattering matrix}
    \mathcal{S}= \begin{pmatrix} r_0 & r_{-1}\\ r_{+1} & r_0 \end{pmatrix},
\end{equation}
where $r_0$ is the specular reflection coefficient and $r_{\pm1}$ denotes the two opposite retroreflection or diffraction channels. The corresponding eigenvalues are
\begin{equation}
    \lambda_{\pm}^{S}=r_0 \pm \sqrt{r_{+1}r_{-1}},
\end{equation}
so that a diffraction channel scattering EP requires $r_{+1}r_{-1}=0$, with only one of the two diffraction channels suppressed. Experimentally, this condition can be reached by combining a phase-gradient design, which redirects incident light into a selected diffraction order, with spatially engineered loss, which cancels one retroreflection channel through destructive interference and absorption while leaving the opposite channel finite.

At microwave frequencies, this mechanism has been demonstrated using the metallic gradient metasurface shown in Fig.~\ref{fig:3.4_scattering_EPs}(E), where split-ring supercells combine phase-gradient control with deliberately engineered local loss~\cite{dong2020loss}. At visible frequencies, the bilayer metasurface shown in Fig.~\ref{fig:3.4_scattering_EPs}(F) demonstrates high-efficiency scattering EPs. The upper $\ce{TiO2}$ metagrating layer provides low-loss directional diffraction, while the lower lossy $\ce{Si}$ subwavelength grating supplies the absorption required to suppress the reverse retroreflection channel~\cite{he2023scattering}.

\subsection{Jones matrix EPs in polarization space}
\label{subsec: 3.5}
Response-matrix EPs also arise in the polarization space. A Jones matrix describes the polarization conversion between two basis states of light. In the circular-polarization basis, it can be written as 
\begin{equation}
    \mathcal{J}= \begin{pmatrix} J_{LL} & J_{LR}\\J_{RL} & J_{RR} \end{pmatrix}
    =\frac{1}{2}
    \begin{pmatrix} 
    J_{xx}+J_{yy}+i(J_{xy}-J_{yx}) & J_{xx}-J_{yy}-i(J_{xy}+J_{yx})
    \\J_{xx}-J_{yy}+i(J_{xy}+J_{yx}) & J_{xx}+J_{yy}-i(J_{xy}-J_{yx}) 
    \end{pmatrix}
\end{equation}
where $J_{ij}$ denotes the complex coefficients that convert an input $j$-polarization state into an output $i$-polarization state. The subscripts $L$ and $R$ refer to left- and right-circularly polarized light, whereas the subscripts $x$ and $y$ refer to two orthogonal linear polarizations. For a general $2\times2$ Jones response matrix, the EP condition is 
\begin{equation}
    (J_{LL}-J_{RR})^2+4J_{LR}J_{RL}=0.
\end{equation}
In many reciprocal planar metasurfaces with normal incidence, the linear-basis Jones matrix is symmetric ($J_{xy}=J_{yx}$), which leads to $J_{LL}=J_{RR}$. Under this additional constraint, the EP condition reduces to the disappearance of exactly one cross-circular conversion channel, $J_{LR}=0, J_{RL}\neq0$ or $J_{RL}=0, J_{LR}\neq0$. Such EPs have coalesced circularly polarized eigenstates and are often referred to as chiral Jones-matrix EPs. More general structures, in which the four Jones-matrix elements can be independently controlled, can instead realize EPs at arbitrary points on the Poincar\'e sphere, with arbitrary coalesced eigenpolarizations~\cite{qin2025sphere}.
\begin{figure}[htbp]
\includegraphics[width=\textwidth]{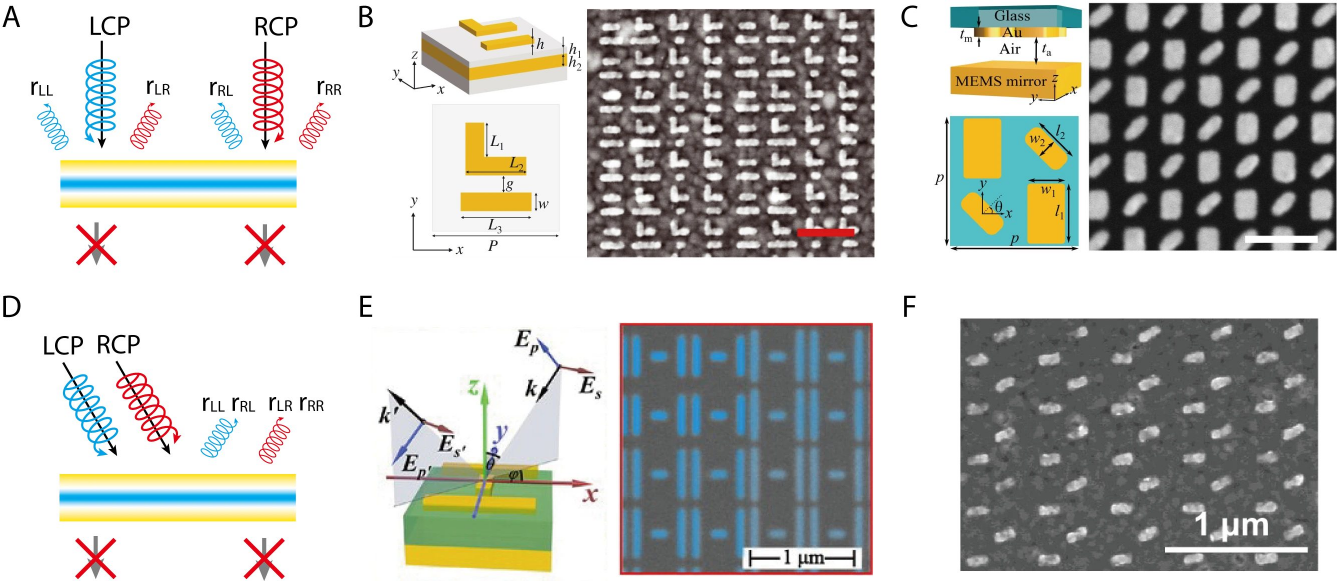}
\centering
\caption{\textbf{Reflection Jones EPs at normal and oblique incidence}
(\textbf{A}) Schematic of normal-incidence circular-polarization reflection basis.
(\textbf{B}) Three-layer reflective chiral plasmonic metasurface, adapted from Ref.~\cite{song2021plasmonic}.
(\textbf{C}) Voltage-tunable MEMS reflection metasurface, adapted from Ref.~\cite{ding2024electrically}.
(\textbf{D}) Schematic of oblique-incidence circular-polarization reflection basis.
(\textbf{E}) Nonlocal $\ce{Au}$ nanorod reflective plasmonic metasurface supporting momentum space chiral EPs, adapted from Ref.~\cite{zhao2024mode}.
(\textbf{F}) Paired $\ce{Al}$ nanorod reflective metasurface for arbitrary polarization EP engineering, reproduced from Ref.~\cite{qin2025sphere}.
}
\label{fig:3.5_Jones_R_EPs}
\end{figure}

The first configuration uses the reflection-based Jones matrix. For normal-incidence reflection, we use the circular-polarization channel convention shown in Fig.~\ref{fig:3.5_Jones_R_EPs}(A). The handedness of each circular state is defined with respect to its local propagation direction, and the reflection Jones matrix is
\begin{equation}
    \mathcal{J_R}= \begin{pmatrix} r_{LL} & r_{LR}\\r_{RL} & r_{RR} \end{pmatrix}.
\end{equation}

Figure \ref{fig:3.5_Jones_R_EPs}(B) presents a three-layer reflective plasmonic metasurface consisting of a patterned chiral metallic layer, a dielectric spacer, and an opaque metallic ground plane~\cite{song2021plasmonic}. The metallic ground plane blocks transmission, so the optical response is governed mainly by the reflection Jones matrix. The top metallic pattern provides anisotropic and chiral coupling between circular polarization channels, whereas the dielectric spacer controls the phase accumulated between the patterned layer and the mirror. In Jones EPs, one conversion coefficient vanishes, for example, $r_{LR}=0$, whereas the opposite coefficient $r_{RL}$ remains finite~\cite{song2021plasmonic}. The mirror-symmetric counterpart of the same unit cell reverses the handedness of the EP and realizes the opposite condition ($r_{RL}=0$, while $r_{LR} \neq 0$), providing a route to paired EPs for arbitrary polarization control~\cite{yang2024creating}. Fig.~\ref{fig:3.5_Jones_R_EPs}(C) shows a dynamically tunable version where a voltage-controlled piezoelectric micro-electromechanical systems (MEMS) mirror changes the cavity spacing above a chiral $\ce{Au}$ meta-atom array, thereby moving the reflection response through Jones EPs of opposite handedness~\cite{ding2024electrically}.

Reflection Jones EPs can also be engineered through extrinsic chirality under oblique incidence, as indicated in Fig.~\ref{fig:3.5_Jones_R_EPs}(D). The nonlocal plasmonic metasurface in Fig.~\ref{fig:3.5_Jones_R_EPs}(E) uses collective guided-mode resonances of a polyatomic $\ce{Au}$ nanorod array to position chiral Jones EPs in momentum space, allowing spin- and angle-selective reflection responses~\cite{zhao2024mode}. Fig.~\ref{fig:3.5_Jones_R_EPs}(F) shows a more general paired aluminum nanorod design where the two rotation angles provide independent control over the off-diagonal Jones-matrix elements. This additional degree of freedom allows the coalesced eigenpolarization to be placed away from the circular-polarization poles of the Poincar\'e sphere~\cite{qin2025sphere}.

\begin{figure}[htbp]
\includegraphics[width=0.85\textwidth]{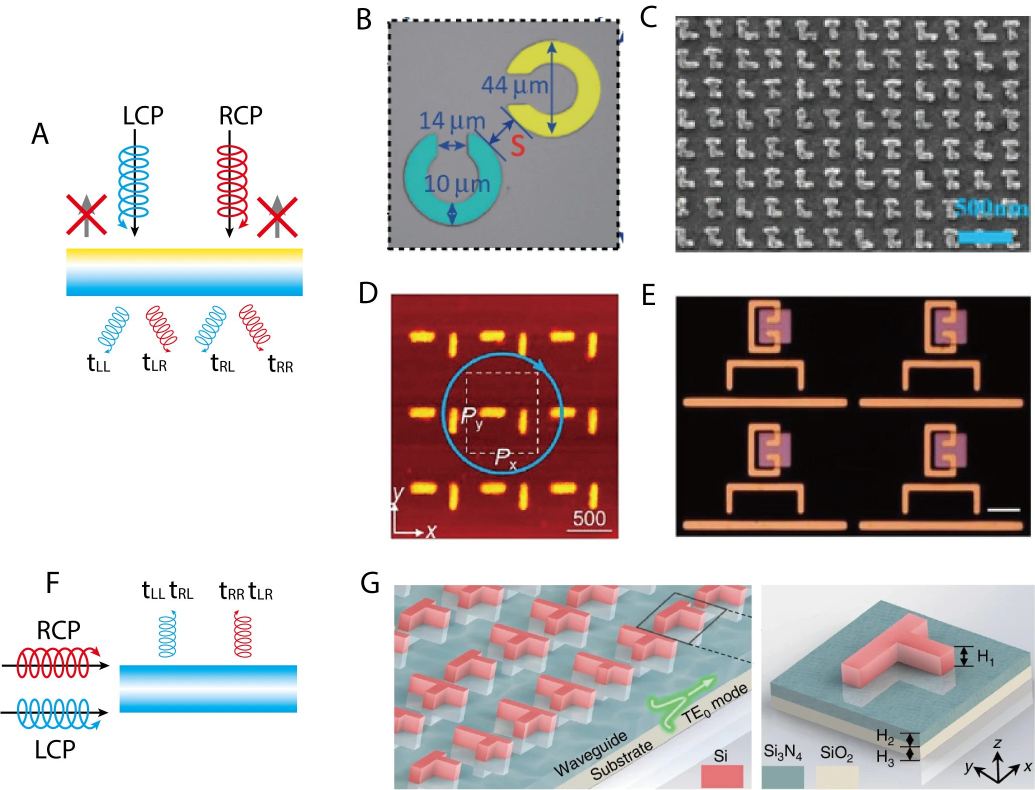}
\centering
\caption{\textbf{Free-space and integrated transmission Jones EPs}
(\textbf{A}) Schematic of the normal-incidence circular-polarization transmission basis.
(\textbf{B}) Terahertz metasurface formed by two orthogonally oriented split-ring resonators with unequal losses, reproduced from Ref.~\cite{lawrence2014manifestation}.
(\textbf{C}) Chiral $\ce{\tau}$-shaped $\ce{Ag}$ meta-atoms supporting twin transmission EPs, reproduced from Ref.~\cite{wu2024twins}.
(\textbf{D}) Nonlocal $\ce{Au}$ nanorod plasmonic metasurface operating at telecom wavelengths, reproduced from Ref.~\cite{li2024nonlocal}
(\textbf{E}) Photoactive terahertz metasurface containing coupled metallic split-ring resonators and an amorphous-$\ce{Ge}$ loss layer, reproduced from Ref.~\cite{yu2024creating}.
(\textbf{F}) Schematic of guided-wave circular-polarization input and free-space output convention.
(\textbf{G}) Waveguide integrated dielectric metasurface consisting of silicon meta-atoms on a $\ce{Si3N4}$/$\ce{SiO2}$ photonic platform, reproduced from Ref.~\cite{yi2025metasurfaceEP}.
}
\label{fig:3.6_Jones_L_EPs}
\end{figure}

The second configuration uses a transmission-based Jones matrix. For free-space transmission at normal incidence, the circular-polarization channel convention is shown in Fig.~\ref{fig:3.6_Jones_L_EPs}(A), and the transmission Jones matrix is
\begin{equation}
    \mathcal{J_T}= \begin{pmatrix} t_{LL} & t_{LR}\\t_{RL} & t_{RR} \end{pmatrix}.
\end{equation}
As with $\mathcal{J_R}$, the EP conditions here require either $t_{LR}=0$ or $t_{RL}=0$, while the other remains finite. Thus, the transmission Jones EP is experimentally realized by engineering asymmetric circular-polarization conversion in transmission while keeping the opposite conversion channel finite.

The first experimental observation of a coalesced polarization state at a Jones EP was reported in a terahertz metasurface operating in polarization space. The meta-atoms were two orthogonally oriented split-ring resonators (SRRs) with matched resonance frequencies but unequal losses, implemented with dissimilar metallic elements on a silicon substrate, as shown in Fig.~\ref{fig:3.6_Jones_L_EPs}(B)~\cite{lawrence2014manifestation}. Such structures use two independent linear resonant responses as the basis for constructing a non-Hermitian Jones matrix. The coupling between the two resonators and the loss contrast determine whether the eigenpolarizations remain distinct or coalesce at an EP. 
A more general terahertz implementation uses anisotropic $\ce{Au}$/$\ce{Gr}$ SRRs with different gap sizes, allowing the radiative and absorptive losses as well as the near-field coupling to be tuned beyond the ideal balanced gain–loss case~\cite{park2020observation}. Without fabricating multiple static samples, active tuning was then introduced using a gated graphene terahertz metasurface, in which two coupled SSRs are bridged by a graphene microribbon whose conductivity is controlled through an ion-gel gate~\cite{baek2023non}.

Another transmission-type implementation uses a subwavelength non-Hermitian metasurface composed of three $\ce{Ag}$ rectangular bars arranged in a chiral $\tau$-shaped unit cell, as shown in Fig.~\ref{fig:3.6_Jones_L_EPs}(C)~\cite{wu2024twins}. By breaking the in-plane mirror symmetry, the phase and amplitude of the transmission channels can be controlled independently. This design enables the formation of twin EPs with opposite chirality, analogously to the paired-EP strategies developed in reflection-type Jones metasurfaces~\cite{song2021plasmonic,yang2024creating}. At telecom wavelengths, chiral transmission EPs have been realized in the nonlocal $\ce{Au}$ nanorod plasmonic metasurface shown in Fig.~\ref{fig:3.6_Jones_L_EPs}(D)~\cite{li2024nonlocal}. Tunable transmission Jones EPs can also be achieved in $\ce{Ge}$-hybrid terahertz non-Hermitian metasurfaces, where coupled metallic SRRs and an additional cut-wire resonator control direct and indirect radiative coupling~\cite{he2023transient,yu2024creating,he2025loss}. In the implementation shown in Fig.~\ref{fig:3.6_Jones_L_EPs}(E), a thin amorphous-$\ce{Ge}$ layer serves as a photoactive lossy element. Optical pumping generates photocarriers and changes its conductivity, thereby tuning the loss of selected resonators and enabling loss-induced anti-chiral EPs~\cite{yu2024creating}.

Transmission Jones EPs can also be implemented in on-chip dielectric photonic platforms. In the integrated metasurface of all-dielectric waveguides shown in Fig.~\ref{fig:3.6_Jones_L_EPs}(G), a fused-silica substrate supports a $\ce{Si3N4}$ planar waveguide and $T$-shaped silicon meta-atoms are patterned on top of the waveguide~\cite{yi2025metasurfaceEP}. The incident wave is injected from the lateral edge of the chip and propagates inside the $\ce{Si3N4}$ layer as a guided mode. As it travels along the waveguide, the silicon meta-atoms act as out-coupling antennas that continuously extract the guided wave into free-space radiation. Their geometry controls the local polarization conversion and allows the transmission-type Jones response to be engineered in an integrated photonic setting.

\subsection{Bloch EPs in spatially periodic photonic structures}
\label{subsec: 3.6}
When a non-Hermitian system is periodic, translational symmetry allows its eigenmodes to be represented by the Bloch wavevector $\mathbf{k}$. The wavefunction can be written as
\begin{equation}
    \psi_{nk}(\mathbf{r})=e^{i\mathbf{k}\cdot\mathbf{r}}u_{nk}(\mathbf{r}),
    \qquad
    u_{nk}(\mathbf{r+R})=u_{nk}(\mathbf{r})
\end{equation}
where $u_{nk}$ has the same periodicity as the lattice and $R$ is a Bravais lattice vector. The band problem is then reduced to a $\mathbf{k}$-dependent effective Hamiltonian $\hat{\mathcal{H}}(\mathbf{k})$. The eigenvalue represents the complex band energy in electronic systems, the complex eigenfrequency or resonance energy in photonic crystal, or the complex propagation constant in waveguide arrays. The eigenvector represents the internal Bloch-mode profile within one unit cell. A Bloch EP occurs when non-Hermitian Bloch bands coalesce in both eigenvalues and eigenvectors in momentum space. This distinguishes Bloch EPs from propagation-Hamiltonian or resonant-Hamiltonian EPs that are obtained by tuning external control parameters. Bloch EPs exist in the intrinsic band structure of a periodic lattice, where the Bloch wavevector itself acts as a continuous parameter. 

For a generic two-band non-Hermitian Bloch Hamiltonian, one may write
\begin{equation}
    \hat{\mathcal{H}}(\mathbf{k})=d_0(\mathbf{k})\mathbb{I}
    +\mathbf{d}(\mathbf{k})\cdot\boldsymbol{\sigma}
\end{equation}
where $d_0$ gives a common complex energy shift of bands, $\boldsymbol{\sigma}$ denotes the Pauli matrices, and $\mathbf{d}(\mathbf{k})=\mathbf{d}_{\rm R}(\mathbf{k})+i\mathbf{d}_{\rm I}(\mathbf{k})$ is a complex vector. The corresponding complex eigenvalues are
\begin{equation}
    E_{\pm}(\mathbf{k})
    =d_0(\mathbf{k})\pm
    \sqrt{\mathbf{d}_{\rm R}^2-\mathbf{d}_{\rm I}^2
    +2i\,\mathbf{d}_{\rm R}\cdot\mathbf{d}_{\rm I}}.
\end{equation}
Therefore, for a non-trivial $\mathbf{d}(\mathbf{k})\neq0$, the exceptional-degeneracy conditions of the band are
\begin{equation}
    \mathbf{d}_{\rm R}^2(\mathbf{k})
    =
    \mathbf{d}_{\rm I}^2(\mathbf{k}),
    \qquad
    \mathbf{d}_{\rm R}(\mathbf{k})\cdot
    \mathbf{d}_{\rm I}(\mathbf{k})=0.
\end{equation}
Since these two EP conditions are real equations, isolated EPs are generic in a two-dimensional Brillouin zone, whereas exceptional lines or rings can appear in higher-dimensional momentum spaces or under additional symmetry constraints.

\begin{figure}[htbp]
\includegraphics[width=0.9\textwidth]{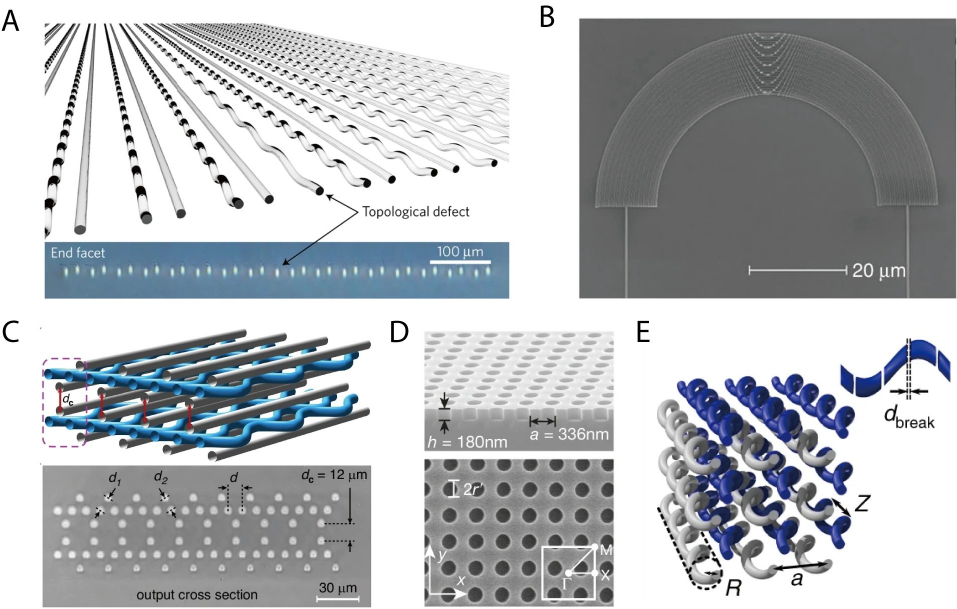}
\centering
\caption{\textbf{Representative implementations of Bloch EPs}
(\textbf{A}) End-facet microscope image of a laser-written fused silica waveguide array with alternating straight and sinusoidally modulated waveguides for site-dependent radiative loss~\cite{weimann2017topologically}. 
(\textbf{B}) A bent silicon waveguide lattice, where alternating waveguides are coated with chromium layers to induce loss~\cite{xu2016experimental}.
(\textbf{C}) Coupled non-Hermitian optical waveguide arrays fabricated via femtosecond laser direct writing in a glass chip~\cite{wang2025coupled}.
(\textbf{D}) Single photonic crystal slab with spawning exceptional rings out of Dirac cones~\cite{zhen2015spawning}.
(\textbf{E}) Bipartite helical waveguide array fabricated in borosilicate glass, one sublattice has periodic breaks to induce controlled loss~\cite{cerjan2019experimental}.
}
\label{fig:3.7_Bloch_EPs}
\end{figure}

In a one-dimensional Brillouin zone, the Bloch wavevector $\mathbf{k}$ provides only one real degree of freedom. Therefore, a second-order Bloch EP is not generic in a one-dimensional lattice unless additional constraints or tuning parameters are introduced. The passive PT-symmetric dimerized waveguide lattice shown in Fig.~\ref{fig:3.7_Bloch_EPs}(A) provides a representative one-dimensional realization. In this platform, femtosecond-laser-written waveguides in fused silica form a non-Hermitian Su-Schrieffer-Heeger (SSH) chain. Alternating loss is introduced by periodically wiggling selected waveguides, and the dimerization controls the transition between PT-broken and PT-unbroken bulk bands. At the transition point, two Bloch bands coalesce at the Brillouin-zone boundary, corresponding to a one-dimensional Bloch EP~\cite{weimann2017topologically}. Another representative realization is a PT-symmetric synthetic silicon photonic lattice on a silicon-on-insulator waveguide array. As illustrated in Fig.~\ref{fig:3.7_Bloch_EPs}(B), chromium stripes deposited on every other waveguide introduce alternating loss and double the unit cell. This configuration generates a two-band non-Hermitian Bloch Hamiltonian with two EPs in the transverse band structure. Curved waveguides were used to sweep the transverse Bloch momentum during propagation, and the field evolution was imaged by scanning near-field optical microscopy~\cite{xu2016experimental}. Another one-dimensional platform links Bloch EPs to non-Hermitian skin physics. In femtosecond-laser-written photonic waveguide arrays, a bipartite lattice with engineered artificial gauge fields and on-site losses was used to create multiple pairs of EPs in the momentum-space band structure. As shown in Fig.~\ref{fig:3.7_Bloch_EPs}(C), the waveguides were fabricated in boroaluminosilicate glass, with non-Hermitian loss introduced by periodic breaks in selected waveguides. In this system, the EPs reshape the quasienergy spectrum under periodic boundary conditions and strongly affect the coupled non-Hermitian skin effect, demonstrating that momentum-space Bloch EPs can act as tunable singularities governing non-Hermitian transport dynamics~\cite{wang2025coupled}.

Compared with one-dimensional lattices, two-dimensional photonic-crystal slabs provide a larger in-plane momentum space, where light is confined through Bragg interference in periodic dielectric nanostructures~\cite{chan2025essay}. A representative experimental platform is the leaky photonic-crystal slab, where non-Hermiticity arises from radiation into free space. Large-area periodic patterns are fabricated in a $\ce{Si3N4}$ slab on a silica layer and immersed in an index-matching liquid. By adjusting the aperture radius and the refractive index of the surrounding liquid, dipole and quadrupole modes with different symmetries are tuned to collide near the $\Gamma$ point. Angle-resolved reflectivity measurements revealed an exceptional ring emerging from a Dirac cone~\cite{zhen2015spawning}, as shown in Fig.~\ref{fig:3.7_Bloch_EPs}(D). Another canonical realization is the observation of a bulk Fermi arc and polarization half charge. In that experiment, a $\ce{Si3N4}$ layer on silica was patterned into rhombic unit cells with elliptical air holes. Angle-resolved scattering was performed using a vertically polarized tunable continuous-wave Ti:sapphire laser. The experiment directly imaged the transition from closed isofrequency contours to an open bulk Fermi arc, and polarimetry measurements reconstructed the associated half-integer polarization winding~\cite{zhou2018observation}. Another realization uses an all-dielectric terahertz metasurface to connect Bloch EP with bound states in the continuum (BICs). The sample consists of a free-standing high-resistivity silicon slab perforated by a square lattice of air holes. By tuning the in-plane wavevector through the incident angle, a quasi-Friedrich-Wintgen BIC evolves into an exceptional ring. Optical pumping injects carriers into silicon, modifies the dissipative perturbation, and dynamically switches the EP state~\cite{wang2025photoswitchable}. Bloch exceptional rings have also been observed in one-dimensional perovskite photonic crystals, whose optical response contains additional degrees of freedom that can support multiple exceptional rings and a bulk Fermi torus in momentum space~\cite{na2025multiple}. Experimentally, cross-polarization momentum-space imaging has further simplified the direct observation of Bloch EPs in leaky photonic-crystal slabs by suppressing the non-resonant background that usually masks guided resonances in conventional reflectivity spectra~\cite{nguyen2023direct}.

Beyond two-dimensional slabs, Bloch EPs can also form higher-dimensional exceptional manifolds in three-dimensional photonic lattices. A representative experiment is the realization of a Weyl exceptional ring in a bipartite helical waveguide array. The structure consists of single-mode helical waveguides written in borosilicate glass by femtosecond direct laser writing. In the Hermitian limit, the helical waveguide array behaves as a three-dimensional photonic crystal supporting a Weyl point. As shown in Fig.~\ref{fig:3.7_Bloch_EPs}(E), controlled breaks added to one waveguide sublattice non-Hermiticity through sublattice-selective loss. As a result, the Weyl point expands into a closed ring of exceptional points in momentum space~\cite{cerjan2019experimental}.

\subsection{Floquet EPs in temporally periodic photonic structures}
\label{subsec: 3.7}
Floquet EPs arise in temporally periodic non-Hermitian systems as exceptional degeneracies of the Floquet Hamiltonian, or equivalently of the one-period evolution operator~\cite{longhi2017floquet}. For a linear system driven by a time-periodic Hamiltonian of period $T$,
\begin{equation}
    i\frac{d}{dt}\psi(t) = \hat{\mathcal{H}}(t)\psi(t),
    \qquad
    \hat{\mathcal{H}}(t+T) = \hat{H}(t)
\end{equation}
the Floquet theorem gives solutions of the kind
\begin{equation}
    \psi_n(t) = e^{-i\varepsilon_nt} u_n(t),
    \qquad
    u_n(t+T) = u_n(t).
\end{equation}
where $\varepsilon_n$ is the normalized quasienergy (i.e., when assuming $\hbar=1$,  that is a complex natural angular frequency) of the system, the one-period evolution operator is 
\begin{equation}
    U(T) = \mathcal{T} e^{-i\int_{0}^{T}\hat{\mathcal{H}}(\tau)d\tau} = e^{-i \hat{\mathcal{H}}_F T}
\end{equation}
where $\mathcal{T}$ executes the time ordering and $U(T)$ defines the Floquet Hamiltonian $\hat{\mathcal{H}}_F$. A Floquet EP occurs when the degenerate quasienergy has fewer independent Floquet eigenstates than its order.

\begin{figure}[htbp]
\includegraphics[width=0.9\textwidth]{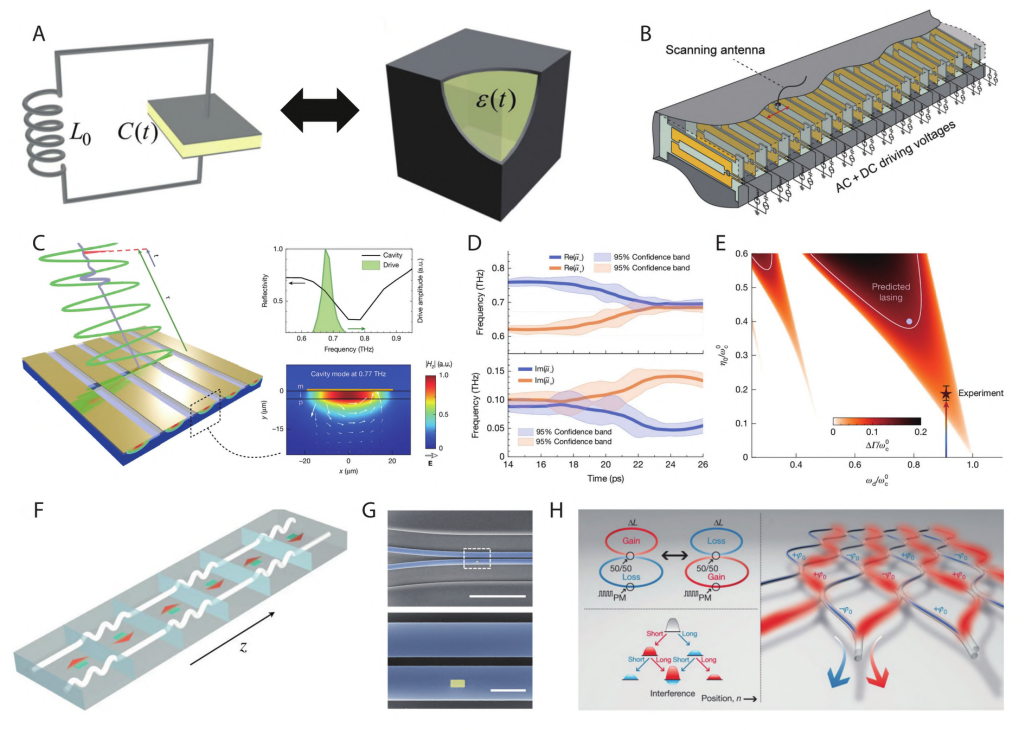}
\centering
\caption{\textbf{Experimental platforms for Floquet and Floquet-type EPs in photonics. }
(\textbf{A}) A single LC resonator with a time-varying capacitor $C(t)$ (left), and a metallic resonator filled with a dielectric of time-varying permittivity $\varepsilon(t)$ (right)~\cite{kazemi2019exceptional}. 
(\textbf{B}) Varactor-loaded split-ring resonator array embedded in a microwave waveguide, realizing a time-modulated photonic Floquet medium~\cite{park2022revealing}. 
(\textbf{C}) All-optically driven surface-plasmon cavity metamaterial realizing a laboratory-time terahertz photonic time crystal. Upper right: equilibrium reflectivity at 290$\mathrm{K}$ (black) and normalized drive spectrum (green); lower right: electric-field vectors and magnetic-field amplitude~\cite{Guo2026Plasmonic}.
(\textbf{D}) Temporal evolution of the complex Floquet eigenvalues (solid curves), with 95\% confidence bands~\cite{Guo2026Plasmonic}.
(\textbf{E}) Calculated parametric-loss-compensation map versus drive frequency and modulation strength, marking the experimental operating point and predicted plasmonic-lasing region~\cite{Guo2026Plasmonic}.
(\textbf{F}) Integrated waveguide array with periodic gain–loss modulation along the propagation direction for Floquet PT symmetry~\cite{liu2024floquet}.
(\textbf{G}) On-chip coupled-waveguide implementation of spatially chirped Floquet PT photonics~\cite{mao2025chip}.
(\textbf{H}) Coupled fiber-loop temporal mesh lattice realizing discrete-time PT-symmetric Floquet evolution~\cite{regensburger2012parity}.
}
\label{fig:3.8_Floquet_EPs}
\end{figure}

In electromagnetics, the simplest realization already occurs in a single time-periodic resonator \cite{kazemi2019exceptional} (see Fig.~\ref{fig:3.8_Floquet_EPs}A). 
Kazemi \textit{et al.} showed that in such a system 
a second-order EP degeneracy arises as a consequence of the temporal modulation~\cite{kazemi2019exceptional}. 
In a classical state-space formulation,
%
%
the evolution of the state vector $\boldsymbol{\Psi}(t)$ over one modulation period is described by the 
state-transition matrix $\mathbf{\Phi}$ as
\begin{equation}
\boldsymbol{\Psi}(t+T)
=
\mathbf{\Phi}\boldsymbol{\Psi}(t),
\end{equation}
Together with the solution form arising from the Floquet theorem,
\begin{equation}
\mathbf{\Phi}\boldsymbol{\Psi}_{n}
=
e^{-i\omega_{n}T}\boldsymbol{\Psi}_{n},
\end{equation}
one can determine the natural complex-valued frequencies $\omega_n$ of a system 
The exceptional point is reached when $\mathbf{\Phi}$ becomes 
non-diagonalizable and develops a nontrivial Jordan block. For a non-diagonal 
$2\times2$ transition matrix, a second-order degeneracy occurs when
\begin{equation}
\frac{\operatorname{tr}(\mathbf{\Phi})}{2}
=
\pm\sqrt{\det(\mathbf{\Phi})}.
\end{equation}
Remarkably, this condition can be realized in a lossless and gainless single 
LC resonator simply by periodically switching its capacitance between two 
values. In this case, $\det(\mathbf{\Phi})=1$, and the {\em degenerate} Floquet 
multiplier is $e^{-i\omega_{E}T}=\pm1$, corresponding to quasifrequencies 
at the center or edge of the "temporal Brillouin zone" \cite{kazemi2019exceptional}. The experimental realization of a simple Floquet EP in a singe resonator has been  demonstrated using a time-modulated capacitor \cite{kazemi2022experimental}. The degeneracy of the two system's natural frequencies, the algebraic growth of the magnitude of the time domain signal were demonstrated, together with the EP-enhanced sensitivity of the natural frequency to a system's perturbation. At the exceptional point, the generalized Floquet eigenvector produces an algebraic response: the resonator current amplitude grows linearly with time, while the stored energy grows 
quadratically. This growth does not originate from a conventional gain element; 
instead, energy is exchanged with the external source responsible for the temporal modulation. Importantly, this single-resonator example demonstrates 
that temporal periodicity itself can provide the Floquet degrees of freedom required for eigenvalue and eigenvector coalescence by simply modifying the modulation frequency, without the need for physically coupled resonators with PT symmetry including the conventional balanced gain and loss.

Floquet EP occur also in the realm of microwave and photonic time-crystal platforms \cite{zurita2009reflection}. Time and space-time modulations have been studied  waveguide systems  \cite{reyes2015observation} and it has been shown that EPs occur there as well \cite{wang2018photonic,rouhi2020exceptional,koutserimpas2020electromagnetic}. A representative experimental realization of Floquet EP has been realized in a customized rectangular waveguide loaded with a one-dimensional array of split-ring LC resonators. The resonators were patterned on low-loss dielectric substrates using a printed-circuit-board process, and each split gap was loaded with a varactor diode. Fig.~\ref{fig:3.8_Floquet_EPs}(B) shows the varactor-loaded resonator array, in which an AC voltage applied to the diodes synchronously modulates the resonator capacitances and resonance frequencies, forming an effective time-periodic photonic medium. The Bloch-Floquet band structure was reconstructed through spatiotemporal field measurements, and exceptional transitions were observed at the edges of the primary momentum gap~\cite{park2022revealing}. More recently, laboratory-time Floquet EP physics was extended from electrically modulated microwave platforms to an all-optically driven terahertz photonic time crystal shown in Fig.~\ref{fig:3.8_Floquet_EPs}(C). This platform was implemented in an $\ce{InSb}$ surface-plasmon cavity metamaterial. As shown in Fig.~\ref{fig:3.8_Floquet_EPs}(D), strong, coherent subcycle modulation of the carrier kinetic energy and effective mass enabled time-resolved reconstruction of two complex Floquet resonances. The resulting parametric amplification compensated for more than 50\% of the plasmonic loss. Plasmonic lasing was predicted under optimized conditions, as shown in Fig.~\ref{fig:3.8_Floquet_EPs}(E), but was not observed experimentally~\cite{Guo2026Plasmonic}.

By contrast, in certain waveguide implementations, the formal equivalence between paraxial Helmholtz equation and the time-dependent Schr\"odinger equation allows evolution along the propagation coordinate $z$ to emulate laboratory-time dynamics. Longitudinally periodic modulation thereby implements a $z$-Floquet evolution analogues in integrated borosilicate-glass waveguide arrays comprising periodically arranged evanescently coupled waveguide sections. In the integrated waveguide array shown in Fig.~\ref{fig:3.8_Floquet_EPs}(F), straight waveguides provide the lower-loss channels, whereas tailored curved waveguides provide the higher-loss channels. By periodically swapping the positions of the lower-loss and higher-loss channels along the propagation direction, the system realizes a Floquet PT-symmetric evolution whose quasienergy spectrum and EPs are controlled by the modulation period~\cite{liu2024floquet}. The on-chip Floquet PT-symmetric photonic waveguides shown in Fig.~\ref{fig:3.8_Floquet_EPs}(G) combine longitudinally chirped dissipative modulation with integrated tunable mode switches, enabling reconfigurable asymmetric transmission~\cite{mao2025chip}.

Broader Floquet-type photonic EPs have been realized in discrete-time synthetic photonic lattices using two coupled optical fiber loops of different lengths, where time-multiplexed optical pulses form a large-scale temporal mesh lattice. The coupled fiber-loop system shown in Fig.~\ref{fig:3.8_Floquet_EPs}(H) incorporates gain and loss to realize a PT-symmetric synthetic lattice operating near EPs~\cite{regensburger2012parity}. Another realization of the discrete Floquet step uses a linear-optical interferometric network in single-photon nonunitary quantum walks. The walker state is encoded in photon spatial modes, while the coin state is encoded in polarization. In this platform, the step operator defines the Floquet evolution, and non-Bloch EPs were observed in the generalized Brillouin zone~\cite{xiao2021observation}.

 \subsection{Exceptional Points in Lossless-Gainless Waveguides}
 \label{sec:EP-DBE-SIP}

\begin{figure}[htbp]
 \includegraphics[width=\textwidth]{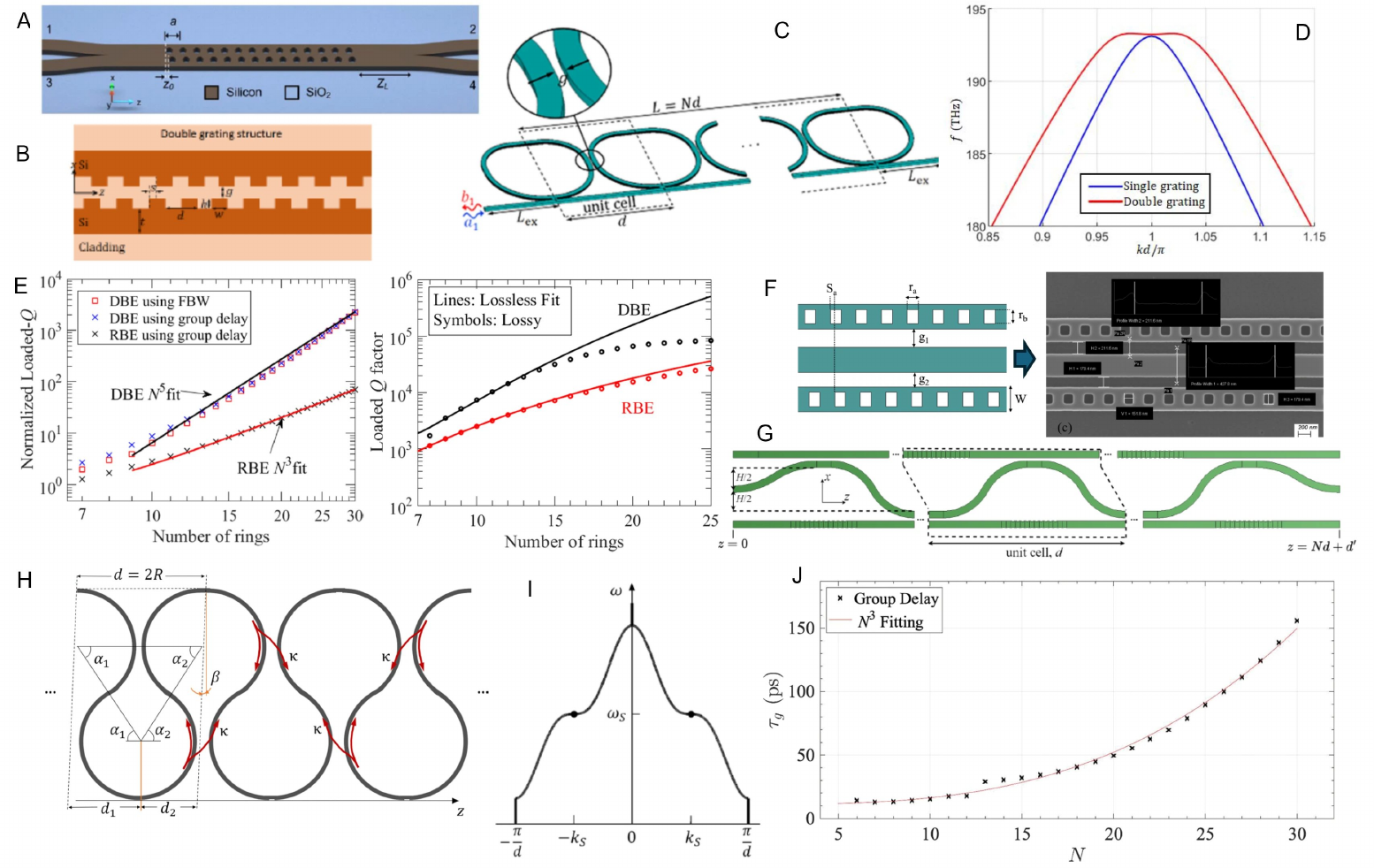}
 \centering
 \caption{\textbf{Lossless and gainless waveguides that support either a degenerate band edge (DBE) or a stationary inflection point (SIP).}
 (\textbf{A}) Multimode silicon waveguides with holes~\cite{burr2016experimental}.
 (\textbf{B}) DBE waveguide made by two coupled gratings~\cite{mealy2022degenerate}. 
 (\textbf{C}) CROW side coupled to a bus supports either a DBE or an SIP~\cite{nada2023design}.
 (\textbf{D}) Very flat dispersion diagram with a DBE (red curve) obtained with the double grating waveguide in (\textbf{B}), compared with the dispersion diagram of a regular band edge obtained with modes in a single grating waveguide (blue curve) ~\cite{herrero2023lasing}. (\textbf{E}) Typical $N^5$  scaling of the $Q$ factor in a DBE cavity as in (\textbf{C}) made of $N$ rings and deviation from such ideal scaling when cavity's internal losses are present.  The SIP is supported by the CROW in  (\textbf{C}), and in other three-path waveguides as in (\textbf{F}) using a SOI platform \cite{paul2024experimental},  in  (\textbf{G}) with distributed Bragg reflectors~\cite{furman2025impact}, and also as in the  serpentine waveguide 
(\textbf{H})~\cite{herrero2023lasing}.
 (\textbf{I}) Very flat dispersion diagram with an SIP at ($\omega_S, k_S$)~\cite{herrero2023lasing}. (\textbf{J}) Group delay of a cavity made of a periodic waveguide as in (\textbf{G}) with $N$ unit cells and the SIP-typical $N^3$ fitting \cite{furman2023frozen}.} 
 \label{fig:DBE-SIP-Waveguides}
 \end{figure}

It is possible to support EPs in lossless and gainless waveguides where more than one mode is propagating in each direction. Though most of the literature has been concentrated on analyzing EPs in the context of quantum mechanics or systems with PT symmetry,  some papers in the early 2000' started to show that EPs exist also in waveguiding systems without gain and loss~\cite{figotin_oblique_2003,figotin_gigantic_2005,figotin2006frozen,figotin_slow_2006, figotin_slow_2011}. In that work, the authors used the term `stationary points' to refer to EPs, and they demonstrated that photonic crystals made of multilayer anisotropic layers support EPs of order three and four, called stationary inflection point (SIP) and degenerate band edge (DBE), respectively. They also explained the concept of "frozen mode" referring to the complex degenerate modes in cavities made of finite-length waveguides supporting the SIP or the DBE. That research was aligned with the topic of slow -light phenomena that started to be very important 20-25 years ago. The analysis of both SIP- and DBE-supporting multimode waeguides was cast in terms of transfer matrix, i.e., transferring the state vector (i.e., the polarization state) from cell to cell of a periodic multimodal waveguide. 
Later on, the original findings on SIP and DBE have been extended conceptually to optical waveguides including gratings using coupled mode theory~\cite{gutman2011degenerate, gutman2012slow}. Due to the high sensitivity of wavenumbers and polarization states (i.e., eigenvectors) to perturbations, realizing EPs of order 3 and 4 in lossless and gainless waveguides is not an easy task. Despite that, the DBE has been demonstrated experimentally in silicon photonics~\cite{burr2016experimental}. The occurrence of the DBE has been demonstrated experimentally also at microwave frequencies in different waveguide technologies~\cite{Othman17ExpDem,  abdelshafy2019TAPexceptionalExper,Mealy2020GeneralConditions,Zheng22MTTSynthMeasDBE}. In early work by Chabanov~\cite{Chabanov08StronResTra}, the resonant transmission properties of split band-edge (SBE) in a circular waveguide filled by layered anisotropic media~\cite{yarga2008degenerate} were explored; the SBE is highly related to the DBE. 
 The DBE is an EP of order four among four waveguide modes and, if only four modes are used, it occurs at the center or at the edge of the Brillouin zone \cite{nada2017theory}. Few DBE structures have been proposed to coupled modes based on multimode waveguides with at least two modes in each direction. Notably, it can be obtained by simply using a double grating waveguide \cite{burr2016experimental,mealy2022degenerate,herrero2025advancements,zamir2026silicon}. The DBE offers various interesting properties: (i) the dispersion diagram is very flat (Fig.~\ref{fig:DBE-SIP-Waveguides}(D), red curve) and in the neighborhood of the DBE at ($\omega_d,k_d$) it is well described by $(\omega-\omega_d)\propto(k-k_d)^4$ \cite{figotin_gigantic_2005,figotin_slow_2011,othman_giant_2016,nada2017theory,nada2018giant,mealy2022degenerate}, whereas the modal dispersion in the neighborhood of a regular band edge at ($\omega_r,k_r$), which can be obtained with a single grating, is well described by $(\omega-\omega_r)\propto(k-k_r)^2$; (ii) the quality factor of a cavity working at a DBE resonance scales as $Q\propto N^5$ where $N$ is the number of unit cells \cite{figotin_gigantic_2005,nada2017theory,nada2018giant,zamir2026silicon}, whereas that of a cavity made by a regular periodic structure working at an RBE scales as $Q\propto N^3$, as shown in Fig.~\ref{fig:DBE-SIP-Waveguides}(E); (iii) the local density of states inside a DBE cavity scales as $N^4$ as shown in \cite{othman_giant_2016}; (iv) the maximum field enhancement at a DBE resonance scales as $N^4$ as shown in \cite{figotin_gigantic_2005,herrero2025advancements,zamir2026silicon}. These trends are ideal, i.e., they occur in lossless and gainless waveguide cavities assuming a perfect realization. Both losses and fabrication tolerances affect such scalings as shown in \cite{nada2018giant,zamir2026silicon}. For example, the effect of internal losses is shown in Fig.~\ref{fig:DBE-SIP-Waveguides}(E). A suitable application of a DBE is in making high $Q$ cavities without using any mirror because the degeneracy is inherently mismatched to any possible load. 

 The SIP is an EP of order three among three  waveguide modes, hence it requires a three-path (three-way) waveguide channel. Some examples are provided in Figs.~\ref{fig:DBE-SIP-Waveguides}(F), (G) and (H). It has been discussed in \cite{ballato_frozen_2005,  figotin_slow_2011, scheuer2011optical, gutman2012slow, paul2021frozen, nada2017theory, li2017frozen, herrero2022frozen,nada2023design, herrero2023lasing,furman2023frozen,furman2025impact}.  
 It has been demonstrated experimentally at microwaves in \cite{apaydin2012experimental,nada2020frozen}, and only recently there have been measurements partially demonstrating some of its properties at near infrared \cite{paul2024experimental,furman2025experimental}.

 It has various interesting properties: (i) the dispersion diagram is very flat showing an inflection point (Fig.~\ref{fig:DBE-SIP-Waveguides}(I)) and in the neighborhood of the SIP at ($\omega_s,k_s$) it is well described by $(\omega-\omega_s)\propto(k-k_s)^3$; (ii) the quality factor of a cavity working at an SIP resonance scales as $Q\propto N^3$ where $N$ is the number of unit cells  \cite{herrero2022frozen,nada2023design} and the associated group delay scales also as $N^3$ \cite{nada2023design,herrero2023lasing,furman2023frozen} as shown Fig.~\ref{fig:DBE-SIP-Waveguides}(J). These trends are ideal, i.e., they occur in lossless and gainless waveguide cavities assuming a perfect realization and both losses and fabrication tolerances affect such scalings as shown in \cite{furman2025impact}.  A suitable application of an SIP is in making high $Q$ cavities without using any mirror because the degeneracy is inherently mismatched to any possible load and also in making devices with large group delays \cite{furman2025experimental}.



        

 
\newpage 
\section{Exceptional Points in Topological Photonics}
This section first reviews the intrinsic topology associated with EPs, including eigenvalue winding and eigenmode monodromy. We focus on the adiabatic regime, where encircling implements analytic continuation and the outcome is fixed by the monodromy. We then extend the discussion to high-dimensional exceptional geometries and multi-EP settings, where encircling paths realize braid-group operations and enable non-commuting (non-Abelian) operations on modes. Finally, we distinguish these static invariants from dynamical encircling protocols, where time-dependent evolution introduces non-adiabatic transitions (NAT) and direction-dependent state selection.
\label{sec: Exceptional Points in Topological Photonics}

\subsection{From exceptional point degeneracies to topological defect singularities}
\begin{figure}[htbp]
\centering
\includegraphics[width=0.85\textwidth]{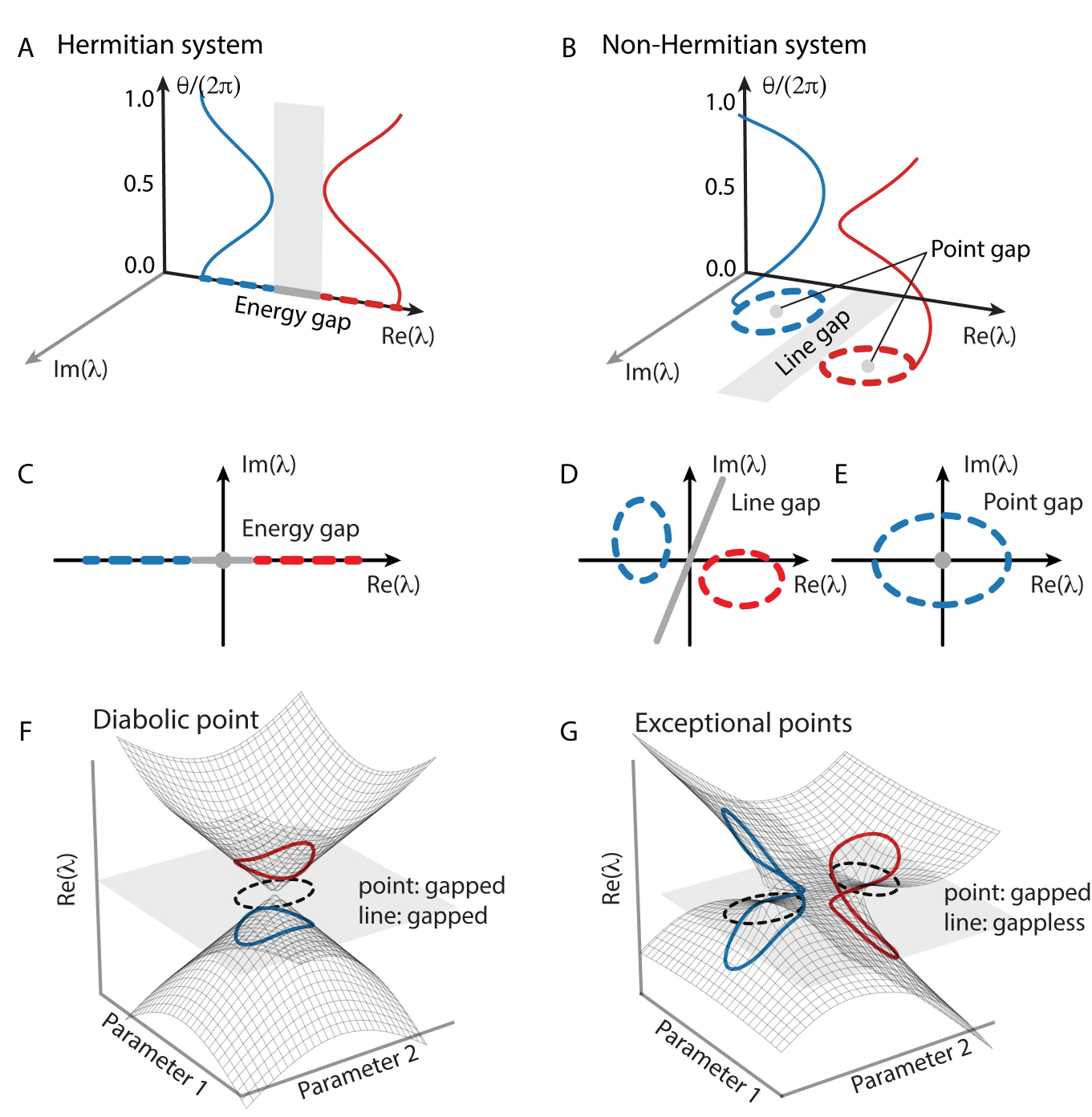}
\caption{
\textbf{Line and point gaps in Hermitian and non-Hermitian spectra}
(\textbf{A,C}) Hermitian spectrum with an ordinary energy gap on the real axis.
(\textbf{B}) Complex spectrum of a non-Hermitian system.
(\textbf{D}) Line gap defined by a reference line that does not intersect the spectrum.
(\textbf{E}) Point gap defined by a reference energy around which the spectrum may wind.
(\textbf{F}) Local eigenvalue surfaces near a Hermitian diabolic point, a small enclosing loop is both line-gapped and point-gapped.
(\textbf{G}) Local eigenvalue surfaces near an EP, the enclosing loop is point-gapped and line-gapless. 
Gray lines and gray points denote the chosen reference line and reference energy, respectively. Blue and red denote the two bands.
Schematic drawn by the authors based on Refs.~\cite{bergholtz2021exceptional,ding2022non,kawabata2019classification}}
\label{Fig:band gap}
\end{figure}

In Hermitian systems, the energy spectrum is restricted to the real axis, and topological phases are defined relative to a reference energy (e.g., the Fermi energy) lying inside a spectral gap, as shown in Fig.~\ref{Fig:band gap}(A). In the complex-energy plane, such a gap corresponds to an interval on the real axis that contains no eigenvalues, as shown in Fig.~\ref{Fig:band gap}(C).
In non-Hermitian systems, eigenvalues are generally complex and trace trajectories in the complex plane as the parameter $\theta$ evolves, as shown in Fig.~\ref{Fig:band gap}(B). The notion of a spectral gap must be generalized~\cite{kawabata2019symmetry,bergholtz2021exceptional,ding2022non}. A line gap is open if, after shifting the spectrum by that reference energy, there exists a line through the reference point that is not intersected by the spectrum, as shown in Fig.~\ref{Fig:band gap}(D). A line gap allows an adiabatic deformation to a Hermitian spectrum without closing the gap and thus represents an extension of Hermitian topology~\cite{kawabata2019symmetry}. 
In contrast, a point gap at a reference energy $E_{\mathrm{ref}}$ is open on a manifold $\mathcal{C}$ if the spectrum avoids that point, as shown in Fig.~\ref{Fig:band gap}(E), i.e., $\det[H(k)-E_{\mathrm{ref}}]\neq 0, \forall k\in\mathcal C$, this type of topology is intrinsic to non-Hermitian systems and has no direct Hermitian counterpart. 
The local distinction between a Hermitian degeneracy and an EP is shown in Figs.~\ref{Fig:band gap}(F) and ~\ref{Fig:band gap}(G). A sufficiently small enclosing loop around a Hermitian diabolic point remains gapped in both the line-gap and point-gap senses, as illustrated in Fig.~\ref{Fig:band gap}(F). In contrast, on a small loop around the EP energy $E_{\mathrm{ref}}$, the spectrum remains point-gapped but is necessarily line-gapless, as shown in Fig.~\ref{Fig:band gap}(G), since the spectrum winds around the reference energy and thus intersects any line passing through it.

This local point-gap winding underlies the interpretation of EPs as topological defects in parameter space. They act as branch-point singularities of the multi-sheeted Riemann surface defined by the complex spectrum. Consequently, adiabatic analytic continuation of eigenmodes along a closed loop $\mathcal{C}$ does not simply accumulate an Abelian geometric phase, as in Hermitian DP case~\cite{berry1984diabolical}, but instead produces a nontrivial monodromy. Eigenvalues and eigenvectors permute, returning to their original configuration only after multiple encircling cycles~\cite{heiss1999phases,heiss2000repulsion}.

\subsection{Non-Hermitian topological invariants of EPs}
Consider a sufficiently small closed loop $\mathcal{C}\simeq S^1$ enclosing an isolated EP in parameter space. With respect to the EP energy $E(k_{\mathrm{EP}})$, the spectrum on $\mathcal{C}$ remains point-gapped, while any line gap is closed. Thus, the complex function $\det[H(k)-E(k_{\mathrm{EP}})]$ defines a map $\mathcal{C}\to\mathbb{C}^*$. This enables the definition of a point-gap winding number~\cite{kawabata2019classification}
\begin{equation}
W=\frac{1}{2\pi i}\oint_{\mathcal{C}} d\mathbf{k}\cdot\nabla_{\mathbf{k}}
\ln\det\!\left[H(\mathbf{k})-E(\mathbf{k}_\mathrm{EP})\right]
\in\mathbb{Z}.
\end{equation}
For an isolated second-order EP involving two eigenvalues $E_m$ and $E_n$, a complementary topological invariant is the vorticity~\cite{shen2018topological,zhou2018observation} 
\begin{equation}
\nu
= -\frac{1}{2\pi}\oint_{\mathcal{C}}
d\mathbf{k}\cdot\nabla_{\mathbf{k}}
\arg\!\bigl[E_m(\mathbf{k})-E_n(\mathbf{k})\bigr],
\end{equation}
This quantity measures the winding of the eigenvalue difference around the origin in the complex-energy plane. Owing to the square-root branch structure of a second-order EP, $\nu$ takes half-integer values (e.g., $\pm1/2$)~\cite{zhou2018observation}. For a two-band model, one finds the relation $W=-2\nu$. Unlike $\nu$, the invariant $W$ remains well-defined even when more than two bands coalesce. Both invariants depend only on the homotopy class of $\mathcal{C}$, provided that the point gap remains open.

\subsection{Static encircling EPs: adiabatic monodromy and mode braiding}
The point-gap winding discussed above becomes operational when system parameters are transported along a closed loop $\mathcal C$ in control space. On such a loop, the spectrum remains point-gapped with respect to $E_{\mathrm{EP}}$, so eigenvalues trace a winding trajectory in the complex plane, while the eigenmodes undergo a nontrivial monodromy because $\mathcal C$ links the branch point of the associated Riemann surface.

A direct experimental route to this static topology is quasi-static (stroboscopic) encircling: one reconstructs the instantaneous eigenvalues and eigenvectors along $\mathcal C$ and performs analytic continuation to identify the sheet permutation. This approach has revealed the branch-point topology and mode exchange in platforms including microwave cavities~\cite{dembowski2001experimental,dembowski2004encircling}, chaotic optical microcavities~\cite{lee2009observation}, photonic systems~\cite{zhou2018observation,zhong2018winding}, exciton--polariton systems~\cite{gao2015observation}, and acoustic structures~\cite{ding2016emergence}.

For a simple second-order EP, like the eigenvalue, each eigenstate also undergoes a permutation only after a twofold cycle $\mathcal C^{(2)}$, where superscript $(2)$ indicates the loop $\mathcal C$ traversed twice  in the same orientation. After two turns, each state returns to its initial branch and accumulates, in addition to the usual dynamical phase, a biorthogonal geometric phase \cite{mailybaev2005geometric,liang2013topological}.
\begin{equation}
\gamma_n = i \oint_{\mathcal C^{(2)}}
\langle \bar{\psi}_n^{L}(\boldsymbol{\lambda})|d \bar{\psi}_n^{R}(\boldsymbol{\lambda})\rangle
,
\end{equation}
where $\ket{\bar{\psi}_n^{R}}$ and $\bra{\bar{\psi}_n^{L}}$ are biorthonormal right and left eigenvectors, respectively. The corresponding Berry connection is $\mathcal A_n(\boldsymbol{\lambda})= i \langle \bar{\psi}_n^{L}(\boldsymbol{\lambda})|\, d \bar{\psi}_n^{R}(\boldsymbol{\lambda})\rangle$, so that $\gamma_n=\oint_{\mathcal C^{(2)}} \mathcal A_n$.
Encircling two EPs can result in a $\pi$ phase shift even without state permutation, depending on the homotopy class of the loop with signed holonomy~\cite{dembowski2004encircling,ryu2024exceptional}.

Considering a system described by an effective $N \times N$ Hamiltonian, the topology becomes richer once more than two modes participate, i.e., when $N>2$~\cite{bergholtz2021exceptional,ding2022non,wang2021topological}. Complex eigenvalues can braid across the control manifold, giving rise to topology associated with the ordered eigenvalue branches. For a maximal $N$-th order EP ($\rm{EP}_N$), the local spectral topology is described by Abelian winding numbers. Non-Abelian behavior arises in $N>2$ multiband spectra when encircling lower-order EPs generates braiding operations whose products can depend on the order of encirclement~\cite{staalhammar2025abelian}. On a generic two-dimensional control manifold $\mathcal{B}$ parameterized by two independent variables, second-order EPs ($\rm{EP}_2$s) appear as isolated degeneracies between various mode pairs, reflecting the codimension-two character of exceptional degeneracies in the absence of additional symmetries~\cite{wojcik2022eigenvalue}. For loops $\mathcal{C}\subset\mathcal{B}$ with a common basepoint, considered in the punctured space with the EPs excluded, the choice of basepoint can itself affect whether two loops are topologically equivalent and hence whether they generate the same braid~\cite{guria2024resolving}. When multiple EPs involving different band pairs coexist, concatenating encircling loops can generate braid-like operations on eigenmodes. These operations can fail to commute, realizing non-Abelian transformations.

\begin{figure}[h]
\centering
\includegraphics[width=0.9\textwidth]{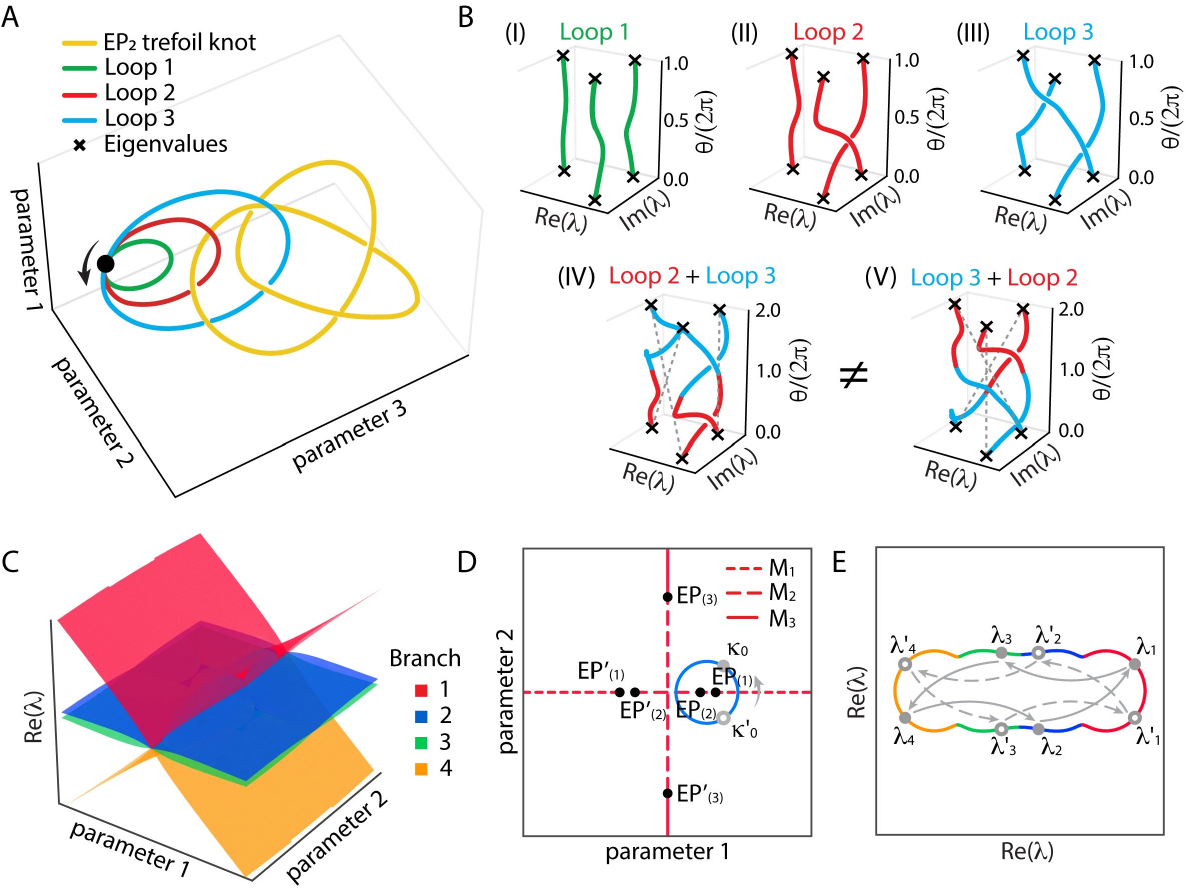}
\caption{\textbf{Complementary descriptions of non-Abelian eigenvalue braiding induced by encircling EPs}
(\textbf{A,B}) Discriminant-locus representation, adapted from Ref.~\cite{patil2022measuring}. 
(\textbf{A}) Trefoil-shaped locus of $\rm{EP}_2$s and three representative loops sharing a common base point belong to distinct homotopy classes.
(\textbf{B}) Eigenvalue braids generated by individual loops and by two oppositely ordered loop compositions.
(\textbf{C-E}) Branch-cut representation, adapted from Ref.~\cite{zhong2018winding}.
(\textbf{C}) Real parts of four eigenvalue sheets in a non-Hermitian four-state model.
(\textbf{D}) Locations of $\rm{EP}_2$s and the associated branch cuts labeled by the permutation matrices $M_i$. A closed loop encircling two $\rm{EP}_2$s yields different eigenvalue permutations when the evolution starts from $\kappa_0$ or $\kappa_0'$.
(\textbf{E}) Eigenvalue braid generated by the ordered sequence of branch-cut crossings. Starting from $\kappa_0$, the eigenvalue ordering evolves as $\{\lambda_1,\lambda_2,\lambda_3,\lambda_4\}\to\{\lambda_3,\lambda_1,\lambda_4,\lambda_2\}$; starting from $\kappa_0'$, it evolves as $\{\lambda_1,\lambda_2,\lambda_3,\lambda_4\}\to\{\lambda_2,\lambda_4,\lambda_1,\lambda_3\}$.}
\label{Fig:nonAbelian_braiding}
\end{figure}

One practical way to characterize these braids is to map the loop into the space spanned by the coefficients of the characteristic polynomial of $\hat{H}$. These coefficients smoothly parametrize the unordered set of eigenvalues and define a control space $\mathcal{L}_N \simeq \mathbb{C}^{N-1}$. Degeneracies correspond to the vanishing of the discriminant, forming a hypersurface $\mathcal{V}_N$ within $\mathcal{L}_N$, as shown in Figs.~\ref{Fig:nonAbelian_braiding}(A) and ~\ref{Fig:nonAbelian_braiding}(B)~\cite{patil2022measuring}. Although the high dimensionality and nontrivial geometry of $\mathcal{L}_N$ make direct visualization difficult, it makes clear that the resulting eigenvalue braid is determined by how the loop winds around the degeneracy hypersurface.

For large $N$, analytical expressions for the eigenvalues are often difficult to obtain. A complementary way to determine the permutation associated with a given loop is to artificially introduce branch cuts (BCs) in $\mathcal{B}$ for the multivalued eigenvalues $\lambda$, and then track how $\mathcal{C}$ crosses these BCs~\cite{ryu2012analysis,zhong2018winding,pap2018non,ryu2022classification}. In a non-Hermitian four-state model defined on a two-dimensional parameter space shown in Fig.~\ref{Fig:nonAbelian_braiding}(C), each BC crossing is associated with a permutation operator, as shown in Fig.~\ref{Fig:nonAbelian_braiding}(D), represented by a permutation matrix acting on the ordered eigenstate vector. The overall monodromy is then obtained by multiplying these matrices in the order in which the loop crosses the BCs. Fig.~\ref{Fig:nonAbelian_braiding}(E) displays the resulting eigenvalue braid. BCs provide a convenient labeling convention for tracking the multivalued spectrum, as it involves only quantities defined on the two-dimensional space $\mathcal{B}$. However, while operationally convenient, a BC-based description may obscure the underlying topology, since the permutation depends on the ordering of BC crossings rather than directly on the homotopy class of the loop~\cite{zhong2018winding}.


\subsection{Dynamical encircling EPs: non-adiabatic transition}
Although the state permutations in adiabatic encircling are governed by topology of the Riemann surface, dynamical encircling introduces a conceptually distinct regime of state selection~\cite{nenciu1992adiabatic}. Although both scenarios have path-dependence and superficially similar phenomena, dynamical encircling is fundamentally governed by non-adiabatic transitions, and the underlying mechanisms should not be conflated.

Dynamical encircling probes non-unitary time evolution under a parameter-dependent, nonself-adjoint Hamiltonian $\hat{\mathcal{H}}(\chi)$, for which the adiabatic theorem breaks down when eigenstates have asymmetric imaginary parts~\cite{nenciu1992adiabatic}. In photonics, this can be realized either by genuine temporal modulation~\cite{uzdin2011observability,xu2016topological} or equivalently by mapping temporal evolution onto spatial propagation in coupled waveguides~\cite{doppler2016dynamically}. Trajectories tend to the least-decaying branch, so slow parameter variation does not guarantee adiabatic following. Instead, the evolution often consists of extended quasi-adiabatic segments interrupted by abrupt non-adiabatic transitions, a behavior that can be understood in terms of stability-loss delay~\cite{uzdin2011observability,milburn2015general} and Stokes phenomena~\cite{berry2011slow}. These mechanisms lead to the widely discussed “chiral mode switching”: for many loops in the vicinity of an EP, the final state depends primarily on the encircling direction, even when the initial state differs~\cite{doppler2016dynamically,hassan2017dynamically}.

Moreover, rather than depending on the static topological features, such as the number or even order of the encircled EPs~\cite{zhang2019dynamically2}, the dynamical output state is decided by the loop’s starting point and traversal direction, i.e. whether the system initially traverses a stable (low-loss) or unstable (high-loss) Riemann sheet along its evolution trajectory. As such, the chiral-state switching behavior was also found for loops excluding the EP, as long as the starting point lies on the branch cut (or PT-symmetric phase) and the loop is in the vicinity of the EP~\cite{hassan2017chiral,liu2021chip,nasari2022observation}. 

When the encircling loop starts and ends in the PT-symmetric phase, where the imaginary parts of the eigenvalues coalesce ($\operatorname{Im}\lambda_+(t)=\operatorname{Im}\lambda_-(t)$), two distinct scenarios emerge.
If the state evolves on the lower-loss Riemann sheet, the dynamics are stable and effectively adiabatic, enabling chiral mode switching~\cite{doppler2016dynamically}, topological energy transfer~\cite{xu2016topological}, on-chip time-asymmetric mode transfer~\cite{choi2017extremely,yoon2018time}, topological single-mode lasing~\cite{schumer2022topological} and chiral orbital angular momentum (OAM) conversion~\cite{qi2024dynamically}. Exact analytical solutions have rigorously clarified how the system evolution funnels into a preferred eigenstate~\cite{hassan2017dynamically}.
By contrast, if the state initially lies on the higher-loss sheet, which is dynamically unstable, adiabaticity breaks down after a finite delay and a non-adiabatic transition occurs. As a result, the final output state can appear to depend solely on the encircling direction, rather than on the starting point of the loop~\cite{doppler2016dynamically}. 

However, when dynamically encircling an EP with a starting point in the PT-broken phase, where the real parts of the eigenvalues coalesce ($\operatorname{Re}\lambda_+(t)=\operatorname{Re}\lambda_-(t)$), the system exhibits non-chiral behavior, such that the output state is the same regardless of the encircling direction~\cite{zhang2018dynamically}. In anti-PT symmetric systems, in contrast, dynamically encircling the EP with a starting point in the symmetry-broken phase can still lead to chiral dynamics~\cite{zhang2019dynamically,feng2022harnessing,liu2022chip}. Chiral and non-chiral behavior has also been investigated in more complex settings, including hybrid EPs~\cite{zhang2018hybrid}, encirclement along homotopic loops~\cite{zhong2018winding,zhang2019distinct}, and multi-EP encirclement process, even extending to periodic systems that possess an EL~\cite{yu2021general}. Furthermore, highly efficient chiral mode conversion has been achieved by engineering evolution trajectories that encircle EPs~\cite{li2020hamiltonian,shu2022fast,li2022riemann,shu2024chiral}. Chiral transmission in the near and mid-infrared~\cite{li2025high-Performance} has also been demonstrated. A moving-EP strategy enables adiabatic parameter variation along a small encircling loop, thereby shortening device length compared with fixed-EP implementations~\cite{liu2020efficient}. 

Recent work~\cite{li2025high} has demonstrated high-dimensional bilateral EP encirclement in birefringent silicon waveguides, where opposite detunings of TE and TM polarizations lead them to encircle the EP in opposite directions, so that the final output states are jointly determined by the propagation direction and the polarization. In addition, controlling adiabaticity enables partitioning of the eigenstate space into subspaces~\cite{li2025multi}, thereby realizing multi-state chiral switching among multiple modes in a four-state silicon waveguide system.

The system's dynamics are governed by the equations of motion
\begin{equation}
\dot{\mathbf{\Psi}} = -i \mathcal{M}(t) \mathbf{\Psi},
\end{equation}
which is formally equivalent to the time-dependent Schr\"odinger equation $i \frac{\partial}{\partial t} |\psi(t)\rangle = \hat{\mathcal{H}}_{\mathrm{eff}}(t)|\psi(t)\rangle$. Introducing the dimensionless time $s := t/T$ and assuming $\mathcal{M}(s)$ is diagonalizable for all $s$, its instantaneous eigenvalues and biorthonormal eigenvectors are defined by
\begin{equation}
\mathcal{M}(s)\ket{\bar{\psi}_i^{R}(s)}=\lambda_i(s)\ket{\bar{\psi}_i^{R}(s)}, 
\qquad 
\bra{\bar{\psi}_i^{L}(s)}\mathcal{M}(s)=\lambda_i(s)\bra{\bar{\psi}_i^{L}(s)}
\end{equation}
The geometric and non-adiabatic effects are included in the biorthogonal Berry connection
\begin{equation}
    \mathcal{A}_{ij}(s):= 
    i\left\langle\bar{\psi}_i^{L}(s)\middle|\frac{d}{ds}\bar{\psi}_j^{R}(s)\right\rangle,
\end{equation}
whose diagonal elements $\mathcal{A}_{ii}(s)$ generate the complex geometric phase
\begin{equation}
\gamma_i^{\mathrm{G}}(s)=\int_0^s \mathcal{A}_{ii}(s')\,ds',
\end{equation}
while the off-diagonal elements $\mathcal{A}_{i\neq j}(s)$ describe non-adiabatic coupling between instantaneous eigenstates. In a parallel-transport eigenbasis one imposes $\mathcal{A}_{ii}(s)=0$, in this gauge no geometric phase appears explicitly and only dynamical phases remain.

The evolution operator defined by $\mathbf{\Psi}(t)=\mathcal{U}(t)\mathbf{\Psi}(0)$ admits the exact expansion
\begin{equation}
\mathcal{U}(t)=
\sum_{i,j}
U_{ij}(t) e^{-i\gamma_i^{\mathrm{D}}(s)}
\ket{\bar{\psi}_i^{R}(s)} \bra{\bar{\psi}_j^{L}(0)}
\end{equation}
where $\gamma_i^{\mathrm{D}}(s)=T\int_{0}^{s}\lambda_i(s')\,ds'$ is the dynamical phase. The ideal adiabatic limit corresponds to neglecting all interlevel couplings, i.e. $U_{ij}(t)=\delta_{ij}$. A commonly used quasi-adiabatic condition  requires that the non-adiabatic coupling be much smaller than the eigenvalue separation
\begin{equation}
\varepsilon_{pi}(s)
:=
\left|
\frac{T^{-1}\,\mathcal{A}_{pi}(s)}
{\lambda_i(s) - \lambda_p(s)}
\right|
\ll 1 .
\end{equation}
Beyond the ideal adiabatic limit, the evolution coefficients satisfy
\begin{equation}
\dot U_{pq}(t)=
i\sum_{i\neq p}T^{-1}\mathcal{A}_{p i}(s)
e^{-i(\gamma_i^{\mathrm{D}}(s)-\gamma_p^{\mathrm{D}}(s))}
U_{i q}(t),
\end{equation}
where the second index $q$ labels the initial component and is propagated throughout the evolution.
\begin{figure}[h]
  \centering
  \includegraphics[width=\textwidth]{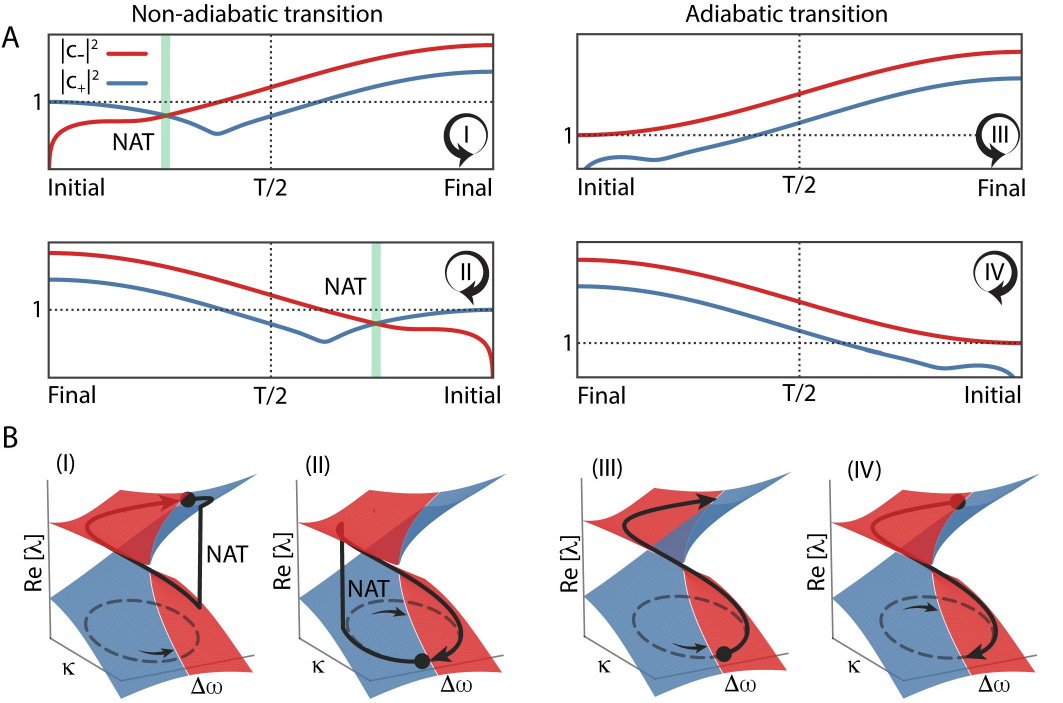}
  \caption{\textbf{Dynamical encirclement and non-adiabatic transitions in a two-level non-Hermitian system.} 
  (\textbf{A}) Time evolution of the biorthogonal modal weights $|c_{-}(t)|^{2}$ (red) and $|c_{+}(t)|^{2}$ (blue) during one closed parameter cycle of duration $T$ around EP. The bottom row is plotted with the time axis reversed to indicate the opposite encircling direction. The insets (I)–(IV) link each trace to the corresponding Riemann-surface panel in (B).
  (\textbf{B}) Schematic trajectories on the two-sheet Riemann surface of the instantaneous eigenvalues $\lambda_{\pm}$ as the parameters trace a closed loop (dashed curve) in the $(\Delta\omega,\kappa)$ plane.
  Red and blue surfaces correspond to stable (low-loss) or unstable (high-loss) eigenvalue branches. Adapted from Ref.~\cite{milburn2015general}}
  \label{fig:dynamical_encircling}
\end{figure}

To make the above discussion concrete, we consider a two-level traceless time-dependent effective Hamiltonian in Eq.~\ref{eq:eff_Hamiltonian}
\begin{equation}
\mathcal{M}(t)=
\begin{pmatrix}
-\Delta\omega(t)-i\Delta\gamma(t) & \kappa(t) \\
\kappa(t) & \Delta\omega(t)+i\Delta\gamma(t)
\end{pmatrix}.
\label{eq:eff_Hamiltonian}
\end{equation}
The instantaneous eigenvalues are $\lambda_{\pm}(t)=
\pm\sqrt{\big[\Delta\omega(t)+i\Delta\gamma(t)\big]^2+\kappa^2(t)}$. Expanding the state in the instantaneous biorthogonal eigenbasis as 
\begin{equation} 
\mathbf{\Psi}(t)=c_{+}(t)\ket{\bar{\psi}_{+}^{R}(t)}+c_{-}(t)\ket{\bar{\psi}_{-}^{R}(t)}. \end{equation} 
The amplitudes $c_{\pm}(t)$ can be reconstructed via $c_\pm(t)=\langle\bar\psi_\pm^L|x(t)\rangle$. The dynamics can exhibit pronounced non-adiabatic transitions (NATs), i.e., a rapid transfer of weight between the two instantaneous eigenstates. Fig.~\ref{fig:dynamical_encircling} illustrates this behavior: the green windows in Fig.~\ref{fig:dynamical_encircling}(A) mark the onset of NAT in the time-domain weights $|c_{\pm}(t)|^{2}$, while Fig.~\ref{fig:dynamical_encircling}(B) provides the geometric interpretation on the eigenvalue Riemann surface, where a NAT appears as an abrupt jump between sheets. 



\subsection{Higher-dimensional exceptional geometries and higher-order exceptional degeneracies}

\begin{figure}[!t]
 \centering
 \includegraphics[width=\textwidth]{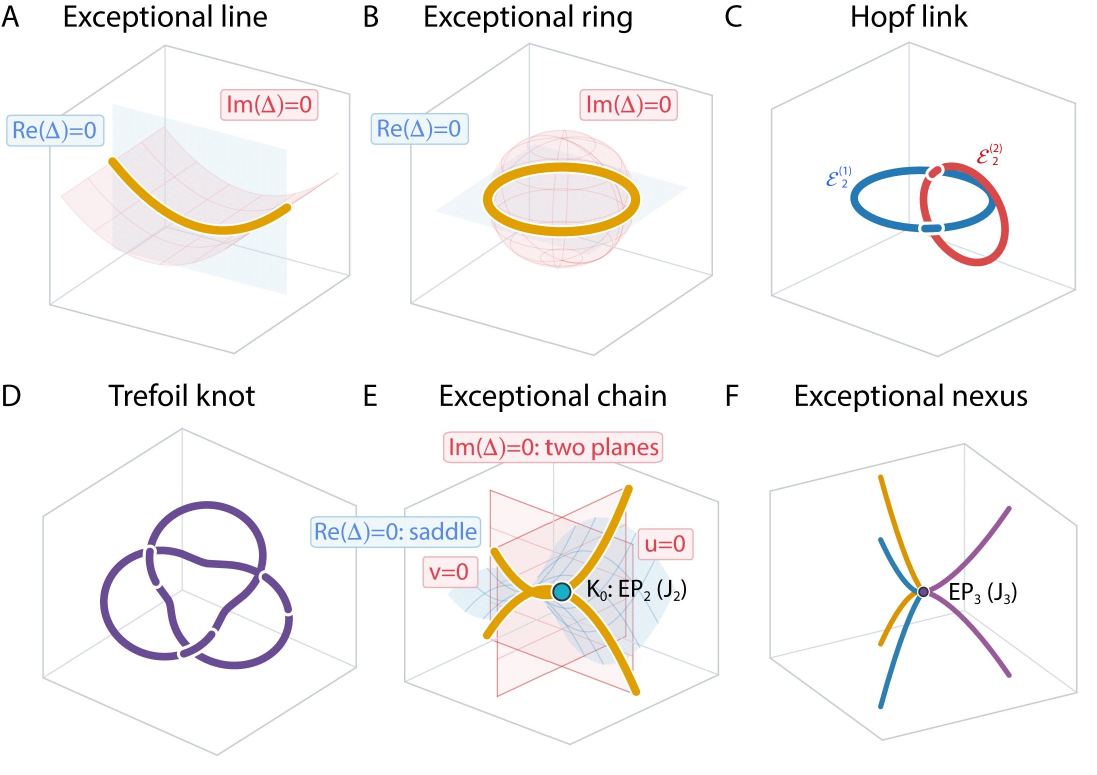}
 \caption{\textbf{Geometry of exceptional loci and local coalescence order.}The boxes represent three-dimensional control or momentum spaces rather than real-space volumes; (\textbf{E}) and (\textbf{F}) show local neighborhoods of the junctions.(\textbf{A}), (\textbf{B}) The blue and red surfaces are the zero sets of $\operatorname{Re}\Delta$ and $\operatorname{Im}\Delta$ in representative two-mode models. Neither surface is generically exceptional by itself. Their gold intersection is an $\mathrm{EP}_2$ locus. Only a finite portion is shown in (\textbf{A}); its ends are plotting limits, not physical terminations. In (\textbf{B}), the locus closes into a ring.(\textbf{C}), (\textbf{D}) Smooth $\mathrm{EP}_2$ curves forming a Hopf link and a trefoil knot, respectively. In (\textbf{C}), $\mathcal E_2^{(1)}$ and $\mathcal E_2^{(2)}$ label the two disjoint linked components; their blue and red colors have no relation to the discriminant surfaces in (\textbf{A}) and (\textbf{B}). White gaps indicate over--under crossings in the projection, not physical endpoints.(\textbf{E}) The saddle $\operatorname{Re}\Delta=0$ and the two planes $\operatorname{Im}\Delta=0$ intersect along two exceptional curves---four incident arms---whose junction retains a two-mode $J_2$ block~\cite{zhang2023symmetry,cui2023experimental}.(\textbf{F}) Leading local geometry of the acoustic nexus in a three-dimensional slice of the control space. Three cusp germs, each comprising two regular $\mathrm{EP}_2$ arms, meet at a three-mode $J_3$ block~\cite{tang2020exceptional}. Each color connects the two arms of one cusp and does not identify a fixed pair of modes.The global embedding of the locus, its junction connectivity, and its local Jordan order are distinct descriptors; none alone determines the others.}
 \label{fig:standalone_global_geometries}
\end{figure}

Beyond isolated EPs, a complete physical characterization separates four pieces of information. The exceptional locus describes where coalescence occurs in momentum or control space; the local order specifies how many modes and eigenvectors coalesce at a given point; a chosen perturbation direction determines the leading spectral splitting; and continuation around a chosen loop determines how the modes are permuted. These properties are related, but none determines the others. A line, ring, link, knot, or surface may consist entirely of $\mathrm{EP}_2$s, whereas an $\mathrm{EP}_3$ may occur at a geometrically simple junction. Figure~\ref{fig:standalone_global_geometries} illustrates the distinction between locus geometry and local order; directional splitting and loop-dependent permutations are discussed below.

Let $D$ denote the number of independent real controls, such as wave-vector components, detunings, coupling strengths, gain, or loss. To describe where degeneracies occur, irrespective of their local order, let $\lambda_\alpha$, with $\alpha=1,\ldots,M$, be the complex eigenvalues of the $M$ modes or bands under consideration; for a resonant Hamiltonian, they are complex eigenfrequencies. Their discriminant is
\begin{equation}
 \Delta=\prod_{1\leq\alpha<\beta\leq M}(\lambda_\alpha-\lambda_\beta)^2 .
 \label{eq:standalone_discriminant}
\end{equation}
The product counts each pair once, so $\Delta=0$ whenever any two spectral branches coincide. Because $\Delta$ is symmetric in the eigenvalues, it remains single valued when the branches exchange. Equation~\eqref{eq:standalone_discriminant} therefore locates the degeneracy set and provides the phase for its winding, but does not determine defectiveness, the number of coalescing modes, or the Jordan structure. These local properties are specified below by Eq.~\eqref{eq:standalone_physical_epn}. In a resonance problem, $\operatorname{Re}\lambda_\alpha$ gives the resonance frequency and $\operatorname{Im}\lambda_\alpha$ the growth or decay rate. A generic pairwise degeneracy therefore imposes the two real conditions $\operatorname{Re}\Delta=\operatorname{Im}\Delta=0$. Thus, denoting the generic $\mathrm{EP}_2$ locus by $\mathcal E_{\mathrm{EP}_2}$,
\begin{equation}
 \operatorname{codim}_{\mathbb R}\mathcal E_{\mathrm{EP}_2}=2,
 \qquad
 \dim\mathcal E_{\mathrm{EP}_2}=D-2 .
 \label{eq:standalone_generic_dimension}
\end{equation}
The codimension counts the independent real matching conditions. When solutions exist, EPs are therefore isolated for $D=2$, form lines for $D=3$, and form surfaces for $D=4$~\cite{bergholtz2021exceptional,ding2022non}. At a regular point of this locus, one pair of eigenfrequencies and eigenvectors coalesces, producing a conventional $\mathrm{EP}_2$.

Figures~\ref{fig:standalone_global_geometries}(a) and \ref{fig:standalone_global_geometries}(b) visualize this counting. The zero sets of $\operatorname{Re}\Delta$ and $\operatorname{Im}\Delta$ form two surfaces, and their intersection is a one-dimensional exceptional locus. In a three-dimensional control space, the locus may cross the experimentally accessible window or close into a ring. A smooth exceptional line cannot terminate at an interior regular point; an apparent endpoint must instead lie at the boundary of the displayed or accessible region, continue through a periodic identification, or enter a singular junction. A closed, unknotted component is conventionally called an exceptional ring~\cite{zhen2015spawning,cerjan2019experimental,na2025multiple}.

Closing and intertwining exceptional curves change their global embedding without changing their local order. Two disjoint closed components can form a Hopf link [Fig.~\ref{fig:standalone_global_geometries}(c)], whereas a single component can form a trefoil knot [Fig.~\ref{fig:standalone_global_geometries}(d)]~\cite{carlstrom2018exceptional,yang2019non,carlstrom2019knotted}. Every regular point of either structure can retain the same local $J_2$ structure. Moreover, the same knot type can arise in physically different parameter spaces. Carlstr\"om \emph{et al.} constructed a trefoil-shaped exceptional line directly in three-dimensional momentum space~\cite{carlstrom2019knotted}. Patil \emph{et al.}, by contrast, enclosed an $\mathrm{EP}_3$ in a four-dimensional control volume; the $\mathrm{EP}_2$ locus on its three-dimensional boundary appears as a trefoil after stereographic projection~\cite{patil2022measuring}.

Mode-selective non-Hermiticity provides a direct physical route from Hermitian nodes to extended exceptional structures. In two control dimensions, unequal modal linewidths can split a Dirac point into two $\mathrm{EP}_2$s joined by a bulk Fermi arc, along which the real parts of the two eigenfrequencies coincide~\cite{zhou2018observation}. In three dimensions, mode-dependent gain, loss, or radiation can inflate a Weyl point into a Weyl exceptional ring~\cite{xu2017weyl,cerjan2019experimental,xu2022observation,liu2022experimental}. The corresponding equal-frequency set becomes a Fermi sheet bounded by the exceptional line. A non-Hermitian term proportional to the identity cannot produce either transformation because it shifts all eigenfrequencies equally and leaves their differences unchanged.

Exceptional curves may also meet at singular, often symmetry-organized junctions, forming an exceptional chain. The number of incident branches does not reveal the number of coalescing modes. Figure~\ref{fig:standalone_global_geometries}(e) shows a four-arm chain junction that remains a two-mode $J_2$ degeneracy, whereas Fig.~\ref{fig:standalone_global_geometries}(f) shows a six-arm exceptional nexus with a three-mode $J_3$ coalescence.

At a chain junction, exceptional branches cannot begin or end arbitrarily: their discriminant windings must balance. Enclose an isolated junction by a small sphere, with each incident half-branch piercing the sphere once. A small loop $\gamma_\ell$ around each puncture, oriented by the common outward-normal convention of the sphere, carries the integer winding
\begin{equation}
 \nu_{\Delta,\ell}
 =\frac{1}{2\pi}\oint_{\gamma_\ell}d\arg\Delta,
 \qquad
 \sum_\ell \nu_{\Delta,\ell}=0 .
 \label{eq:standalone_source_free_physical}
\end{equation}
The sum vanishes because $\Delta$ is single valued and nonzero on the remainder of the sphere. For simple branches, $\nu_{\Delta,\ell}=\pm1$, so positive- and negative-winding branches balance~\cite{zhang2023symmetry}. In the active mechanical system of Cui \emph{et al.}, particle--hole symmetry renders the full discriminant real and its winding trivial. A real line gap at $\operatorname{Re}\lambda=0$ separates the positive- and negative-frequency bands. The measured invariant was instead the winding of the positive-frequency discriminant $\Delta_+$, formed only from pairs of bands with $\operatorname{Re}\lambda>0$~\cite{cui2023experimental}.

Periodic momentum space imposes a related global neutrality condition. In two dimensions, opposite edges of the complete Brillouin zone are identified, so the zone is a torus $T^2_{\mathrm{BZ}}$. Each isolated degeneracy carries a discriminant winding $\nu_{\Delta,i}$, and the absence of a boundary requires
\begin{equation}
 \sum_{\mathbf k_i\in T^2_{\mathrm{BZ}}}\nu_{\Delta,i}=0 .
 \label{eq:standalone_doubling_physical}
\end{equation}
If all zeros are simple $\mathrm{EP}_2$s, their charges $\nu_{\Delta,i}=\pm1$ therefore have zero net sum~\cite{yang2021fermion}. This doubling law applies to the complete periodic Brillouin zone, not to an arbitrary noncompact control plane.

\FloatBarrier

For the non-Kramers antiunitary classes considered here, symmetry can change the codimension count by making the two degeneracy conditions dependent. On a control subspace that preserves $\mathcal{PT}$ symmetry, the spectrum is invariant under $\lambda\mapsto\lambda^*$; for the particle--hole-type $\mathcal{CP}$ constraint, it is invariant under $\lambda\mapsto-\lambda^*$. In either case, $\Delta$ is real throughout the symmetry-preserving subspace. We denote the number of independent real controls in this subspace by $D_{\mathrm{sym}}$ and use the superscript ``sym'' for exceptional loci evaluated within the corresponding symmetry-restricted Hamiltonian family. A regular zero then requires only one real condition:
\begin{equation}
 \operatorname{codim}_{\mathbb R}\mathcal E_{\mathrm{EP}_2}^{\mathrm{sym}}=1,
 \qquad
 \dim\mathcal E_{\mathrm{EP}_2}^{\mathrm{sym}}=D_{\mathrm{sym}}-1 .
 \label{eq:standalone_symmetry_dimension}
\end{equation}
A three-dimensional symmetry-preserving control space can therefore support a two-dimensional exceptional surface~\cite{zhou2019exceptional,okugawa2019topological}. The symmetry does not remove non-Hermiticity; it locks together the frequency- and linewidth-matching conditions. Nor does it increase the local order: a smooth point of the resulting surface remains an $\mathrm{EP}_2$. Away from the symmetry-preserving family, the two conditions generically become independent again. Cusps and self-intersections may occur where the surface ceases to be smooth, but their visual morphology alone does not establish higher-order coalescence~\cite{sayyad2023symmetry,hu2023non}.

A distinct, symmetry-independent route is provided by structural constraints. As Eq.~\eqref{eq:generic_unidirectional_EP} illustrates, effectively one-way coupling between frequency-degenerate modes can keep a restricted Hamiltonian family defective while other parameters vary. The architecture then remains defective throughout this restricted parameter family instead of reaching an EP by satisfying a symmetry-reduced degeneracy condition. This mechanism has been explored in sensing, coherent absorption, and microlasers~\cite{zhong2019sensing,soleymani2022chiral,liao2023chip}.

The intrinsic dimension of an exceptional locus should not be inferred from the number of axes in a plot. In the magnon--polariton experiment of Ref.~\cite{zhang2019experimental}, zero detuning and critical coupling impose two conditions in a four-dimensional control space. Once the detuning is fixed, the remaining solutions appear as a surface in three plotted coordinates, although the underlying locus still has codimension two. Conversely, the Klein bottle and spun trefoil of Ref.~\cite{zhang2024exploring} are surfaces traced by complex band eigenvalues in spectral space rather than exceptional surfaces in the space of experimental controls. What matters is the number of independently varied physical controls.

Higher-order coalescence concerns a different question: how many resonances collapse to the same complex frequency and the same eigenmode. Let $\hat H_N$ be the effective Hamiltonian restricted to the $N$-dimensional invariant subspace involved. A genuine $\mathrm{EP}_N$ involving all $N$ modes satisfies
\begin{equation}
 \lambda_1=\lambda_2=\cdots=\lambda_N=\lambda_*,
 \qquad
 \dim\ker(\hat H_N-\lambda_*I_N)=1 .
 \label{eq:standalone_physical_epn}
\end{equation}
Here $\lambda_*$ is the common complex eigenfrequency and $I_N$ is the identity on this subspace. The first condition gives an $N$-fold degenerate eigenfrequency, so the resonance positions and growth or decay rates coincide. The second states that only one linearly independent eigenmode remains. Eigenvalue multiplicity alone is not enough: a fourfold eigenfrequency may correspond to one $J_4$ chain or, for example, to two independent $J_2$ chains, and only the former is an $\mathrm{EP}_4$~\cite{Kato1966Perturbation,ashida2020non,ryu2022classification}. In calculations, the distinction can be established from the rank or nullity of $\hat H_N-\lambda_*I_N$. Experimentally, reconstructed mode profiles and spectra measured along independent perturbation directions provide complementary evidence of eigenvector coalescence.

The required tuning follows from the same physical count used for an $\mathrm{EP}_2$. The common eigenfrequency $\lambda_*$ is free to move, whereas the $N-1$ complex frequency differences must vanish. Each additional coalescing mode therefore contributes two real conditions:
\begin{equation}
 \operatorname{codim}_{\mathbb R}\mathcal E_{\mathrm{EP}_N}=2(N-1),
 \qquad
 \operatorname{codim}_{\mathbb R}\mathcal E_{\mathrm{EP}_N}^{\mathrm{sym}}=N-1 .
 \label{eq:standalone_higher_order_count}
\end{equation}
The first result applies to a generic complex Hamiltonian family. A local $\mathcal{PT}$ or $\mathcal{CP}$ constraint makes the independent root-matching conditions real and gives the second count within the symmetry-preserving family. Provided that the available controls span these conditions independently, a generic $\mathrm{EP}_3$ requires four real control combinations and an $\mathrm{EP}_4$ requires six; the corresponding symmetry-reduced counts are two and three. These are codimensions, not the total numbers of experimental controls. The reduced count applies only when the symmetry maps the same set of $N$ coalescing eigenvalues onto itself. With the spectral origin chosen at the symmetry center, $\lambda_*$ must be real in the $\mathcal{PT}$ class or purely imaginary in the $\mathcal{CP}$ class~\cite{delplace2021symmetry,staalhammar2021classification,mandal2021symmetry}. Otherwise, a separate symmetry-related set of eigenvalues accompanies the coalescence, and the generic local codimension is retained. Additional coupling relations may further reduce the number of independently adjusted controls.

Physical realizations make higher-order EPs accessible by locking together conditions that would otherwise require independent tuning. Symmetry-constrained gain and loss enable an experimental $\mathrm{EP}_3$ in coupled microrings and underlie related optomechanical proposals~\cite{hodaei2017enhanced,jing2017high}. Waveguide-mediated unidirectional coupling can impose longer Jordan chains~\cite{wang2019arbitrary}, while inverse design and radiative-channel engineering can bring several photonic bands to the same complex frequency~\cite{lin2016enhanced,fu2025achieving}. These platforms differ in implementation, but the diagnostic requirement is the same: several modes must converge to one complex eigenfrequency and one eigenmode.

The familiar fractional-power response is an important signature of this coalescence, but it is not by itself a proof of EP order. Equation~\eqref{eq:Puiseux expansion} already gives the full Puiseux expansion for a size-$N$ Jordan block and the generic splitting $\Delta\lambda_{\mathrm{EP}}\propto\varepsilon^{1/N}$. The point relevant here is that the observed exponent depends on the direction of the perturbation in control or Hamiltonian space. For a generic direction, the leading Puiseux coefficient is nonzero. A symmetry-preserving or otherwise special direction can make this coefficient vanish and reveal a different leading power. The same $J_3$ point may then exhibit one regular, integer-power branch together with a square-root pair~\cite{demange2011signatures,mandal2021symmetry}. The measured splitting exponent must therefore be interpreted together with the perturbation direction and independent evidence of eigenvector coalescence.

Finally, even the local Jordan order does not uniquely determine the outcome of encircling. Around the $J_3$ nexus of Tang \emph{et al.}, loops drawn in two different parameter planes generate either a three-mode cycle or a transposition that exchanges two modes while leaving the third unchanged~\cite{tang2020exceptional}. A complete physical characterization must therefore state where the exceptional locus lies, how many modes coalesce locally, how the spectrum splits along the chosen perturbation, and which mode exchange is generated by the specified loop.

\subsection{Coexistence of EP states and bound states in the continuum}

A natural route toward hybrid non-Hermitian singularities is provided by photonic systems in which EPs interact with BICs. The two constrain different properties of an open-system spectrum. A BIC lies in the radiation continuum yet remains decoupled from every open radiation channel, whereas an EP is a non-diagonalizable degeneracy at which complex eigenvalues---here frequencies or propagation constants---and eigenmodes coalesce. An exceptional bound state in the continuum (EP-BIC) requires both conditions within the same modal subspace: the modes must coalesce defectively, and the coalesced eigenstate must remain decoupled from every open radiation channel.

Radiative darkness does not imply the absence of all loss. An exact BIC has zero radiative linewidth and hence $Q_{\mathrm{rad}}\rightarrow\infty$, whereas absorption and other intrinsic losses can keep its total quality factor finite~\cite{hsu2016bound,koshelev2023bound}. When the radiative and intrinsic decay channels are independent, $Q_{\mathrm{tot}}^{-1}=Q_{\mathrm{rad}}^{-1}+Q_{\mathrm{int}}^{-1}$, where $Q_{\mathrm{int}}$ accounts for nonradiative loss. In periodic structures, finite extent and disorder introduce additional leakage, while breaking the protecting symmetry converts a symmetry-protected BIC into a radiative quasi-BIC~\cite{koshelev2023bound}.

\begin{figure}[!t]
\centering
\includegraphics[width=\textwidth]{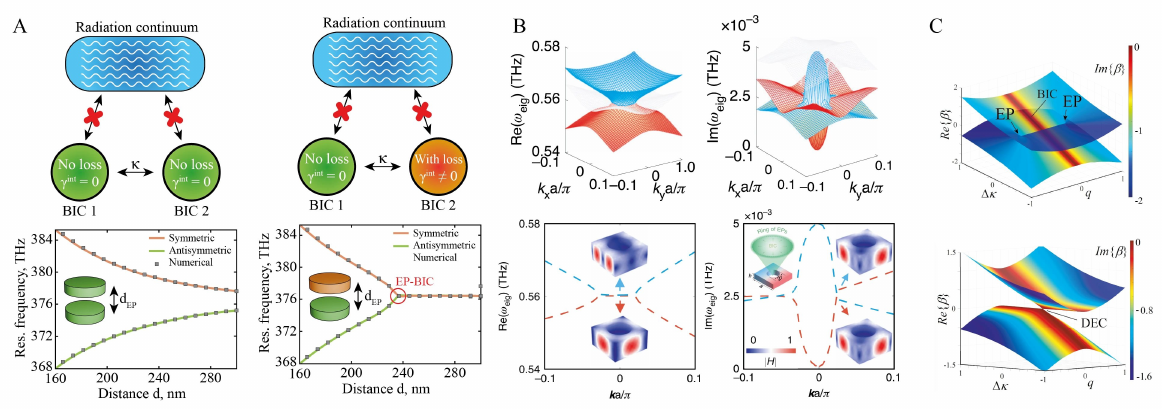}
\caption{\textbf{Representative relations between BICs and exceptional degeneracies.}
(\textbf{A}) Two radiatively dark resonances, coupled with strength $\kappa$ controlled by their separation $d$, are shown without intrinsic loss (left) and with intrinsic loss $\gamma^{\mathrm{int}}$ in BIC~2 (right). Red crosses denote vanishing radiative coupling. The lossy system reaches a second-order EP-BIC at $2|\kappa|=\gamma^{\mathrm{int}}$ and $d=d_{\mathrm{EP}}$; in the lossless plot, $d_{\mathrm{EP}}$ is shown only for reference. Curves and gray symbols are coupled-mode and full-wave results, respectively~\cite{canos2025exceptional}.
(\textbf{B}) Real and imaginary eigenfrequencies over the in-plane Bloch wave vector $(k_x,k_y)$, where $a$ is the lattice period; the lower panels show representative cuts and normalized $|H|$ mode profiles. A Friedrich--Wintgen BIC at $\Gamma$ connects to an off-$\Gamma$ exceptional ring of leaky modes~\cite{wang2025photoswitchable}.
(\textbf{C}) The surfaces show $\operatorname{Re}\beta$ versus the coupling $q$ and propagation-constant mismatch $\Delta\kappa$, with color encoding $\operatorname{Im}\beta$. A BIC on one band leaves the EP pair intact (upper). When the BIC conditions of both bands coincide and their loss-suppression regions overlap, the EPs disappear and a Dirac point embedded in the continuum (DEC) remains (lower)~\cite{pujol2023dirac}.}
\label{Fig:coexistence_EP-BIC}
\end{figure}

The mechanism underlying Fig.~\ref{Fig:coexistence_EP-BIC}(A) is transparent in the minimal two-mode model
\begin{equation}
 \hat H_{\mathrm{BIC}}
 =\begin{pmatrix}
   \omega_0 & \kappa\\
   \kappa & \omega_0-i\gamma^{\mathrm{int}}
  \end{pmatrix},
 \qquad
 \widetilde\omega_{\pm}
 =\omega_0-\frac{i\gamma^{\mathrm{int}}}{2}
 \pm\sqrt{\kappa^2-\frac{(\gamma^{\mathrm{int}})^2}{4}} .
 \label{eq:epbic_effective_hamiltonian}
\end{equation}
Here the two BICs have the same uncoupled frequency $\omega_0$, remain decoupled from radiation, and interact through a real reciprocal coupling $\kappa$; BIC~2 carries an intrinsic decay rate $\gamma^{\mathrm{int}}$. Without intrinsic loss, the restricted Hamiltonian is Hermitian and the degeneracy at $\kappa=0$ is diabolic. At $2|\kappa|=\gamma^{\mathrm{int}}$, the complex eigenfrequencies and eigenvectors coalesce without opening a radiation channel, producing a second-order EP-BIC~\cite{canos2025exceptional}.

Equation~\eqref{eq:epbic_effective_hamiltonian} also illustrates the general result of Ref.~\cite{canos2025exceptional}: in passive reciprocal systems with radiation as the only decay channel, exact BICs can meet only at a diabolic point because radiative decoupling removes the anti-Hermitian radiation term from their modal subspace. Unequal intrinsic loss removes this restriction while preserving radiative darkness; outside this reciprocal setting, Ref.~\cite{canos2025exceptional} also identifies unidirectional coupling as a possible alternative. Full-wave calculations for stacked bilayer and trilayer dielectric metasurfaces predict second- and third-order EP-BICs, respectively. These states remain nonradiating even though intrinsic absorption may keep $Q_{\mathrm{tot}}$ finite.

\FloatBarrier

The EP-BIC in Fig.~\ref{Fig:coexistence_EP-BIC}(A) cannot be excited directly by an incident far-field wave in the passive reciprocal setting considered here. It may instead be inferred by weakly opening a radiation channel and following the resulting quasi-BIC resonances toward the singular point. A generic perturbation of amplitude $\varepsilon$ unfolds an $N$th-order EP-BIC with a complex-eigenfrequency splitting proportional to $\varepsilon^{1/N}$, as described by Eq.~\eqref{eq:Puiseux expansion}. This root law, however, does not by itself determine the radiative linewidth. In the bilayer calculation of Ref.~\cite{canos2025exceptional}, the square-root contribution to the total linewidth arises from redistribution of intrinsic absorption and is absent from the radiative linewidth. Comparisons with an ordinary symmetry-protected quasi-BIC must therefore distinguish radiative and intrinsic losses.

BICs and EPs more commonly remain distinct while occupying the same parameter space. Longhi showed theoretically that a $\mathcal{PT}$-symmetric defect can support a localized state in a waveguide-lattice continuum; in one engineered lattice, the algebraically localized BIC at the $\mathcal{PT}$-breaking threshold is an exceptional point in the continuous spectrum, explicitly distinct from a finite-dimensional EP~\cite{longhi2014}. Song \emph{et al.} found that balanced gain and loss instead split an ordinary BIC into a nonradiating $\mathcal{PT}$-induced BIC and a spatially extended lasing-threshold mode~\cite{song2020coexistence}. Zhang \emph{et al.} proposed a passive anti-$\mathcal{PT}$ microring network supporting Fabry--P\'erot BICs through common-bus dissipative coupling. Conventional anti-$\mathcal{PT}$ EPs occur in the same platform but do not coincide with a BIC in the same modal subspace~\cite{zhang2023realization}.

Spatial proximity is likewise insufficient to establish an EP-BIC. A dielectric waveguide beneath a metal grating supports a Friedrich--Wintgen BIC and, for selected grating thicknesses, neighboring EPs at different Bloch wave vectors~\cite{kikkawa2020bound}. An ideal coupled-split-ring metasurface has a BIC line between two EPs of opposite chirality; the parameter-space path studied between them crosses the BIC line~\cite{niu2024metasurface}. Aligning the loss minima of two coupled quasi-BIC resonances in a proposed dual-waveguide circuit pushes an EP into an ultralow-loss regime without eliminating its linewidth~\cite{qin2022exceptional}. A dielectric ring provides a related geometry with two EPs and two adjacent Friedrich--Wintgen quasi-BICs~\cite{Solodovchenko2024Feb}. These configurations realize coexistence or a BIC-assisted low-loss EP, not an exact EP-BIC.

Beyond coexistence, a BIC can reorganize the exceptional structure itself. In Fig.~\ref{Fig:coexistence_EP-BIC}(B), calculations for an all-dielectric terahertz metasurface show that a Friedrich--Wintgen BIC at $\Gamma$ becomes leaky away from $\Gamma$ and coalesces with its partner mode on an exceptional ring. Angle-resolved measurements along $\Gamma$--$X$ resolved the corresponding BIC-to-EP evolution, while photocarrier loss introduced by optical pumping lifted the EP degeneracy~\cite{wang2025photoswitchable}.

Figure~\ref{Fig:coexistence_EP-BIC}(C) shows that the outcome depends on whether one or both bands become dark. A BIC on only one band crosses the Fermi arc while leaving the EP pair intact (upper). When the BIC conditions of two distinct bands coincide on the arc and their associated loss-suppression regions overlap sufficiently, both modes decouple from radiation; the EPs disappear and a gap opens along the former arc except at their crossing (lower). This crossing has a real degenerate eigenvalue, two orthogonal eigenstates, and conical dispersion, and is therefore a locally Hermitian Dirac point embedded in a non-Hermitian continuum~\cite{pujol2023dirac}.

The examples collected in Fig.~\ref{Fig:coexistence_EP-BIC} illustrate that radiative decoupling can reshape a non-Hermitian spectrum in qualitatively different ways. As shown here, it may coincide with defective coalescence, organize nearby exceptional loci, or suppress exceptional degeneracies. An EP-BIC is defined specifically by the simultaneous occurrence of exact radiative decoupling and defective modal coalescence within the same modal subspace; a large total quality factor alone is neither sufficient nor defining.

\newpage 
\section{Exceptional Points in Quantum Optics and Photonics}
\label{sec: Exceptional Points in Quantum Optics and Photonics}
\subsection{Open quantum systems and the Lindblad equation}
In the standard formulation of pure state quantum mechanics, the state of an isolated system is represented by a state vector $|\Psi(t)\rangle$ on a Hilbert space. The state vector satisfies the normalization condition $\langle \Psi(t)|\Psi(t)\rangle =1$ in any moment of quantum evolution that is governed by the Schr\"odinger equation,
\begin{equation}\label{Q_Schr}
i\frac{\partial}{\partial t}|\Psi(t)\rangle=H|\Psi(t)\rangle.
\end{equation}
where $H$ is the Hamiltonian operator and Planck's constant is set to unity $\hbar=1$. In the simplest case when $H$ is independent of time, the formal solution of Eq.~\eqref{Q_Schr} can be written as
\begin{equation}\label{Q_evol}
|\Psi(t)\rangle=e^{-iH(t\!-\!t_0)}|\Psi(t_0)\rangle, 
\end{equation}
where $|\Psi(t_0)\rangle$ is the initial state. One can immediately see from Eq.~\eqref{Q_evol} that $H$ has to be Hermitian. The Hermiticity is required for the evolution operator $e^{-i(t\!-\!t_0)H}$ to be unitary, so that the evolution operator is unitary and the norm of the state vector is conserved. At this point it is obvious that pure state quantum mechanics can not be non-Hermitian as non-Hermiticity violate the axiomatic principle of unitary evolution.

Non-Hermitian Hamiltonis can be introduced for a broader class of quantum systems, namely, open quantum systems. An open quantum system is a subsystem of a larger quantum system. The key feature of an open system is that it is allowed to exchange matter, energy, and entropy with its environment. Unlike their closed counterparts, open quantum systems are to be described by the density matrix $\rho(t)$, rather than the state vector $|\Psi(t)\rangle$. Quantum description in terms the state vector is a special case in quantum mechanics of open systems,
when
\begin{equation}\label{Q_pure}
\rho=|\Psi\rangle\langle\Psi|.
\end{equation}
In a more generic case the density matrix has more than one non-zero eigenvalue and can only be written as
\begin{equation}\label{Q_mix}
\rho=\sum_n\rho_n|\Psi_n\rangle\langle\Psi_n|,
\end{equation}
where eigenvalues $\rho_n$ and eigenvectors $|\Psi_n\rangle$ are obtained by solving the eigenvalue problem
\begin{equation}
\rho|\Psi_n\rangle=\rho_n|\Psi_n\rangle.
\end{equation}
When the density matrix with more than one non-zero eigenvalue, Eq.~\eqref{Q_mix} describes so-called mixed states. The entanglement in a mixed state can be quantified by the von Neumann entropy 
\begin{equation}
S=-\sum_n \rho_n\ln(\rho_n),
\end{equation}
which is equal to zero for pure states. If the eigenvalues $\rho_n$ are conserved in quantum evolution one can solve the Schr\"odinger equation separately for each state $|\Psi_n\rangle$. This leads to von Neumann equation 
\begin{equation}\label{Q_vonN}
\frac{\partial \rho}{\partial t}=-i[H,\rho], 
\end{equation}
where $[H,\rho]=H\rho-\rho H$ is the commutator.
Equation~\eqref{Q_vonN} describes unitary evolution of the density matrix and, in this sense, is equivalent to the Schr\"odinger equation.

An open quantum system arises when the system of interest is coupled to an environment and only the reduced state of the system is observed. Figure~\ref{fig:5.1_quantum}(A) illustrates decoherence induced by system-environment coupling in an open quantum system. Although the total system plus environment may evolve unitarily, the reduced density matrix generally follows nonunitary dynamics because correlations, energy, particles, or information can be exchanged with the surroundings. In this situation, a single normalized state vector is generally insufficient, and the reduced density matrix is the appropriate object for describing both pure and mixed states. The reduced dynamics may change the purity and the von Neumann entropy of the subsystem. In particular, an initially pure reduced state can become mixed through entanglement with the environment. 

A broad and physically consistent class of Markovian reduced dynamics is generated by quantum dynamical semigroups and is described by the Gorini-Kossakowski-Sudarshan-Lindblad master equation~\cite{gorini1976completely,Lindblad1976generators},
\begin{equation} \label{Lindblad}
\frac{\partial \rho}{\partial t}
=
-i[H,\rho]
+
\sum_m \gamma_m
\left(
L_m\rho L_m^\dagger
-\frac{1}{2}\{L_m^\dagger L_m,\rho\}
\right).
\end{equation}
The first term is the coherent von Neumann contribution generated by the system Hamiltonian, while the second term describes irreversible environmental channels. The Lindblad (jump) operators $L_m$ specify the dissipative processes induced by the environment, and the non-negative coefficients $\gamma_m$ give the corresponding rates. The Lindblad equation preserves the trace, Hermiticity, and complete positivity, and therefore provides the standard effective description of Markovian open quantum systems. In microscopic derivations, this form typically relies on weak system-environment coupling and bath correlation times much shorter than the characteristic system timescales.

\subsection{Hamiltonian and Liouvillian EPs}

The Lindblad equation contains two different spectral objects relevant to EP physics. The first one is obtained by isolating the effective non-Hermitian Hamiltonian in Eq.~\eqref{Lindblad}. The master equation can be rewritten as
\begin{equation}\label{Lindblad2}
\frac{\partial \rho}{\partial t}
=
-i\left(H_{\mathrm{eff}}\rho-\rho H_{\mathrm{eff}}^{\dagger}\right)
+
\sum_m \gamma_m L_m\rho L_m^\dagger ,
\end{equation}
where
\begin{equation}\label{Heff}
H_{\mathrm{eff}}
=
H-\frac{i}{2}\sum_m\gamma_m L_m^\dagger L_m .
\end{equation}
The anti-Hermitian part of $H_{\mathrm{eff}}$ describes the loss of norm associated with unresolved quantum jumps. In the quantum-trajectory interpretation, $H_{\mathrm{eff}}$ generates the conditional, non-unitary pure state evolution of the system without jump trajectories, while the recycling term $\sum_m \gamma_m L_m\rho L_m^\dagger$ represents the stochastic quantum jumps induced by coupling to the environment. Neglecting the jump term therefore corresponds to a conditional or post-selected approximation, in which only trajectories with no detected jump events are retained. 

A Hamiltonian exceptional point (HEP) occurs when two or more eigenvalues and eigenvectors of $H_{\mathrm{eff}}$ coalesce. HEPs are spectral singularities of an effective non-Hermitian Hamiltonian and are widely used in semiclassical, conditional, or post-selected descriptions of photonic and quantum optical systems. However, $H_{\mathrm{eff}}$ alone does not generate the full trace-preserving density-matrix dynamics, because the quantum-jump terms are excluded.

The second spectral object is the full Liouvillian superoperator. Equation~\eqref{Lindblad} can be written in the compact form
\begin{equation}\label{Liouvillian}
\frac{\partial \rho(t)}{\partial t}
=
\mathcal{L}\rho(t),
\end{equation}
where $\mathcal{L}$ is a Liouvillian superoperator which acts as a linear map on the space of density matrices. The solution is given by
\begin{equation}
\rho(t) = e^{\mathcal{L}t}\rho(0)
\end{equation}
where $e^{\mathcal{L}t}$ represents the quantum evolution semigroup. Equation \eqref{Liouvillian} can be written in a vectorized form leading to the eigenvalue problem \cite{minganti2018spectral,hatano2019exceptional} for the matrix representing the superoperator
\begin{equation}
\mathcal{L}|\rho\rangle=\lambda|\rho\rangle
\end{equation}
Liouvillian exceptional points (LEPs) represent singularities of the full Liouvillian generator and therefore affect density-matrix relaxation modes, transient dynamics, and the approach to the stationary state~\cite{minganti2019quantum,minganti2018spectral}. Depending on the reservoir channels, they may be induced by decoherence~\cite{chen2022decoherence} and detected through coherence or spectral correlation functions~\cite{arkhipov2020liouvillian}. 

The distinction between HEPs and LEPs is controlled by quantum jumps. While $H_{\mathrm{eff}}$ describes only the conditional no-jump evolution, the full Liouvillian also contains the recycling terms $\sum_m\gamma_m L_m\rho L_m^\dagger$, which couple different density-matrix modes and modify the degeneracy conditions of the generator. Consequently, HEPs and LEPs are not generally synchronized: a system may display a LEP without any corresponding HEP, or both may occur at shifted parameter values~\cite{minganti2019quantum}. This relation can be continuously tuned in the hybrid-Liouvillian formalism, where detector efficiency and trajectory postselection interpolate between the no-jump Hamiltonian limit and the full Lindblad ensemble~\cite{minganti2020hybrid}.

The full Liouvillian spectrum is richer than the Hamiltonian spectrum because it retains the quantum-jump terms~\cite{arkhipov2020quantum}. Higher-order LEPs provide a particularly clear manifestation of this difference. In dissipative linear bosonic systems, an $n$th-order HEP of an effective non-Hermitian Hamiltonian can generate LEPs of higher order in the larger superoperator eigenspace~\cite{arkhipov2020liouvillian}. In this framework, the Hamiltonian spectrum forms only a substructure of the Liouvillian spectrum, while coherence functions and spectral functions provide experimentally accessible probes of the corresponding Liouvillian symmetry regimes.

The experimental distinction between HEPs and LEPs is clearly exposed in the reconstructed quantum object. Postselected quantum-state tomography in a continuously monitored superconducting transmon reconstructs conditioned no-jump trajectories governed by an effective non-Hermitian Hamiltonian. The resulting signatures, including symmetry breaking, enhanced decoherence, and eigenstate non-orthogonality, correspond to eigenstate coalescence in a trajectory-conditioned HEP~\cite{naghiloo2019quantum}. By comparison, quantum process tomography reconstructs the density-matrix dynamical map and hence the Liouvillian generator. In this case, an LEP appears as a coalescence of Liouvillian relaxation modes and eigenmatrices, and it can occur even when the associated effective Hamiltonian has no exceptional degeneracy~\cite{abo2024experimental}. The observable distinction is therefore between state-level coalescence in postselected Hamiltonian dynamics and relaxation-mode coalescence in the unconditional Lindblad evolution.

Quantum EPs are also identified through quantum observables beyond complex eigenfrequency spectra. In lossy photonic couplers, crossing an EP reshapes two-photon output correlations and switches the Hong-Ou-Mandel interference response, showing that multiphoton interference can resolve EP physics in quantum correlations~\cite{klauck2025crossing}. In trapped-ion platforms with engineered dissipation, suppressing quantum jumps isolates the effective Hamiltonian dynamics, and quantum-state tomography resolves the coalescence of three eigenstates at a third-order HEP. Under the full dissipative evolution, quench dynamics of the full open system reveals an intrinsic third-order LEP~\cite{chen2025quantum}. With both decay and dephasing channels retained, quantum jumps can generate high-order LEPs and displace Liouvillian exceptional lines and points~\cite{wu2026experimental}. The physical origin of these differences is the quantum jump term in the Lindblad equation, which is removed by postselection but retained in unconditional dynamics.

\subsection{Quantum topology and dynamic encircling of EPs}

The topology of an EP becomes observable when system parameters are transported along a closed loop around the singularity. In classical wave systems, this loop is usually described through eigenfrequency sheets and mode amplitudes, whereas in quantum systems the quantum measurements select which topology is observed. Postselection isolates conditional no-jump dynamics, whereas unconditional evolution retains dephasing, relaxation, and quantum jumps that drive the density matrix toward a steady state. Quantum chiral transfer is therefore clearest either in conditioned trajectories or within a finite transient window before relaxation dominates~\cite{sun2024encircling}.

In postselected quantum dynamics, HEP encircling implements a time-dependent loop in the parameter space of an effective non-Hermitian Hamiltonian,
\begin{equation}
H_{\mathrm{eff}}(t)
=
H(t)-\frac{i}{2}\sum_m\gamma_m L_m^\dagger(t)L_m(t).
\end{equation}
The loop changes the instantaneous eigenvectors of $H_{\mathrm{eff}}(t)$ and can convert the branch-point topology into conditional quantum-state transfer. In a single nitrogen-vacancy center in diamond, Hamiltonian dilation realizes the time-dependent non-Hermitian loop in an enlarged Hermitian system, and encircling the HEP produces asymmetric or symmetric mode switching depending on the path~\cite{liu2021dynamically}. In a dissipative superconducting transmon, real-time loops in a non-Hermitian qubit submanifold generate nonreciprocal quantum state transfer and chiral geometric phases, showing that EP proximity can be used for path-dependent quantum state-vector control~\cite{abbasi2022topological}.

In unconditional open-system dynamics, LEP encircling is formulated for the full Liouvillian generator,
\begin{equation}
\frac{\partial \rho(t)}{\partial t}
=
\mathcal{L}(t)\rho(t).
\end{equation}
The loop acts on eigenmatrices and relaxation modes of the density matrix. Dephasing and quantum jumps change the Liouvillian gaps, suppress chiral transfer in the long-time adiabatic limit, and recover chirality only at intermediate evolution times through non-adiabatic pathways~\cite{sun2023chiral}. Single-photon interferometry has
reconstructed this transient LEP chirality by simulating quantum Langevin dynamics and tracking density-matrix evolution. Figure~\ref{fig:5.1_quantum}(B,C) shows the driven-system level scheme and its dephasing-dependent transient chiral response. The disappearance of chirality after relaxation to the steady state confirms that LEP encircling is a finite-time open-system effect rather than a purely eigenvector permutation~\cite{gao2025photonic}.

Higher-order EPs extend quantum encircling from two-mode exchange to multi-sheet braiding. Figure~\ref{fig:5.1_quantum}(D-F) shows control loops and eigenvalue braiding near a third-order EP in a superconducting quantum system. Quasistatic loops resolve the braiding, whereas fast loops generate quantum state transfer. Together, these protocols provide a controllable platform for higher-order braid dynamics~\cite{zhang2025topological}. In integrated photonic chips, single-photon encircling transfers non-Hermitian chirality to the quantum optical regime and provides an on-chip route for path-dependent single-photon control~\cite{tian2023chip}. In photonic quantum walks, an engineered quadruple-degeneracy EP enables chiral switching between Bell states, and the associated Riemann-surface structure protects the entangled-state transfer against path perturbations~\cite{tang2024topologically}. In fully quantum Liouvillian protocols, dissipative encircling can convert singlet and triplet Bell states, while postselection improves the fidelity by removing jump-induced mixing~\cite{khandelwal2024chiral}. In a single-ion quantum heat engine, dynamical LEP encircling enhances positive net work through topology-assisted Landau-Zener-St\"uckelberg processes, extending EP topology from state transfer to energy-cycle control~\cite{bu2023enhancement}. These developments show that quantum EP topology is selected by the dynamical generator being transported. HEP loops control conditioned state-vector evolution, whereas LEP loops control transient mixed-state relaxation and its use in quantum information and thermodynamic protocols.

\subsection{Frontier directions and open questions}

Current frontier research on quantum EPs is turning toward the conditions under which exceptional singularities remain meaningful in realistic open quantum devices.

\begin{figure}[htbp]
\centering
\includegraphics[width=0.85\textwidth]{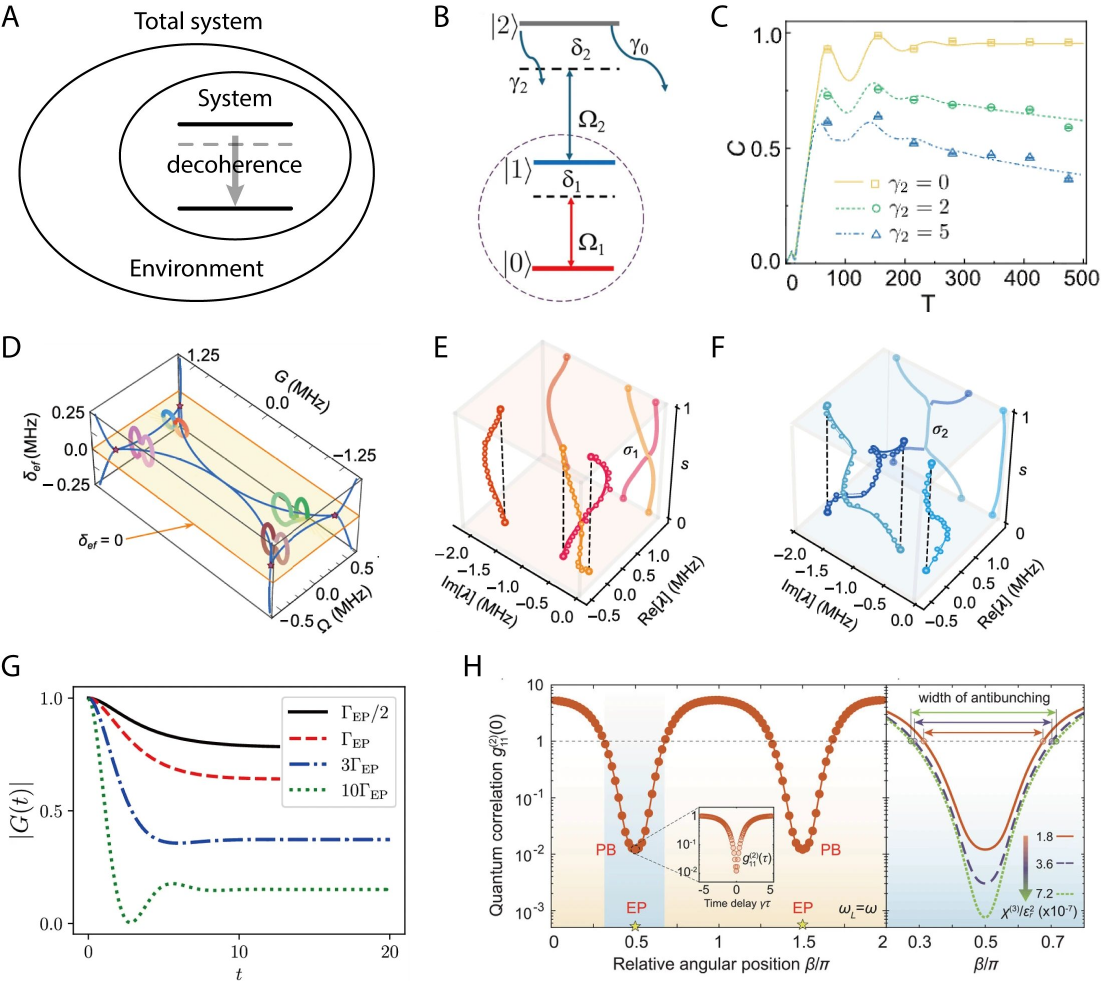}
\caption{
\textbf{Quantum exceptional points in open quantum systems.}
(\textbf{A}) Open-system picture of decoherence induced by system-environment coupling.
(\textbf{B}, \textbf{C}) Liouvillian EP dynamics in a driven dissipative system, showing the level scheme and dephasing-dependent chiral response~\cite{gao2025photonic}.
(\textbf{D}-\textbf{F}) Control loops and eigenvalue braiding near a third-order EP in a superconducting quantum system~\cite{zhang2025topological}.
(\textbf{G}) Non-Markovian decoherence function across the EP, from monotonic decay to memory-induced oscillations~\cite{lin2025non}.
(\textbf{H}) EP-assisted photon blockade revealed by the second-order correlation $g^{(2)}_{11}(0)$~\cite{huang2022exceptional}.
}
\label{fig:5.1_quantum}
\end{figure}

A major frontier direction concerns EPs beyond Markovian Lindblad dynamics. Standard Lindblad descriptions assume short bath correlation times and memoryless reservoirs, but this assumption does not hold in many quantum-optical platforms. Non-Markovian reservoirs introduce memory kernels or structured bath spectra into the open-system generator, enlarging the spectral space in which exceptional degeneracies can occur. Recent non-Markovian EP theory uses pseudomode mappings and hierarchical equations of motion to identify memory-induced degeneracies inaccessible to a time-local Markovian generator~\cite{lin2025non}. Figure~\ref{fig:5.1_quantum}(G) illustrates the associated decoherence crossover from monotonic decay to memory-induced oscillations. Waveguide-quantum electrodynamics systems with retardation provide a concrete implementation, where EPs can separate monotonic relaxation from oscillatory relaxation and can be tuned through propagation delay and emitter geometry~\cite{longhi2026non}. Quantum-channel EPs extend defective spectra from continuous generators to discrete completely positive maps. Interpolating between quantum channels with real spectra and channels with complex-conjugate spectra produces second- and third-order EPs, which has been simulated on a nuclear magnetic resonance quantum processor~\cite{wong2026non}. The open problem is to turn reservoir memory into a controllable spectral resource while preserving readout fidelity and device stability.

The sensing debate around quantum EPs remains an open frontier. Root-law eigenvalue splitting near an EP can enhance spectral response, but enhanced splitting alone is not enough to guarantee improved parameter estimation when assessing EP sensors~\cite{langbein2018no}. Quantum treatments sharpen this point. The fundamental limits are set by the signal-to-noise ratio, quantum Fisher information, detection efficiency, and resource accounting, not by eigenvalue splitting alone~\cite{lau2018fundamental,ding2023fundamental,loughlin2024exceptional,zeng2025non}. However, this does not rule out useful EP sensing. Heisenberg-Langevin~\cite{zhang2019quantum} and Fisher-information analyses~\cite{arkhipov2026achieving} show that appropriate input-output protocols, including heterodyne detection in amplifier configurations, can reach the relevant estimation bounds under specified resources. The unresolved issue is therefore whether a complete quantum measurement protocol can convert spectral amplification into metrological gain after noise and discarded information are counted.

The application outlook is strongest where EP physics is combined with few-photon nonlinearity and quantum photonic functionality. As shown in Fig.~\ref{fig:5.1_quantum}(H), chiral EPs in Kerr resonators reshape the two-photon spectrum and produce exceptional photon blockade. Antibunched emission then appears under conditions that would otherwise favor photon-induced tunneling in Hermitian systems~\cite{huang2022exceptional}. Few-photon photonic devices then test whether EP functionality persists outside the classical-field limit. In superconducting coupled cavities, EP-assisted control has been realized in the regime of a few microwave photons, producing on-chip nonreciprocal transmission controlled by tunable coupling and low-photon-number nonlinear response~\cite{song2024experimental}.

Entanglement control provides a quantum application connected to the topology of multipartite quantum states. Non-Hermitian criticality can reorganize entanglement structure and produce exceptional entanglement transitions with no direct classical-wave analogue~\cite{han2023exceptional}. EP encircling and Riemann-surface topology can then serve as design principles for entangled-state switching, especially in photonic quantum walks and multipath interference platforms where path topology, polarization, and mode structure are jointly controllable. If Bell-state switching protocols can be made robust against loss and imperfect postselection, EP topology may become a practical tool for routing, converting, or protecting entangled states in integrated quantum photonics.

These directions define a more stringent criterion for quantum EP applications. A useful quantum EP must specify the defective object, such as a non-Markovian generator, a Liouvillian, a scattering channel, or an effective Hamiltonian. It must also identify the observable that carries the singularity, such as photon statistics, correlation functions, Fisher information, or entanglement fidelity. Finally, it must show that the advantage remains after reservoir noise, nonlinear saturation, and postselection resources are included.

\newpage 
\section{Applications of Exceptional Points}
\label{sec: Applications of Exceptional Points}
As discussed above, EPs are non-Hermitian degeneracies at which both eigenvalues and eigenvectors coalesce. However, different operators have different EPs with distinct properties for different applications. Hamiltonian pole EPs are used in eigenvalue-response applications, such as sensing and lasing threshold engineering. Absorbing zero EPs can be used for coherent perfect absorption. Real-frequency scattering EPs are used for unidirectional reflection, angular scattering, and diffraction channel control. Jones matrix EPs can be used to manipulate polarization and wavefront. 

\subsection{Eigenvalue response: sensing and metrology}
\label{Section:sensing}

\begin{figure}[htbp]
\centering
\includegraphics[width=0.85\textwidth]{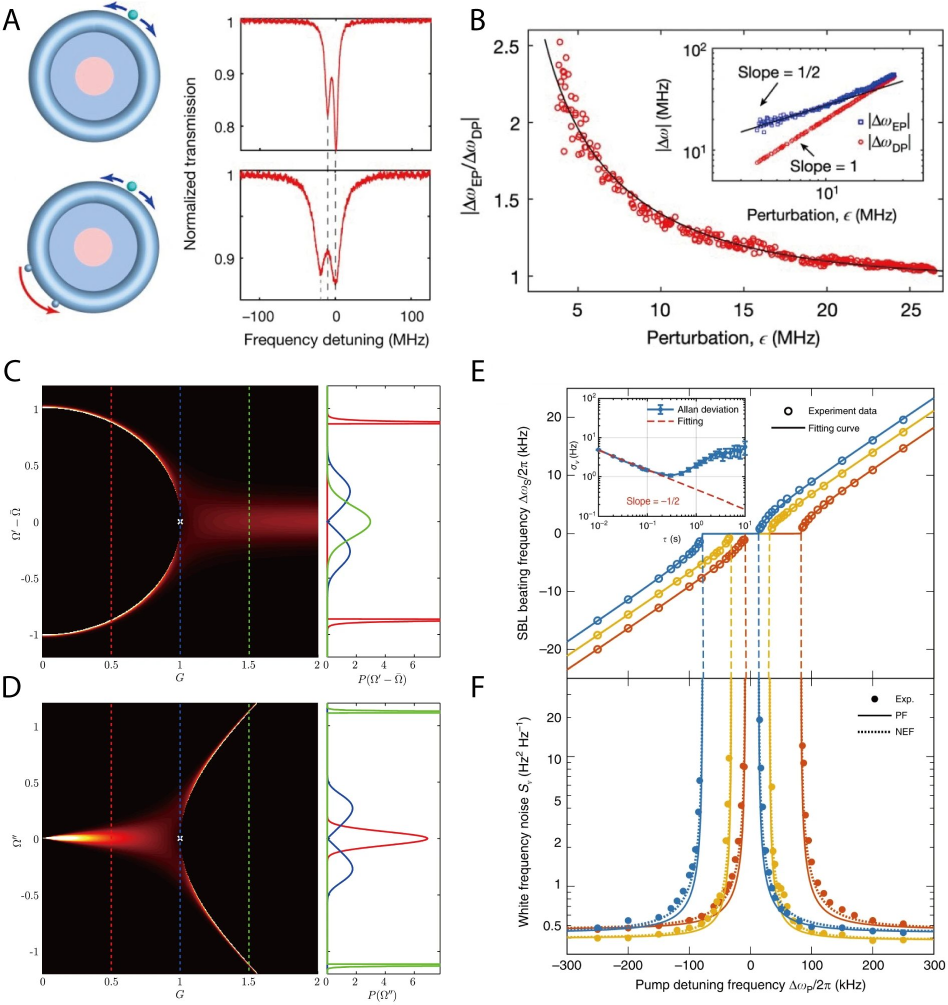}
\caption{\textbf{Enhanced spectral responsivity and noise-limited precision in EP sensing.}
(\textbf{A}) Whispering-gallery-mode microcavity sensing near a conventional degeneracy and near an EP. A weak scatterer-induced perturbation produces a much larger resolvable mode splitting when the resonator is biased at the EP, as seen from the transmission spectra~\cite{chen2017exceptional}.
(\textbf{B}) Measured ratio between EP-induced and DP-induced frequency splitting as a function of perturbation strength~\cite{chen2017exceptional}.
(\textbf{C}, \textbf{D}) Effect of unavoidable detuning fluctuations on the complex eigenfrequency distribution near an EP. The real part (\textbf{C}) and imaginary part (\textbf{D}) of the eigenfrequencies are strongly broadened as the gain--loss parameter approaches the EP, and the right panels show representative spectral distributions at different values of the control parameter~\cite{mortensen2018fluctuations}.
(\textbf{E}) Beat-frequency response of a Brillouin ring-laser gyroscope biased at different distances from the EP. The enhanced scale factor reflects the increased spectral responsivity associated with modal coalescence near the EP~\cite{wang2020petermann}.
(\textbf{F}) Measured white frequency noise of the same gyroscope. The noise increases as the system approaches the EP and follows the Petermann-factor prediction, showing that the enhanced response in (\textbf{E}) can be compensated by excess noise associated with modal nonorthogonality~\cite{wang2020petermann}.}
\label{fig:6.1_sensing}
\end{figure}

Early EP-sensor proposals used the defective spectrum as a perturbation amplifier, with the most developed examples appearing in WGM microcavities. Fig.~\ref{fig:6.1_sensing}(A) shows the canonical nanoparticle sensing configuration, in which a particle introduces an additional backscattering pathway between clockwise and counterclockwise modes. It perturbs the mode coupling and lifts the EP degeneracy~\cite{wiersig2014enhancing,wiersig2016sensors,chen2017exceptional}. Other photonic sensors probe perturbations that occur mainly as diagonal detuning, including thermal shifts \cite{zhao2018exceptional}, refractive-index changes \cite{park2020symmetry}, and Sagnac frequency shifts in ring gyroscopes \cite{lai2019observation,hokmabadi2019non}. However, the central question in EP sensing is whether enhanced spectral responsivity leads to a lower noise-normalized detection limit under a specified measurement protocol. As illustrated in Fig.~\ref{fig:6.1_sensing}(B), early work predicted a sensitivity enhancement from divergent spectral responsivity under idealized conditions near EPs~\cite{chen2017exceptional,hodaei2017enhanced}. In practice, sensor precision and efficacy depend on the signal-to-noise ratio (SNR) after accounting for system-added noise~\cite{loughlin2024exceptional}.

In classical optical sensing platforms, one major concern is operational fragility. Operation exactly at an EP usually requires the simultaneous adjustment of several control parameters. Technical noise, including fabrication imperfections, mechanical vibrations, and thermal drifts, can displace the system from the target EP. As shown in Fig.~\ref{fig:6.1_sensing}(C,D), stochastic detuning fluctuations also broaden the measured response and smear the ideal root-law splitting~\cite{mortensen2018fluctuations}. This issue has motivated robustness engineering strategies, including exceptional surfaces that relax fine-tuning requirements~\cite{zhong2019sensing,zhou2019exceptional,qin2021experimental}, anti-PT-assisted relocation of the operating EP~\cite{mao2023enhanced}, modular sensing architectures that separate the EP-control unit from the external sensor~\cite{mao2024exceptional}, and passive non-resonant EP designs with reduced sensitivity to nuisance perturbations~\cite{landers2026noise}. A second controversy concerns the excess noise associated with open-system modal nonorthogonality. Near a resonant Hamiltonian EP, the Petermann factor grows and the mode distinguishability decreases~\cite{chen2019sensitivity}, producing linewidth broadening and stronger fluctuations in frequency-based readout~\cite{langbein2018no}. This trade-off is especially clear in EP gyroscopes. As shown in Fig.~\ref{fig:6.1_sensing}(E,F), the increase in Sagnac responsivity near the EP is accompanied by Petermann-factor-induced noise that limits the attainable rotation sensitivity in Brillouin ring-laser gyroscopes~\cite{wang2020petermann}. SNR analyses of linear reciprocal sensors further indicate that the same open-system channels responsible for strong spectral response can also amplify thermal or quantum fluctuations, leading to fundamental detection limits comparable to standard resonant sensors under common readout assumptions \cite{loughlin2024exceptional}.

In quantum optical settings, the debate becomes sharper because the effective non-Hermitian Hamiltonian is only one component of a complete Lindblad description. Gain and loss channels bring quantum jumps or Langevin noise, and these additional noise operators must be included in any precision bound. An exact Lindblad treatment of driven non-Hermitian resonators further showed
that quantum noise can strongly modify the response in the $\mathcal{PT}$-symmetric regime, whereas outside this regime a stable parameter domain exists in which the noise contribution to the signal-to-noise ratio can be mitigated by increasing the external field \cite{Maksimov2025QuantumNoise}. Analyses of EP amplifying sensors have therefore tracked both the enhanced response and the amplified-spontaneous-emission or heterodyne noise associated with the same dissipative channels \cite{lau2018fundamental,zhang2019quantum,duggan2022limitations}. Quantum Fisher information studies place stronger constraints on the claimed advantage by accounting for complete information budget including post-selection probability and information carried by environmental channels \cite{ding2023fundamental,zeng2025non}. A direct experimental counterpart has been provided in a waveguide-QED platform, where a passive-PT dimer was emulated by two coupled superconducting-qubit modes, the experiment found no sensitivity enhancement near the quantum EP~\cite{almanakly2026probing}. A related positive direction is pseudo-Hermitian sensing, where a covariant quantum Fisher information formulation in a metric-deformed Hilbert space can reach the corresponding quantum Fisher bound for sensors compared at fixed dimensionality \cite{arkhipov2026achieving}.

These limitations do not imply that EP-inspired sensing is ineffective. Rather, they delineate the specific conditions under which a practical advantage may arise. EP-inspired sensing remains attractive when the dominant noise is technical, when the perturbation channel is engineered to suppress nuisance drifts, or when the readout uses a response singularity separated from exact resonant eigenbasis collapse. Representative approaches include transmission-peak degeneracies in accelerometers and dissipatively coupled gyroscopes~\cite{kononchuk2022exceptional,de2024dissipative}, CPA-EPs whose scattering singularity can be decoupled from a resonant Hamiltonian EP, enabling enhanced readout contrast in a different measurement channel~\cite{wang2026enhancement}, and broader non-Hermitian transduction schemes operating away from EPs~\cite{xiao2024non}. Nonlinear dynamics also offer a possible way to retain strong perturbation response with a finite Petermann factor~\cite{smith2022beyond}. Nonlinear EPs with a complete dynamical basis were theoretically formulated and experimentally observed in optical systems~\cite{bai2023nonlinear,bai2024observation}. Higher-order nonlinear singularities~\cite{bai2023nonlinearity}, oscillation-quenching sensors~\cite{suntharalingam2023noise}, nonlinearity-induced degeneracy lifting~\cite{li2024enhanced}, nonlinear PT phase sensing~\cite{chen2026nonlinear}, and bistable hybrid quantum systems~\cite{wang2026exceptional} further extend this sensing strategy, but the associated noise analysis requires separate treatment, since nonlinear fluctuations can shift the operating point, lower the effective EP order, and generate Bogoliubov-de Gennes noise divergences \cite{zheng2025noise}.

\subsection{EP-enabled lasers}
Lasers are intrinsically described by non-Hermitian modal dynamics, since the balance between optical gain and radiative or material loss and the intermodal coupling jointly determine the complex spectrum~\cite{liertzer2012pump}. Near an EP, the imaginary parts of the relevant eigenfrequencies can bifurcate, redistributing modal effective gain and loss and thereby reshaping which mode reaches threshold, and how effectively it suppresses the competing ones. The EP physics idea explains counterintuitive effects such as pump-induced laser self-termination~\cite{liertzer2012pump,el2014exceptional}, reversal of pump dependence~\cite{brandstetter2014reversing}, loss-induced suppression and revival of lasing in coupled resonator systems \cite{peng2014loss}, and enhancement of LDOS and quality factor \cite{othman_giant_2016}. However, once the laser operates above threshold, gain saturation and population dynamics reshape the effective operator in the nonlinear regime~\cite{benzaouia2022nonlinear}. EP-enabled chiral and directional emission form a related but functionally distinct branch~\cite{peng2016chiral}.

Recent studies explored the effect of nonlinear gain on EP-based steady-state oscillations focusing on the saturated regime and the enhanced sensitivity of the oscillation frequency to perturbations \cite{zhou2016ptSymmBreaking,bai2023nonlinearity,darcie2025responsivity, khurgin2020exceptional, Bradshaw2026Steady-State}.

\subsubsection{Single-frequency and single-spatial-mode lasers}
Laser cavities typically exceed optical wavelengths in size, allowing them to support many closely spaced modes. This leads to strong mode competition for limited gain, causing unstable and fluctuating outputs. By achieving EPs through symmetry breaking in the gain-loss distribution, either by nonuniform pumping~\cite{liertzer2012pump} or deliberate structural arrangements~\cite{feng2014single}, one can suppress unwanted modal competition either by eliminating competing longitudinal modes spectrally or by suppressing higher-order transverse modes spatially. Therefore, EPs provide a route to enhancing overall laser performance, with higher monochromaticity and superior beam quality.

\begin{figure}[htbp]
\centering
\includegraphics[width=0.85\textwidth]{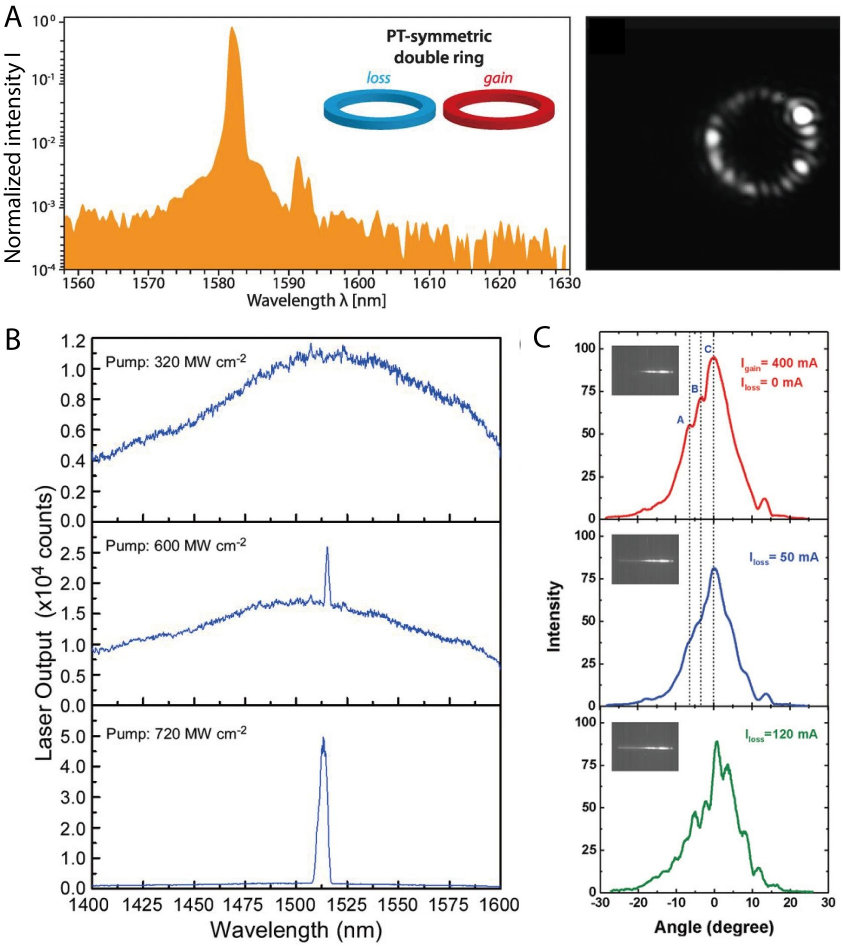}
\caption{\textbf{Threshold engineering for single-longitudinal- and single-transverse-mode EP lasers.}
(\textbf{A}) Single-longitudinal-mode lasing in a PT-symmetric coupled-microring laser. Selective gain-loss contrast drives the target supermode into the broken-PT regime while competing longitudinal resonances remain below threshold~\cite{hodaei2014parity}.
(\textbf{B}) Pump-dependent emission spectra of a PT microring laser with azimuthal gain-loss modulation. Increasing the pump density drives the device from broadband photoluminescence to amplified spontaneous emission and finally to a single WGM lasing peak~\cite{feng2014single}.
(\textbf{C}) Electrically injected single-transverse-mode operation in a PT-symmetric coupled-waveguide laser. Tuning the current in the lossy waveguide controls the gain-loss contrast and suppresses the higher-order far-field transverse mode~\cite{yao2019electrically}.}
\label{fig:6.2_laser1}
\end{figure}

For single-frequency emission, longitudinal modes arise from phase quantization along the cavity's primary propagation axis and correspond to discrete resonance frequencies. In ring or WGM cavities, the azimuthal number quantizes the phase accumulated along the circular propagation path, and plays a role analogous to the longitudinal mode index~\cite{vahala2003optical}. As illustrated in Fig.~\ref{fig:6.2_laser1}(A,B), PT-symmetric microring lasers with engineered gain and loss can select a single azimuthal or longitudinal resonance. Neighboring modes remain suppressed by non-Hermitian modal splitting~\cite{hodaei2014parity,feng2014single}. The same threshold-selection principle was later extended to integrated wavelength-tunable microring lasers, where the EP-related gain landscape can be combined with chip-scale tuning to obtain single-mode emission over a controllable spectral range~\cite{liu2017integrated}. Isolating a single longitudinal mode ensures single-frequency operation with high spectral purity, making it indispensable for high-capacity optical communications, precision spectroscopy, and microwave photonics.

Conversely, transverse modes govern the laser beam's spatial intensity profile perpendicular to the propagation axis. In circular or cylindrical cavities, they are described by radial mode numbers, which represent intensity variations from the center to the boundary. Broad-area lasers and large microring cavities can support many transverse or radial modes, which usually improves output power at the cost of beam quality. In large-area PT-symmetric laser amplifiers, unwanted modes can be driven into less favorable gain-loss branches, allowing the fundamental spatial mode to dominate~\cite{miri2012large}. Transversely multimode microring resonators implement the same idea in a compact resonator geometry, where modal overlap with the lossy region gives higher-order modes a stronger effective suppression~\cite{hodaei2016single}. The electrically injected PT-symmetric lasers~\cite{yao2019electrically} shown in Fig.~\ref{fig:6.2_laser1}(C), together with high-power quasi-PT implementations~\cite{cseker2023single}, demonstrate that this strategy extends beyond optically pumped devices to current-driven operation.

\subsubsection{Above-threshold nonlinear EP lasers and combs}

Early studies were largely confined to pump levels below or near the lasing threshold, with the discussion of lasing mode selection. However, above threshold EP lasing should be treated as a solution of a nonlinear eigenvalue problem, and it can influence how the device behaves once excited. Nonlinear EP lasing can produce characteristic kinks in output power and lasing frequency, while remaining dynamically stable when the inversion relaxation rate is sufficiently large~\cite{benzaouia2022nonlinear}.

\begin{figure}[htbp]
\centering
\includegraphics[width=0.95\textwidth]{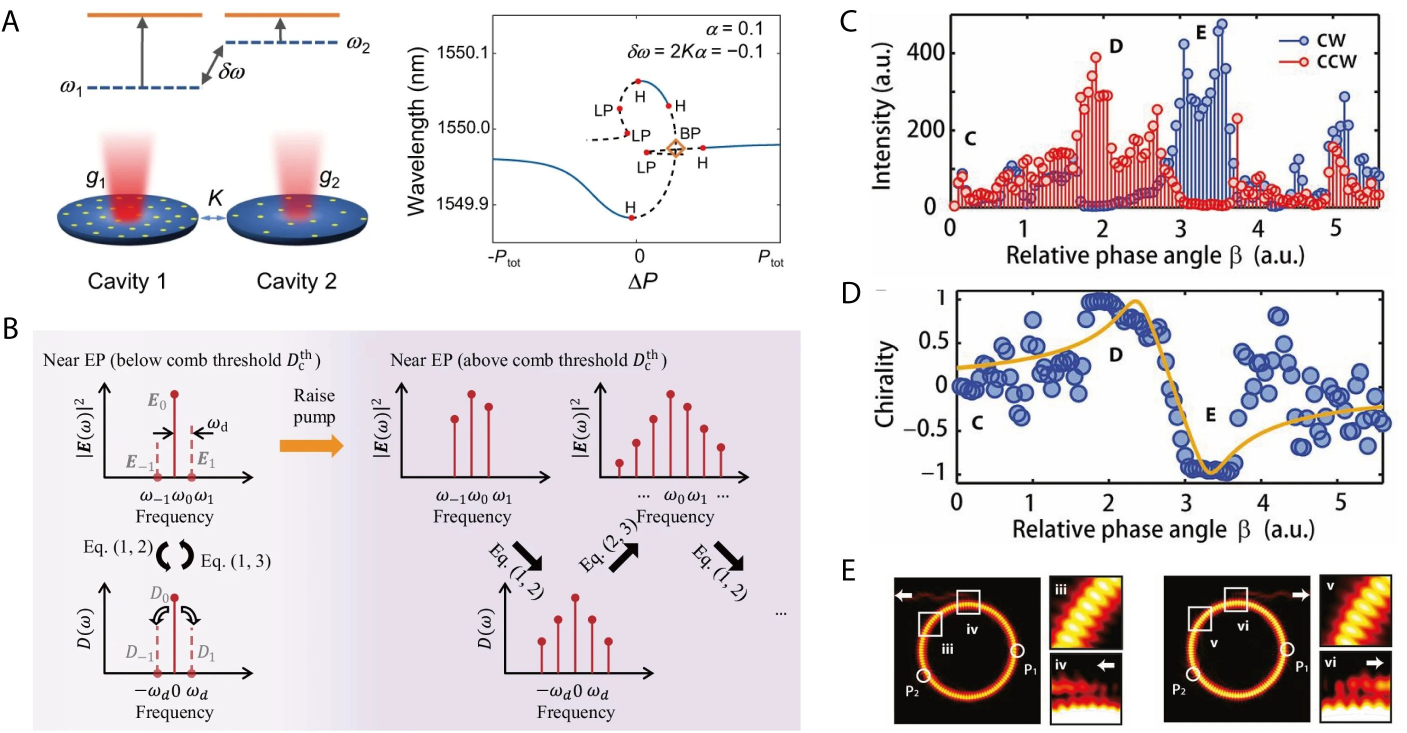}
\caption{\textbf{Above-threshold nonlinear EP lasers and chiral directional emission.}
(\textbf{A}) Above-threshold nonlinear EP in coupled semiconductor nanolasers. Pump imbalance tunes the nonlinear lasing branches and produces a branch point (BP), accompanied by Hopf bifurcations (H) and limit points (LP)~\cite{ji2023tracking}.
(\textbf{B}) EP-laser frequency-comb formation driven by dynamic gain. Increasing the pump above the comb threshold generates multiple sidebands around the lasing mode~\cite{gao2024dynamic}.
(\textbf{C}, \textbf{D}) Chiral lasing in a WGM microlaser. The relative phase angle $\beta$ between two scatterers controls the CW/CCW intensities and reverses the intracavity chirality~\cite{peng2016chiral}.
(\textbf{E}) Directional out-coupling from the same chiral WGM laser under opposite circulation states~\cite{peng2016chiral}.}
\label{fig:6.2_laser2}
\end{figure}

As shown in Fig.~\ref{fig:6.2_laser2}(A), EPs can be tracked above threshold in coupled semiconductor nanolasers by varying the total pump power, with nearby instabilities producing self-pulsing dynamics~\cite{ji2023tracking}. Programmable optical excitation in coupled nanolasers further enables control of lasing gaps far above threshold, indicating that pump landscapes can serve as active control knobs for nonlinear EP operation~\cite{fischer2024controlling}. Fig.~\ref{fig:6.2_laser2}(B) illustrates how dynamic gain generates frequency combs in EP lasers through population inversion oscillations, without requiring a separate external comb generator~\cite{gao2024dynamic}.

Several important issues remain open in current EP-lasing research. One major direction is the realization of robust on-chip integrated EP lasers. For example, an exceptional surface–tailored topological microlaser improves tolerance to parameter drift in an on-chip laser platform~\cite{liao2023chip}. In addition, threshold-order analysis also shows that the intuitive rule connecting the first lasing mode with the highest-Q passive mode can fail near EPs. A lower-Q mode may lase first when the non-Hermitian coupling and gain saturation reshape the modal competition~\cite{kullig2025exceptional}. This result gives EP lasers a broader design role in active photonics, especially where output dynamics and mode competition are more important than passive-cavity quality factors.

\subsubsection{Chiral and directional EP lasers}

Some EP-laser applications focus on eigenstate coalescence rather than threshold control. In WGM cavities, CW and CCW traveling-wave modes can coalesce into a chiral eigenstate when asymmetric backscattering is engineered by scatterers or boundary perturbations. As shown in Fig.~\ref{fig:6.2_laser2}(C-E), the resulting intracavity circulation biases resonator outcoupling, enabling directional lasing from a compact, reciprocal integrated photonic microcavity~\cite{peng2016chiral}. 

The same mechanism can be combined with angular momentum extraction. In EP-engineered microring microlasers, unidirectional intracavity circulation and angular grating outcoupling can generate vortex laser beams with designed orbital angular momentum (OAM)~\cite{miao2016orbital}. Here the EP is used to control the handedness and modal content of the lasing field, while the outcoupling structure converts that modal chirality into a structured free-space beam. This route is relevant to compact vortex sources, mode-division multiplexing, integrated beam shaping, and vectorial laser emission.

Recent coupled-nanolaser systems move this idea toward reconfigurable directionality. When two distant nanolasers are coupled through an integrated waveguide, the waveguide can mediate both frequency and loss coupling, placing the active system near an EP under suitable pumping. The emission direction can then be tuned by pump power rather than fixed solely by static scatterer placement \cite{madiot2024harnessing}. A related topological route uses exceptional-state transfer inside a laser cavity, allowing the gain medium to select modes generated by non-Hermitian state conversion \cite{schumer2022topological}.

\subsubsection{Lasing with DBE and SIP cavities}
It has been proposed that lasers can be conceived also in cavities supporting the DBE \cite{othman_giant_2016,veysi2018DBELaser,herrero2025advancements,guo2025degenerate} or the SIP \cite{herrero2023lasing,zamir2023low,furman2025impact}, whose EP properties have been explained in Sec.~\ref{sec:EP-DBE-SIP}. The introduction of small gain in such cavities perturbs the systems slightly away from their ideal EP condition though it does not modify significantly the DBE or SIP modal dispersion diagram, i.e., most of the DBE ad SIP properties are retained. These lasing principles are not based on conventional cavity concepts and present interesting properties. One of the most remarkable ones is that the lasing gain threshold scales in unusual ways with the cavity length. For example, the gain threshold in DBE-based cavities scales ideally as $N^{-5}$, where $N$ is the number of unit cells making the cavity, as shown in \cite{othman_giant_2016, veysi2018DBELaser,herrero2025advancements}. The lasing threshold in SIP-based cavities scales ideally as $N^{-3}$, as shown in \cite{herrero2023lasing,zamir2023low,furman2025impact}. Another remarkable property is that such DBE and SIP cavities do not require mirrors and they can be connected directly to a single mode waveguide; this is a consequence that the EP degeneracies are highly mismatched to terminations so any load can be attached to such degenerate multi-mode waveguides without significantly affecting the resonance frequency and the quality factor. It has been discussed and partially shown via numerical experiments that any variation of a cavity load does not affect significantly the lasing frequency and the lasing threshold \cite{veysi2018DBELaser,herrero2025advancements}. 

\subsection{Coherent absorption and waveform capture}
The absorbing EPs introduced in Sec.~\ref{subsec: 3.3} provide a device principle for controlling how coherent incident radiation is dissipated. In this setting, the relevant singularities are associated with scattering zeros, namely input states that produce no outgoing field. Conventional critical coupling corresponds to bringing a single scattering zero to the real-frequency axis. Coherent perfect absorption extends this condition to multichannel illumination, where a prescribed superposition of incident beams matches the zero eigenchannel of the scattering matrix and deposits all incident energy into the lossy region~\cite{chong2010coherent,baranov2017coherent}. The applications discussed here use the degeneracy structure of these zero channels to relax spectral, spatial, or temporal matching constraints that limit ordinary CPA.

\subsubsection{Coherent perfect absorption and arbitrary wavefront absorption}
Traditional absorbers rely on intrinsic material dissipation together with impedance matching or multiple internal reflections to suppress reradiation and lengthen the dwell time of light in the lossy region. 
In the single-mode case, perfect absorption is achieved when the radiative leakage is balanced by the non-radiative loss, whereas coherent perfect absorption (CPA) generalizes this idea to multichannel scattering, in which a prescribed coherent superposition of input waves excites a scattering zero on the real-frequency axis and is therefore completely absorbed as the time-reversed counterpart of lasing at threshold~\cite{chong2010coherent,baranov2017coherent}. However, conventional CPA requires a well-defined input wavefront and phase relation and exhibits a narrow Lorentzian absorption lineshape. 

\begin{figure}[htbp]
\centering
\includegraphics[width=0.90\textwidth]{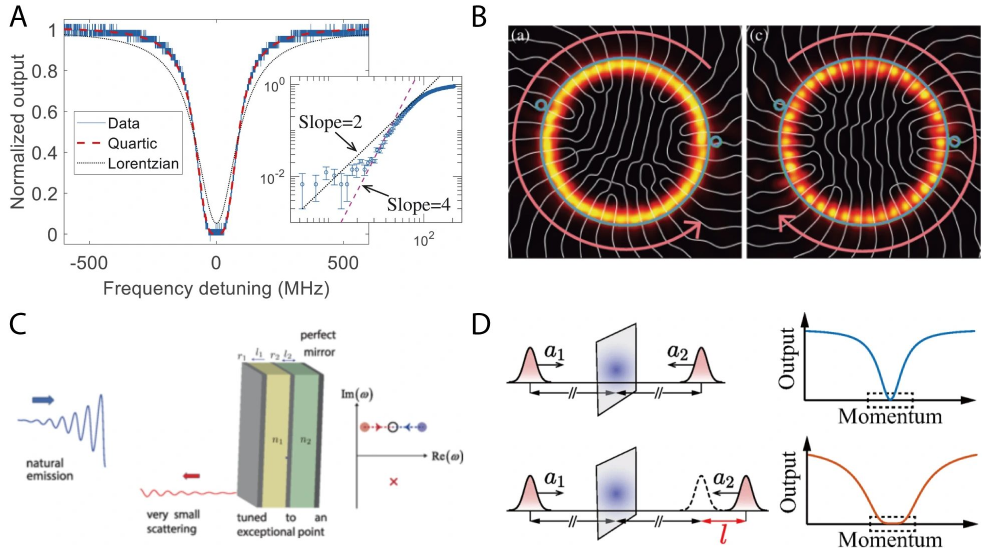}
\caption{\textbf{Coherent absorption at and beyond scattering-zero EPs.}
(\textbf{A}) CPA-EP in an optical microcavity. The normalized output exhibits a flattened quartic lineshape near perfect absorption, in contrast to the ordinary Lorentzian response~\cite{wang2021coherent}.
(\textbf{B}) Chiral perfect absorption at an absorbing EP. The coalesced incoming zero channel selects one circulation state in a disk resonator, enabling strongly direction-dependent absorption~\cite{sweeney2019perfectly}.
(\textbf{C}) Temporal waveform capture using an absorbing EP cavity~\cite{farhi2024efficient}.
(\textbf{D}) High-order coherent absorption through asynchronous inputs~\cite{xu2026high}.}
\label{fig:6.3_CPA}
\end{figure}

At a CPA-EP, two absorbing channels coalesce with their zero eigenvalue, so the zero has higher algebraic order and the residual scattering near the absorption frequency changes its local lineshape. As shown in Fig.~\ref{fig:6.3_CPA}(A), this change converts the usual Lorentzian absorption feature into a flatter quartic, or squared-Lorentzian, response, improving tolerance to frequency detuning~\cite{sweeney2019perfectly,wang2021coherent}. The same zero-channel degeneracy can also introduce chiral absorption. Fig.~\ref{fig:6.3_CPA}(B) illustrates chiral absorption when the coalesced incoming eigenchannel favors one circulation or input direction. Illumination from that direction can satisfy the CPA condition, whereas reverse illumination remains only partially absorbed~\cite{sweeney2019perfectly}. The main practical problem is the stability of operating at an isolated EP. Exceptional surfaces address this issue by replacing a discrete degeneracy with a continuous manifold of absorbing EPs. A waveguide-coupled resonator with engineered directional feedback has demonstrated chiral and degenerate perfect absorption on such an exceptional surface, together with a squared-Lorentzian absorption lineshape~\cite{soleymani2022chiral}.

The next obstacle for coherent absorption is spatial matching. A small mismatch in transverse profile can reroute energy into outgoing channels, even when the frequency and material loss are correctly tuned. By combining the spectral degeneracy of an absorbing EP with a massively degenerate cavity, CPA can be achieved for arbitrary incident wavefronts within the supported angular and spatial acceptance~\cite{horner2024coherent}. The wavefront-insensitive absorbers have potential use in imaging-compatible light harvesting and multimode photonic circuits with complex-wavefront control.

\subsubsection{Temporal waveform capture and photon catching}
A complementary limitation appears in the time domain. Perfect capture by a passive cavity usually requires the incoming waveform to match the time reverse of the cavity emission, which corresponds to an exponentially rising envelope for a simple decaying resonance. Naturally emitted photons, pulsed signals, and wave packets generated in realistic photonic circuits often have temporal profiles that deviate from this ideal shape, producing partial reflection or incomplete state transfer.

An absorbing EP modifies this temporal matching condition by increasing the order of the absorbing zero. As illustrated in Fig.~\ref{fig:6.3_CPA}(C), the cavity response captures additional temporal orders of a general incoming waveform. Imperfect wave packets therefore scatter less strongly than in a simple critically coupled cavity~\cite{farhi2024efficient}. This concept is promising for passive state transfer, single-photon detection, and efficient capture of spontaneously emitted photons, where active waveform shaping is technically costly or introduces additional noise. However, this application remains theoretical, with open questions about optical-frequency integration, and compatibility with single-photon coherence requirements.

\subsubsection{Nonlinear and high-order coherent absorption}
Nonlinear coherent absorption provides another route to adaptive matching. In nonlinear resonators, the scattering zeros depend on the incident intensity as well as on the relative phase and amplitude of the input beams. A microwave resonator experiment showed that coherent excitation can drive the system into a self-induced near-perfect absorption state in the proximity of scattering-zero EPs~\cite{suwunnarat2022non}.

High-order absorbing EPs extend the same zero-coalescence idea to several incoming channels or modes. Their flatter absorption minima can improve bandwidth and, when used as a sensing readout, can separate the absorption singularity from the resonant Hamiltonian EP that often carries strong modal-noise penalties. High-order CPA-EP sensing has recently been used to improve signal-to-noise performance in passive cavity-magnonic systems~\cite{wang2026enhancement}. Fig.~\ref{fig:6.3_CPA}(D) presents an EP-free route to high-order perfect absorption, where asynchronous coherent inputs provide momentum-dependent phase control and reshape the absorption lineshape without coalescing scattering zeros~\cite{xu2026high}. In selected geometries, this approach may offer greater tolerance or simpler tuning.

\subsection{Directional asymmetric scattering at scattering EPs}
\label{sec:Unidirectional characteristics at scattering EPs}
The scattering response-matrix EPs introduced in Sec.~\ref{subsec: 3.4} provide a method for directional wave control at a fixed real frequency. In the present section, the emphasis is placed on reciprocal directional scattering, one-sided reflection cancellation, angularly asymmetric diffraction, and nanoscale directional launching.

\begin{figure}[htbp]
\centering
\includegraphics[width=0.95\textwidth]{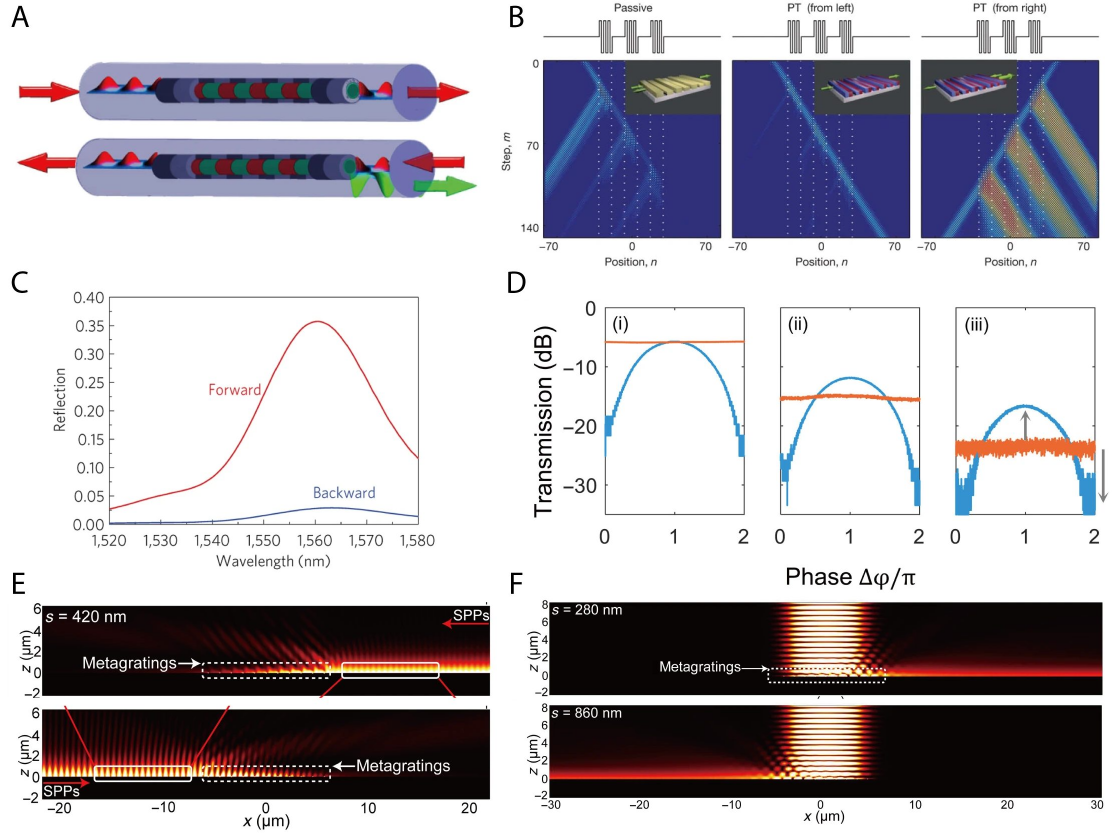}
\caption{\textbf{Directional asymmetric scattering at response-matrix EPs.}
(\textbf{A}) Unidirectional invisibility in a PT-symmetric Bragg grating with longitudinal gain-loss modulation~\cite{lin2011unidirectional}.
(\textbf{B}) Unidirectional invisibility in a synthetic PT photonic lattice near a scattering EP~\cite{regensburger2012parity}.
(\textbf{C}) One-sided reflection suppression in a PT-symmetric optical metamaterial, showing strongly asymmetric reflection for opposite incidence directions~\cite{feng2013experimental}.
(\textbf{D}) Phase-controlled asymmetric transmission in a non-Hermitian metawaveguide~\cite{zhao2016metawaveguide}.
(\textbf{E}, \textbf{F}) EP-assisted subwavelength control of surface-plasmon-polariton (\textbf{E}) reflection and (\textbf{F}) excitation using non-Hermitian metagratings~\cite{xu2023subwavelength}.}
\label{fig:6.4_unidirection}
\end{figure}

\subsubsection{Unidirectional reflectionless and invisibility}
For reciprocal scalar two-port systems, Lorentz reciprocity constrains the transmission amplitudes, while the reflection channels can still be highly asymmetric. Eq.~\ref{eq: permuted scattering matrix} expresses this response by rewriting the two-port permuted scattering matrix. The earliest EP-based directional scattering proposals were developed in longitudinally modulated PT-symmetric media. In a complex Bragg grating, the real and imaginary parts of the refractive index are arranged along the propagation direction so that Bragg reflection from one side is cancelled while reflection from the opposite side remains finite or is enhanced. As illustrated in Fig.~\ref{fig:6.4_unidirection}(A), unidirectional invisibility occurs when the transmission amplitude approaches unity and its phase approaches that required through the background medium~\cite{lin2011unidirectional}. This mechanism directly links a defective scattering eigenchannel to a practical optical function. A structure can be almost invisible for one incidence direction and reflective for the reverse direction.

The experimental development followed a clear path from synthetic and integrated photonic platforms toward planar and nanophotonic implementations. The time-multiplexed fiber-loop network shown in Fig.~\ref{fig:6.4_unidirection}(B) realizes a PT-symmetric synthetic photonic lattice with one-way reflectionless behavior. Distributed attenuation and optical amplification emulate the required gain-loss contrast~\cite{regensburger2012parity}. The integrated silicon waveguide shown in Fig.~\ref{fig:6.4_unidirection}(C) achieves optical-frequency unidirectional reflectionlessness through interference between guided-mode Bragg scattering and distributed loss from periodically arranged $\ce{Ge}$-$\ce{Cr}$ absorptive sections~\cite{feng2013experimental}. The metawaveguide architecture shown in Fig.~\ref{fig:6.4_unidirection}(D) exploits the same directional interference to enable asymmetric interferometric light-light switching~\cite{zhao2016metawaveguide}. Large-area multilayers further showed that single-sided reflection cancellation can also be implemented in thin-film coatings, where layer thickness and material absorption control the phase and amplitude of the two reflection pathways~\cite{feng2013demonstration}.

Recent work has moved this functionality toward tunability and subwavelength integration. Spatially chirped Floquet PT-symmetric waveguides on chip use a longitudinally varying dissipative modulation to produce reconfigurable forward and backward transmission responses near non-Hermitian phase transitions~\cite{mao2025chip}. Fig.~\ref{fig:6.4_unidirection}(E,F) shows how non-Hermitian metagratings apply the same response-matrix principle to enable unidirectional excitation and reflection of surface plasmon polaritons within a subwavelength footprint~\cite{xu2023subwavelength}. Mid-infrared graphene plasmonic metasurfaces offer an electrically tunable material platform for one-sided reflection cancellation~\cite{li2023analytical}, while plasmonic systems supporting an EP-induced Huygens dipole provide a route to switchable unidirectional radiation in deeply subwavelength emitters~\cite{moritake2023switchable}. These developments shift the application target from proof-of-principle one-way invisibility toward integrated routing, plasmonic launching, compact antennas, and active directional emitters.

\subsubsection{Angular retroreflection}
Directional scattering EPs are not restricted to one-dimensional two-port geometries. In free space, a phase-gradient metasurface can map different illumination angles into different diffraction channels, and engineered loss can suppress one of these channels while preserving the opposite retroreflection channel. The relevant response matrix is then written in a diffraction-channel basis, as shown in Eq.~\ref{eq: diffraction scattering matrix}. The resulting device can retroreflect efficiently for one incident angle while suppressing the reverse angular channel through destructive interference and absorption.

This angular response was first developed in the broader context of loss-assisted phase-gradient metasurfaces, where localized dissipation was used to create extreme angular asymmetry in reflection and absorption~\cite{wang2018extreme,li2022controlling, cuesta2022}. The EP interpretation became explicit in acoustic and electromagnetic metasurface studies, where the coalescence of diffraction-channel eigenvectors was connected to the suppression of one retroreflection coefficient~\cite{wang2019extremely,dong2020loss}. It allows the metasurface to operate as an angularly selective mirror, one-way retroreflector, or directional absorber while remaining planar and passive. A visible-frequency implementation based on a bilayer metagrating has brought this concept closer to nanophotonic applications, where one incidence direction becomes retroreflected with high efficiency and the opposite direction is strongly absorbed~\cite{he2023scattering}. This separation of diffractive and dissipative functions is an important design principle for future scattering EP devices, because it allows high contrast without sacrificing all useful output power to material loss.

Directional scattering devices need large angular contrast, robustness to fabrication tolerances and operation over useful spectral and angular bandwidths. Tunable loss elements, phase-change materials, graphene, MEMS actuation, or spatiotemporal modulation could add active control, while multilayer dielectric architectures may reduce parasitic absorption. For applications requiring optical isolation, scattering EPs should be combined with genuine reciprocity-breaking mechanisms.



\subsection{Jones EP metasurfaces for polarization and wavefront control}
\label{sec:Polarization selection at Jones EPs}
The Jones response-matrix EPs introduced in Sec.~\ref{subsec: 3.5} provide a polarization-space approach to non-Hermitian wavefront engineering. When two eigenpolarizations coalesce, the metasurface acquires a singular polarization channel. One cross-polarized conversion coefficient can vanish while the opposite conversion channel remains finite. This singular channel allows a specific polarization state to be selected and mapped into a designed wavefront. Reflection and transmission can convert polarization singularity into a programmable optical function, and their distinction mainly determines which external channel is used to collect the signal. Reflection-type Jones EP metasurfaces have demonstrated nearly one-way circular-polarization conversion, where a selected handedness is converted with a finite amplitude while the reverse cross-conversion channel is driven to zero~\cite{song2021plasmonic}. Transmission-type Jones EPs provide the corresponding functionality for transmitted fields, enabling polarization-selective transmission~\cite{park2020observation}.

\subsubsection{Exceptional topological phase for wavefront engineering}
\begin{figure}[htbp]
\centering
\includegraphics[width=0.85\textwidth]{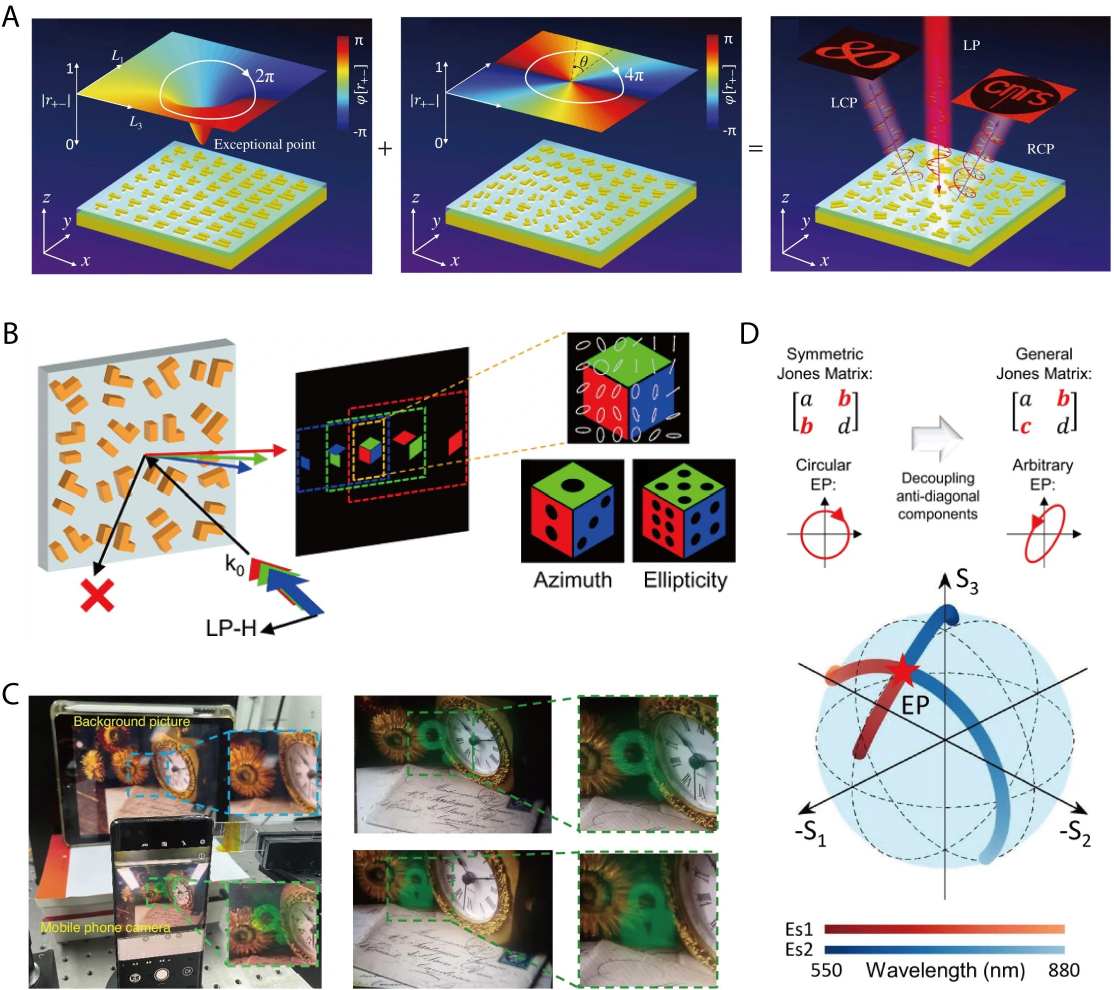}
\caption{\textbf{Jones EP metasurfaces for topological phase and vectorial wavefront control.}
(\textbf{A}) Exceptional topological phase generated by encircling a Jones EP and its combination with geometric phase for spin-dependent wavefront engineering~\cite{song2021plasmonic}.
(\textbf{B}) Arbitrary-polarization Jones EPs for asymmetric vectorial meta-holography, where intensity, azimuth, and ellipticity information are encoded into the output field~\cite{yang2024creating}.
(\textbf{C}) On-chip all-dielectric Jones EP metasurface for integrated holographic and augmented-reality wavefront projection~\cite{yi2025metasurfaceEP}.
(\textbf{D}) Single planar metasurface supporting arbitrarily polarized Jones EPs on the Poincar\'e sphere~\cite{qin2025sphere}.}
\label{fig:6.5_polarization}
\end{figure}

Encircling a Jones EP in metasurface design space creates an exceptional topological phase in a selected polarization-conversion channel. This phase can span a full 2$\pi$ range while the channel amplitude is controlled by the proximity to the singularity. As shown in Fig.~\ref{fig:6.5_polarization}(A), combining the exceptional topological phase with geometric-phase design enables independent phase programming across circular-polarization channels for compact vectorial wavefront control~\cite{song2021plasmonic}. 

This principle has already moved beyond a single polarization-conversion demonstration. Exceptional topological phase coding metasurfaces use the EP-induced phase winding to independently address circularly polarized wavefronts, enabling spin-resolved phase patterns with high channel contrast~\cite{li2024independent}. Full-color vectorially encrypted meta-holography extends the same mechanism to wavelength-dependent image formation, where the EP-controlled polarization channel supplies an asymmetric vectorial degree of freedom for multiplexed holographic reconstruction~\cite{yang2024asymmetric}. Reversible and polarization-independent wavefront control further shows that the EP-induced phase can be combined with additional phase mechanisms to reduce the dependence on a fixed incident spin state~\cite{hu2024dynamic}. These developments make Jones EP metasurfaces especially relevant for compact holography, display optics, polarization-multiplexed imaging, and spin-dependent beam shaping.

A major recent step is the transfer of this concept from plasmonic free-space platforms to integrated dielectric photonics. The on-chip device shown in Fig.~\ref{fig:6.5_polarization}(C) modulates the topological phase around a Jones EP, decoupling orthogonal circular polarizations to project independent, switchable "Key" and "Lock" virtual images~\cite{yi2025metasurfaceEP}. Captured directly in a real-world environment, these holographic images float over the background with suppressed zero-order diffraction noise, demonstrating a highly efficient, compact architecture for next-generation displays and information storage.

\subsubsection{Arbitrary-polarization EPs on the Poincar\'e sphere}
Most early Jones EP metasurfaces were designed around circular polarization, so the coalesced eigenstates occupied the poles of the Poincar\'e sphere. This restriction has recently been lifted. Fig.~\ref{fig:6.5_polarization}(B) shows pairs of EP created for arbitrary fully polarized states, enabling asymmetric vectorial wavefront modulation beyond the left- and right-handed circular basis~\cite{yang2024creating}. In this regime, the target eigenpolarization can be chosen as a design variable, and the metasurface can address elliptical or linear polarization states with the same singular-response strategy previously used for circular polarization.

The broader implication is that a Jones EP can become a programmable point on the Poincar\'e sphere. As illustrated in Fig.~\ref{fig:6.5_polarization}(D), a single planar metasurface supports arbitrarily polarized EPs whose coalesced eigenpolarizations span the Poincar\'e sphere rather than remaining pinned to circular polarization states~\cite{qin2025sphere}. Transmission platforms are following a similar trajectory. Ellipticity-controlled EPs in multilayer silicon guided-mode resonant metasurfaces provide cross-polarized phase singularities for transmitted light and expand Jones EP control to arbitrary ellipticity~\cite{goldberg2025ellipticity}. These advances change the application landscape from spin-selective optics to general polarization-state engineering, with direct relevance to polarization-multiplexed holography, compact polarimetry, vectorial beam synthesis, and polarization-encoded optical information processing.

The most promising future devices will likely exploit both dimensions of the Poincar\'e sphere and real-space wavefront simultaneously. In such a device, each spatial position of the metasurface could assign a local coalesced eigenpolarization and an associated topological phase, producing a spatially varying map from incident polarization to output wavefront. This concept would allow Jones EP metasurfaces to function as compact vectorial processors rather than fixed circular-polarization filters.

\subsubsection{Novel engineered Jones EP metasurfaces}
Static Jones EP metasurfaces already demonstrate strong polarization selectivity, but reconfigurable operation is required for modulators, adaptive optics, and dynamically switchable holography. Electrical tuning provides one route. A MEMS-based non-Hermitian metasurface has realized voltage-controlled topological phase transitions, allowing the system to move between Jones EPs of opposite handedness and thereby switch the polarization response without fabricating a new sample~\cite{ding2024electrically}. Graphene-based terahertz metasurfaces offer another electrically tunable platform, where the conductivity of graphene adjusts the non-Hermitian polarization response and allows active control of chiral degeneracy~\cite{baek2023non}.

Optical control has recently enabled faster switching. Photoinduced loss in hybrid terahertz metasurfaces can generate transient non-Hermitian degeneracies, providing ultrafast access to polarization-selective EP responses~\cite{he2023transient}. Loss-enabled chirality inversion further shows that the handedness of a Jones EP can be switched in situ by controlling dissipative channels, giving picosecond-scale control of the chiral response~\cite{he2025loss}. Anti-chiral EP metasurfaces extend this active-control strategy to polarization-selective terahertz switching with large modulation depth~\cite{yu2024creating}. These results place Jones EPs in the functional space of active flat optics, where the target is no longer only a high-contrast static polarization converter, but a fast and reconfigurable polarization-wavefront element.

For flat-optics applications, the most competitive Jones EP devices will be those that combine arbitrary-polarization control with high optical efficiency and active tuning, preferably in dielectric or hybrid platforms compatible with photonic integration.

\subsection{Emerging applications beyond canonical EP functionalities}
Beyond sensing, lasing, coherent absorption, directional scattering, and Jones-matrix wavefront control, recent work is moving EP physics toward device-oriented photonic and electromagnetic functionalities. In these emerging EP directions, the relevant performance metrics attract more attention. Routing devices are judged by conversion efficiency and crosstalk, nonlinear processors by switching energy and conversion efficiency, storage devices by delay-bandwidth product and retrieval fidelity, and quantum photonic platforms by state fidelity, photon statistics, and decoherence tolerance.

\subsubsection{Integrated mode converters and routers}
Integrated photonic circuits require compact elements that can transform one optical mode into another with high selectivity. The EP-driven guided-wave system can be steered through a designed trajectory in parameter space so that an input state evolves toward a selected output mode or polarization channel.

Early encircling-based devices faced a tradeoff between modal selectivity and propagation loss, because slow evolution improves state discrimination while increasing attenuation. Hamiltonian-hopping protocols reduced this constraint by constructing segmented evolution paths that preserve chiral transfer with a shorter device length~\cite{li2020hamiltonian}. Fast-encirclement designs pushed the same idea toward compact integrated mode converters, where the EP trajectory is chosen to preserve conversion efficiency without relying on a fully adiabatic loop~\cite{shu2022fast}. Riemann-encircling devices further connected EP monodromy with polarization-locked conversion, allowing asymmetric transformation between polarization-defined channels~\cite{li2022riemann}.

A recent direction is reconfigurable routing. Hybrid silicon non-Hermitian switching uses EP-assisted state transfer to control routing between integrated photonic pathways~\cite{feng2025non}. It shows that the EP trajectory improves the device metric compared with conventional adiabatic interferometric switches or mode multiplexers of similar footprint and loss.

\subsubsection{Nonlinear frequency conversion and all-optical modulation}
Nonlinear photonic applications use EPs to reshape the optical field distribution and the phase relation between interacting waves. The target is enhanced conversion, modulation, or signal processing at lower pump power or shorter interaction length. This use of EPs is closely connected to active and resonant platforms, where the same gain-loss or radiative-coupling landscape that creates the EP can also intensify nonlinear response. Optical nonlinearity can also provide a tuning mechanism for the exceptional degeneracy itself: in coupled ring resonators, Kerr-induced frequency shifts can compensate an initial resonance mismatch and restore second- and third-order EP conditions \cite{Ramezanpour2021Kerr}.

Frequency conversion near non-Hermitian phase boundaries has shown that PT-symmetric modal control can increase second-harmonic generation efficiency by maintaining favorable modal overlap and phase evolution in the nonlinear process~\cite{hou2024enhanced}. In optical signal processing, parity-time symmetry has enabled highly efficient nonlinear transformations with reduced energy cost, suggesting a route toward compact wavelength conversion and all-optical computing primitives~\cite{kim2024parity}. Exciton-polariton perovskite metasurfaces provide a complementary ultrafast regime, where operation near an EP produces a large transient modulation response suitable for femtosecond all-optical control~\cite{masharin2024giant}. Near-exceptional coupling can also strongly reshape nonlinear steady-state response: a compact photonic-crystal microcavity with two ultrahigh-$Q$ resonances approaching an EP exhibited thermo-optical tristability and enabled a proof-of-concept three-state optical memory \cite{Liu2026Multistability}.

For practical optical processing, the most competitive devices will likely combine an EP-enhanced response with a material platform that already supports strong optical nonlinearity, rather than relying on the EP alone to compensate for weak material response.
\subsubsection{Slow light for optical storage}
Slow-light and optical-storage applications exploit the strong dispersion and modified temporal dynamics that arise near EPs. The desired function is to delay, hold, and release an optical pulse in a compact resonant system while preserving pulse integrity. This direction is especially relevant for optical buffering, synchronization, and chip-scale temporal signal control, where long delay lines are difficult to integrate.

Light storage near an EP has recently been realized by using the modified modal dynamics around the exceptional degeneracy to slow and recover optical pulses in a compact cavity platform~\cite{zhu2024storing}. It converts EP dispersion into a time-domain functionality. An incoming pulse can be temporarily stored and then released by controlling the system near EP. Compared with ordinary slow-light resonances, EP-based storage offers a sharper temporal response and a tunable route to delay control.

The remaining challenge is to move from proof-of-principle delay to a storage element with competitive system-level performance. Future EP-based storage can use an actively tunable EP buffer that can be integrated with modulators, detectors, and nonlinear processors on the same photonic chip.
\subsubsection{Nonreciprocal photonics near EPs}
Strict nonreciprocal photonics requires a physical mechanism that breaks Lorentz reciprocity, such as dynamic biasing, nonlinear operation or active modulation. EPs can enhance this class of devices by amplifying direction-dependent state selection, increasing modal contrast, or widening the operating bandwidth around a designed non-Hermitian trajectory. 

A representative EP-assisted route is broadband on-chip nonreciprocity based on non-Hermitian dynamical evolution, where the device mimics nonlinear anti-adiabatic quantum jumps near an EP and produces direction-selective transmission over a broad optical band~\cite{choi2017extremely}. Chiral EPs in microresonators provide another route in which active tuning modifies the coupling between clockwise and counterclockwise modes, allowing enhanced control of asymmetric response in a compact resonant platform~\cite{lee2025chiral}.

\subsubsection{Quantum photonic functionalities}
Quantum photonic applications use EPs to control open-system dynamics, photon statistics, and transient state transfer. Exceptional photon blockade illustrates how chiral EP physics can be used to engineer nonclassical emission. By tailoring the non-Hermitian coupling of a nonlinear photonic system, the presence of one photon suppresses the transmission or generation of subsequent photons, providing a route toward single-photon sources with enhanced blockade behavior~\cite{huang2022exceptional}. Liouvillian EPs can shape the relaxation spectrum of the density matrix and can generate transient chiral state transfer in photonic platforms~\cite{gao2025photonic}. Non-Markovian quantum EPs show that memory effects in the environment can create additional or higher-order exceptional degeneracies inaccessible in a purely Markovian model~\cite{lin2025non}. This opens a route to quantum photonic devices in which the reservoir is treated as a design component that controls the exceptional spectrum. Quantum heat engines based on Liouvillian EP encircling provide another indication that EPs may influence energy exchange and thermodynamic cycles in open quantum devices~\cite{bu2023enhancement}.

\newpage 
\section{Conclusion and Outlook}
\label{sec: Conclusion and Outlook}

Today EP photonics has moved far beyond its early connection with balanced gain and loss. EPs can now be engineered through radiative leakage, material absorption, asymmetric coupling, and time modulation, and they have been realized in waveguides, open resonators, coherent absorbers, metasurfaces, periodic systems, and open quantum platforms. This progress shows the generality of EP physics, but it also shows why a single universal description is not sufficient. Although every EP is a defective degeneracy of a non-Hermitian operator, the relevant operator depends on the physical problem. The same structure may support distinct EPs in Hamiltonians, S matricies, incoming-wave operators, Jones matrices, Bloch and Floquet operators, or a Liouvillian. A unified understanding of EP photonics therefore requires identifying not only the degeneracy itself, but also the operator in which it occurs and the physical observables governed by that operator. Accordingly, the experimental signatures of an EP are not universal, but depend on the underlying operator, the excitation and detection scheme, and the observables accessible in a given platform. 

In this view, one of the most important challenges is the experimental observation of spectral EPs. Resonant EPs are defined by the coalescence of complex poles under outgoing boundary conditions, while absorbing EPs are defined by complex zeros under incoming boundary conditions. Most optical experiments measure reflection or transmission spectra along the real-frequency axis. Such spectra do not give direct access to the complex eigenvalues. Peaks and dips can be shifted or hidden by interference, background scattering, and overlapping modes. The reconstruction of a spectral EP therefore requires phase-sensitive measurements, a reliable model of the relevant channels, or an analytic continuation of the measured $S$ matrix into the complex-frequency plane. Time-domain decay and modal-field measurements can provide further evidence. The situation is different for a scattering EP defined directly in the real-frequency $S$ matrix. It can be observed by measuring the matrix elements of $S$ and following the coalescence of its eigenvalues and eigenvectors. This experimental distinction is especially important for sensing. The root-law splitting of complex eigenvalues gives enhanced responsivity, but it does not by itself give better measurement precision. Mode nonorthogonality, linewidth broadening, gain noise, quantum jumps, thermal drift, and limited measurement time can reduce or remove the apparent advantage. A useful comparison should fix the input power, measurement time, detection efficiency, and dynamic range. It should then evaluate the signal-to-noise ratio, estimation error, or Fisher information rather than eigenvalue splitting alone. EP sensing remains promising when the readout channel gives high contrast without requiring two nearly merged resonances to be resolved. Modular sensing structures, phase and waveform measurements, and systems dominated by technical noise are particularly interesting. We believe these routes offer a more realistic path than operating as close as possible to an isolated resonant EP.

Another major challenge is tuning and stabilization under realistic conditions. An isolated EP may require several frequencies, loss rates, and coupling coefficients to be matched at the same time. Fabrication disorder, temperature drift, pump fluctuations, and material aging can move the system away from the desired point. Exceptional surfaces and symmetry-constrained manifolds can reduce the number of parameters that must be controlled. Post-fabrication trimming, MEMS actuation, carrier injection, phase-change materials, and optical pumping can provide active tuning. Practical devices will probably operate near a calibrated EP region and use feedback to maintain the required response. The noise, speed, and energy cost of this feedback must be included in the device assessment. A related open question is whether a nonlinear system can create and stabilize its own EP. Intensity-dependent refractive index, gain saturation, or carrier redistribution may move the system toward an EP without an external control loop. It remains unclear when such a self-induced EP is stable and reproducible, and when it is only a nonlinear bifurcation with similar spectral features.

In our assessment, the most promising near-term applications are integrated mode conversion, polarization and wavefront control, laser mode selection, and coherent absorption. Guided-wave EP devices can select an output mode or polarization in a compact structure, although propagation loss and conversion efficiency remain important limits. Jones EP metasurfaces can combine polarization selection with phase control. Access to arbitrary states on the Poincar\'e sphere makes them attractive for vectorial holography, compact polarimetry, and polarization-encoded optical processing. This direction is especially promising in low-loss dielectric platforms with electrical or optical tuning. Lasers are also natural EP systems because gain, loss, and mode competition are already part of their operation. EP engineering can select the lasing frequency, spatial mode, circulation, and emission direction. Above threshold, however, the relevant EP must be studied in the nonlinear saturated system rather than inferred from the passive cavity. Absorbing EPs can support coherent absorption and waveform capture with improved tolerance to frequency, spatial profile, or pulse shape.

EPs can also enhance nonreciprocal photonic response, but an important distinction must be kept. An EP alone does not break Lorentz reciprocity. A reciprocal EP structure can show strongly asymmetric reflection, transmission, or polarization conversion without being a true isolator. Genuine nonreciprocity requires an additional mechanism such as spatiotemporal modulation, nonlinear biasing, or magneto-optical coupling. When such a mechanism is present, operation near an EP can strengthen direction-dependent state selection, increase isolation contrast, and in some designs broaden the useful bandwidth. This combination is promising for compact isolators, circulators, and nonreciprocal mode converters. Its value should be judged by isolation, insertion loss, bandwidth, power dependence, and stability rather than by asymmetry alone.

Non-Hermitian band topology and hybrid EP singularities provide a longer-term direction. The main challenge is to connect complex band structures with measurements in finite and disordered samples. Bulk Fermi arcs, polarization half charges, non-Bloch spectra, and mode braiding need experimental protocols that clearly specify the boundary conditions and the measured states. EP-BIC systems require similar care. An exact EP-BIC must be both defective and radiatively dark in the same modal subspace. A high total quality factor or a nearby quasi-BIC is not enough. The most useful opportunity is the independent control of radiative and intrinsic loss. Such control may give strong field buildup and a tunable non-Hermitian response without optical gain. We expect these systems to become important platforms for radiative-channel engineering, nonlinear optics, and narrowband light control.

Time modulation and nonlinear response may make EP devices more programmable. Floquet platforms allow the EP and the encircling path to be changed without fabricating a new structure. Nonlinear materials can support switching, frequency conversion, photon blockade, and frequency comb generation near an EP. They also bring saturation, multistability, noise, and dynamical instabilities. The linear EP must therefore be connected to the stable nonlinear states and fluctuation spectrum of the driven device. Inverse design may help control geometry, radiation channels, modulation, and material dispersion at the same time. The useful target is not the EP condition alone, but a measurable improvement in efficiency, speed, bandwidth, or energy use.

Quantum photonics is the most demanding and potentially the most far-reaching direction. HEPs describe conditioned no-jump evolution, while LEPs describe the full density-matrix dynamics. Postselection, detector loss, decoherence, and discarded trajectories must be included when sensing, state transfer, or entanglement control is evaluated. LEP chirality is often a finite-time effect that disappears during relaxation, which makes the preparation and readout window part of the protocol. Non-Markovian reservoirs add memory and retardation as new control parameters, but they also make the generator harder to define and reconstruct. We regard few-photon nonlinear resonators, integrated quantum walks, and engineered reservoirs as the most useful test platforms in the near future. They can show whether EP physics gives a reproducible improvement in photon statistics, state fidelity, or transient control. Large-scale quantum processing based on EPs remains a more distant goal.

We expect the field to move from isolated demonstrations toward tunable systems in which the EP, excitation, and readout are designed together. Foundry-compatible fabrication, active calibration, time modulation, nonlinear control, and reservoir engineering can turn a fragile spectral point into a useful operating region. The strongest experiments will compare an EP device with the best non-EP design under the same conditions and will report noise, efficiency, stability, and resource cost together with the spectrum. If these standards are met, EPs can become practical tools for mode and polarization control, laser engineering, coherent capture, enhanced nonreciprocity, nonlinear processing, and selected quantum operations. Their main value will not be singular response by itself. It will be the broader ability to control the spectra and dynamics of open photonic systems.

\section{Acknowledgments}
\label{sec: acknowledgements}

The authors acknowledge useful comments and suggestions from Yuri Kivshar, Shuang Zhang, Yongsop Hwang, and Mihail Petrov. F.Z. acknowledges support from China Scholarship Council program. C.T.C. acknowledges support from RGC Hong Kong (AoE/P-502/20). A.B. acknowledges the Russian Science Foundation (project \#25-42-10025), National Natural Science Foundation of China (W2532010), and the academic leadership program Priority 2030.


\newpage
\makeatletter
\renewcommand{\bibfont}{\footnotesize}  
\makeatother
\setlength{\bibsep}{2pt}  

\bibliographystyle{elsarticle-num}
\bibliography{references-AB-corr}
\FloatBarrier
\clearpage

\end{document}